\documentclass[aip,graphicx]{revtex4-1}

\usepackage{url} 
\usepackage{lineno}
\usepackage{soul}
\usepackage{graphicx}
\usepackage{colortbl}
\usepackage{color}
\usepackage{amssymb}
\usepackage{amsmath}
\usepackage{caption}

\newcommand{\jep}{\vec{J}\cdot\vec{E}'}

\newcommand{\Se}{\mathrm{S_e}}

\draft 

\begin{document}


\title{On the origin of ``patchy'' energy conversion in electron diffusion regions} 



\author{Kevin J. Genestreti}
\email[]{kevin.genestreti@swri.org}
\affiliation{Southwest Research Institute, Durham, NH, USA}

\author{Xiaocan Li}
\affiliation{Dartmouth College, Hanover, NH, USA}

\author{Yi-Hsin Liu}
\affiliation{Dartmouth College, Hanover, NH, USA}

\author{James L. Burch}
\affiliation{Southwest Research Institute, San Antonio, TX, USA}

\author{Roy B. Torbert}
\affiliation{University of New Hampshire, Durham, NH, USA}
\affiliation{Southwest Research Institute, Durham, NH, USA}

\author{Stephen A. Fuselier}
\affiliation{Southwest Research Institute, San Antonio, TX, USA}
\affiliation{University of Texas San Antonio, San Antonio, TX, USA}

\author{Takuma Nakamura}
\affiliation{Space Research Institute, Austrian Academy of Sciences, Graz, Austria}
\affiliation{Institute of Physics, University of Graz, Graz, Austria}

\author{Barbara L. Giles}
\affiliation{NASA Goddard Space Flight Center, Greenbelt, MD, USA}

\author{Daniel J. Gershman}
\affiliation{NASA Goddard Space Flight Center, Greenbelt, MD, USA}

\author{Robert E. Ergun}
\affiliation{University of Colorado, Boulder, CO, USA}

\author{Christopher T. Russell}
\affiliation{University of California Los Angeles, Los Angeles, CA, USA}

\author{Robert J. Strangeway}
\affiliation{University of California Los Angeles, Los Angeles, CA, USA}


\date{\today}

\begin{abstract}

During magnetic reconnection, field lines interconnect in electron diffusion regions (EDRs). In some EDRs the reconnection and energy conversion rates are controlled by a steady out-of-plane electric field. In other EDRs the energy conversion rate $\vec{J}\cdot\vec{E}'$ is ``patchy'', with electron-scale large-amplitude positive and negative peaks. We investigate 22 EDRs observed by NASA's Magnetospheric Multiscale (MMS) mission in a wide range of conditions to determine the cause of patchy $\vec{J}\cdot\vec{E}'$. The patchiness of the energy conversion is quantified and correlated with seven parameters describing various aspects of the asymptotic inflow regions that affect the structure, stability, and efficiency of reconnection. We find that (1) neither the guide field strength nor the asymmetries in the inflow ion pressure, electron pressure, reconnecting magnetic field strength, and number density are well correlated with the patchiness of the EDR energy conversion, (2) the out-of-plane axes of the 22 EDRs are typically fairly well aligned with the ``preferred'' axes, which bisect the time-averaged inflow magnetic fields and maximize the reconnection rate, and (3) the time-variability in the upstream magnetic field direction is best correlated with the patchiness of the EDR $\vec{J}\cdot\vec{E}'$. A 3-d fully-kinetic simulation of reconnection with a non-uniform inflow magnetic field is analyzed; the variation in the magnetic field generates secondary X-lines, which develop to maximize the reconnection rate for the time-varying inflow magnetic field. The results suggest that magnetopause reconnection, for which the inflow magnetic field direction is often highly variable, may commonly be patchy in space, at least at the electron scale. 

\end{abstract}


\maketitle 


\section{Introduction}

\subsection{Background and motivation}

Magnetic reconnection in plasmas plays an important role in converting magnetic energy to particle kinetic energy \cite{Sonnerup.1979,BurchandDrake.2009}. At the heart of magnetic reconnection is an electron diffusion region (EDR), where inflowing sheared magnetic fields merge, changing their connectivity \cite{Vasilyunas.1975,Sonnerup.1979,Hesse.2011}. The reconnection electric field $E_R$ cycles magnetic flux through the EDR, thereby mediating the local reconnection rate, maintaining the out-of-plane current density $J_M$, and contributing to the energy conversion rate through $J_ME_R>0$\cite{Hesse.2018}. 

NASA's Magnetospheric Multiscale (MMS) mission investigates reconnection in Earth's magnetosphere \cite{Burch.2016a}. In one EDR observed by MMS, a clear steady reconnection electric field, $E_R$, showed remarkable agreement with both the reconnection rate \cite{Torbert.2018,Genestreti.2018c,NakamuraTKM.2018} and electron energization rate \cite{Bessho.2018} (e.g., Fig. \ref{example}a). MMS has observed other EDRs with electron-scale peaks in the energy conversion rates that can exceed what is expected from typical values of $E_R$, in some cases by several orders of magnitude \cite{Burch.2016b,Burch.2018a,Burch.2018b,Burch.2020,Cassak.2017a,Genestreti.2017,Genestreti.2018a} (e.g., Fig. \ref{example}b). These large-amplitude energy conversion rates often originate from spatially and/or temporally oscillatory electric fields, such that $\vec{J}\cdot\vec{E}'$ displays both positive and negative values (where $\vec{E}'\equiv\vec{E}+\vec{v}_e\times\vec{B}$ is the electric field in the electron frame). We refer to such events as having ``patchy'' energy conversion rates. 

Patchy EDR energy conversion has been observed by MMS more commonly at Earth's magnetopause than in the magnetotail – though far fewer MMS magnetotail EDRs have been yet been identified. Magnetopause reconnection occurs between the highly variable shocked solar wind plasma and Earth's magnetospheric plasma. Magnetotail reconnection occurs within the magnetosphere between similar plasmas. Whereas magnetopause reconnection often has pronounced asymmetries between the two inflow regions and may occur for a wide range of magnetic shear angles \cite{Fuselier.2017}, magnetotail reconnection is often more symmetric with large magnetic shear angles \cite{Eastwood.2010}. While a wide range of conditions of a reconnecting plasma may plausibly influence the structure of an EDR and its energy conversion rate, the seven parameters investigated here focus on conditions that typically differ for magnetopause and magnetotail reconnection.

\subsection{Potential causes of patchy energy conversion}

Asymmetries of upstream densities and magnetic field strengths can displace the inflow stagnation line and X-line \cite{CassakandShay.2007,Burch.2016b}. When the momenta of the two inflowing plasmas are imbalanced a normal-directed current $J_N$ crosses the X-line, which is unique to asymmetric reconnection \cite{CassakandShay.2007,CassakandShay.2008,PritchettandMozer.2009,Burch.2016b}. Heavier ions penetrate farther past the X-line than lighter electrons. Negative charge accumulation occurs as electrons converge on the electron inflow stagnation point, or $\Se$ point for brevity. As these bunched electrons are deflected into the outflow, they meander back and forth across the low-density-side separatrix. The resulting oscillatory $J_N$ and the strongly positive co-located $E_N$ lead to spatially oscillatory $J_NE_N$ \cite{Swisdak.2018,Burch.2018a,Pritchard.2019}. Separation between the X and $\Se$ lines may therefore lead to spatially patchy $\vec{J}\cdot\vec{E}'$ in EDRs.

Asymmetries of upstream densities and pressures can enable cross-field density and pressure gradients at the X-line \cite{Swisdak.2003}. Lower-hybrid or electron drift instabilities may promote the growth of waves and turbulence around the EDR \cite{Price.2016,Ergun.2017,Ergun.2019a,Ergun.2019b,Graham.2017,Le.2017,Wilder.2019}, which may alter the local energy conversion rate in and near the EDR \cite{Price.2016,Le.2017}. The most common form of drift wave found in/near MMS-observed asymmetric EDRs\cite{Wilder.2019} is a 3-d corrugation-like surface wave that originates near the separatrices and ultimately results from an ion pressure gradient \cite{Ergun.2019a,Ergun.2019b}. Alternatively, the corrugation-like surface waves may be a branch of the lower-hybrid drift instability, in which case they are expected to be driven by electron density or pressure gradients \cite{Graham.2017,Wilder.2019}. Thus the degree of asymmetry in the density, ion pressure and/or electron pressure may lead to spatially and temporally patchy $\vec{J}\cdot\vec{E}'$.

During high-magnetic-shear reconnection, highly non-gyrotropic electron velocity distribution functions form as a result of cross-field meandering motions \cite{Hesse.2014,Burch.2016b,Torbert.2018} and the energy conversion is primarily from perpendicular-to-the-magnetic-field currents and electric fields \cite{Wilder.2017}. During low-shear reconnection, electrons are free to stream along a guide magnetic field \cite{Eriksson.2016,BurchandPhan.2016,Genestreti.2017,Genestreti.2018a} and the energy conversion is primarily from parallel currents and electric fields \cite{Wilder.2017}. These unstable velocity distribution functions in low and high magnetic shear EDRs have been shown to act as a free energy source for wave growth, which may modify the energy conversion rate within EDRs \cite{Burch.2018b,Burch.2019,Dokgo.2019,Khotyaintsev.2019}. Alternatively, the guide field may stabilize the EDR against the lower-hybrid drift instability \cite{Huba.1982,Price.2020}.

\begin{figure}
\noindent\includegraphics[width=35pc]{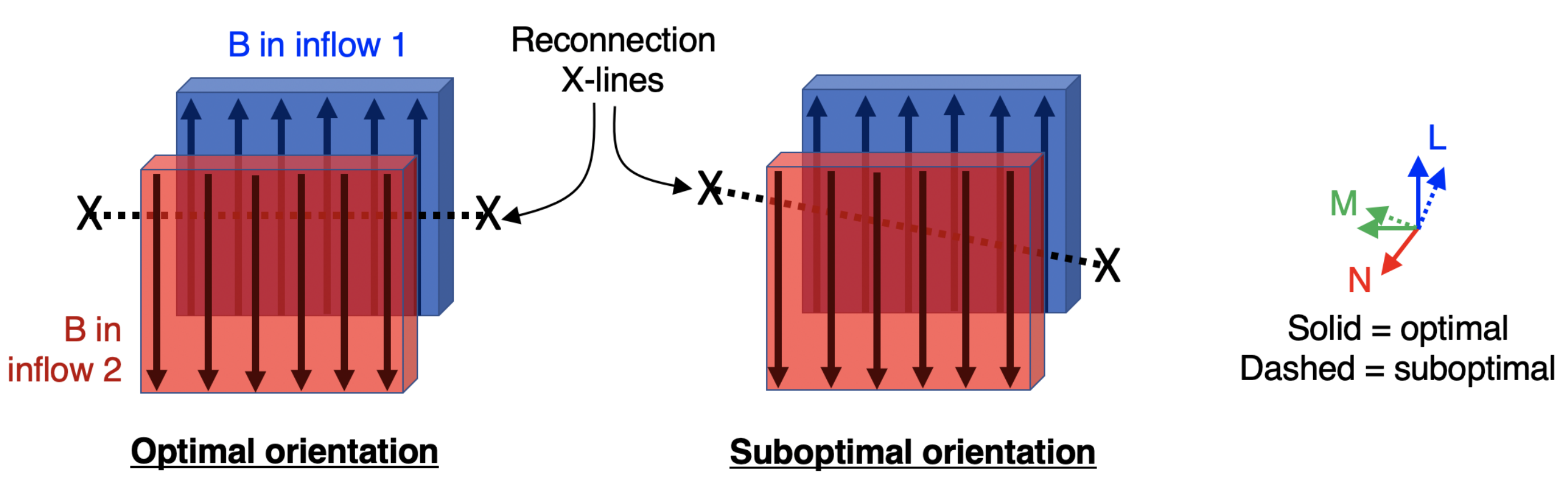}
\caption{Left: a reconnection X-line with the optimal orientation (solid-line $M$ direction) that maximizes the upstream free magnetic energy and the reconnection rate. Right: reconnection with a suboptimal orientation (dashed-line $M$ direction) reconnects the free magnetic energy inefficiently, leading to the growth of secondary reconnection lines that form with optimal orientations\cite{Liu.2018b} (i.e., solid-line $M$ direction).}
\label{optimal_m_sketch}
\end{figure}

Reconnection X-lines have preferred orientations, which optimize the reconnection rate \cite{Hesse.2013,Liu.2018b}. This optimum orientation, corresponding to the solid-line $M$ direction in Figure \ref{optimal_m_sketch}, bisects the upstream magnetic fields \cite{Hesse.2013}. If reconnection is forced to occur in an orientation that is not able to efficiently reconnect the inflowing magnetic energy (dashed $M$ direction, Fig. \ref{optimal_m_sketch}), then secondary reconnection lines may develop along the optimal orientation \cite{Liu.2018b}. This can occur when the reconnecting magnetic field is time-varying or has turbulent fluctuations, which will lead to flux pileup and flux rope generation in the outflow and modulations of the reconnection and flux transport rates \cite{NakamuraTKM.2021,Spinnangr.2021}. In 3-d kinetic simulations, flux ropes often become entangled \cite{Daughton.2011,Lapenta.2015}; it has been proposed that reconnection between entangled flux ropes may be the origin of patchy parallel electric fields observed by MMS\cite{Ergun.2016c}. Therefore, the time-varying upstream magnetic field could result in the patchy EDR J.E'.

\subsection{Outline of this study}

To identify conditions in which patchy EDR energy conversion is most likely to be driven, we perform a multi-event study of 22 MMS-observed EDRs and correlate upstream parameters with the patchiness of the energy conversion. We find that the upstream parameter best-correlated with the patchiness of the energy conversion is the time variability of the upstream magnetic field direction. We then perform a large, 3-d, and fully-kinetic particle-in-cell (PIC) simulation of reconnection with a time-varying upstream field. We find that the current sheet develops secondary tearing lines that have orientations that maximize the reconnection rate of varying inflow fields.

Parameter definitions, methodologies for their identification, and a description of the relevant capabilities of MMS are found in section II. In section III.A we present results of the multi-event study, finding that the strongest correlation is between the patchiness of the EDR energy conversion rate and time variability of the upstream magnetic field direction. In section III.B we analyze a three-dimensional fully-kinetic particle-in-cell (PIC) simulation of reconnection with an unsteady upstream magnetic field. Finally, in section IV, we summarize and interpret these results.

\section{Methodology and MMS dataset}

\subsection{Overview of methodology}

We seek to understand whether one or more of several of the following descriptors of the upstream plasma conditions, enumerated below, may play a predominant role in controlling the patchiness of the EDR energy conversion rate $\sigma_{J\cdot E'}$. 

\begin{enumerate}

\item Distance along the normal direction between the X and $\Se$ lines normalized by the thickness of the EDR, which is estimated as \\
\begin{equation} \delta_{XSe}/2\delta_e=\frac{n_1B_{L2}^2-n_2B_{L1}^2}{\left(B_{L1}+B_{L2}\right)\left(n_1B_{L2}+n_2B_{L1}\right)} \label{eq:dxse} \end{equation} \\
\noindent where $n$ is the number density, $B_L$ is the reconnecting component of the magnetic field, and subscripts 1 and 2 indicate the parameter is associated with one or the other inflow region\cite{Cassak.2017a}.

\item Ion thermal pressure asymmetry $(\left<P_{i1}\right>-\left<P_{i2}\right>)/P_{i0}$, where subscripts 1 and 2 denote the asymptotic pressures in the two inflow regions and the normalization parameter $P_{i0}$ is hybrid asymptotic scalar ion pressure, assumed to follow
\begin{equation} P_{i0}=n_{0}T_{i0}=\left(\frac{n_1B_2+n_2B_1}{B_1+B_2}\right)\left(\frac{n_1T_{i1}B_2+n_2T_{i2}B_1}{n_1B_1+n_2B_2}\right), \label{eq:pi0} \end{equation} \\
\noindent based on previously derived expressions for the hybrid asymptotic number density \cite{CassakandShay.2007} and temperature \cite{Cassak.2017a}.

\item Electron thermal pressure asymmetry $(\left<P_{e1}\right>-\left<P_{e2}\right>)/P_{e0}$, where $P_{e0}$ follows the form of equation \ref{eq:pi0}, where angular brackets indicate time averages

\item Number density asymmetry $(\left<n_{1}\right>-\left<n_{2}\right>)/n_0$, where $n_0$ is given by the left-most parenthetical term in equation \ref{eq:pi0}.

\item Normalized guide magnetic field strength $B_G/B_{L0}$, $B_{L0}$ is the hybrid reconnecting magnetic field component, which follows\cite{CassakandShay.2007} \\
\begin{equation} B_{L0}=\frac{2B_{L1}B_{L2}}{B_{L1}+B_{L2}} \label{eq:bl0} \end{equation} \\
\noindent and the hybrid asymptotic guide field $B_{G}$ is assumed to follow the same form.

\item Angle between the actual and optimal ($M_{opt}$) X-line orientations in the $L$-$M$ plane, where $M_{opt}$ bisects the time-averaged inflow magnetic fields \cite{Hesse.2013}.

\item Angular variability in the upstream magnetic fields $\delta\theta=\left<\mathrm{acos}(\hat{B}\cdot\left<\hat{B}\right>)\right>$.

\end{enumerate}

\begin{figure}
\noindent\includegraphics[width=25pc]{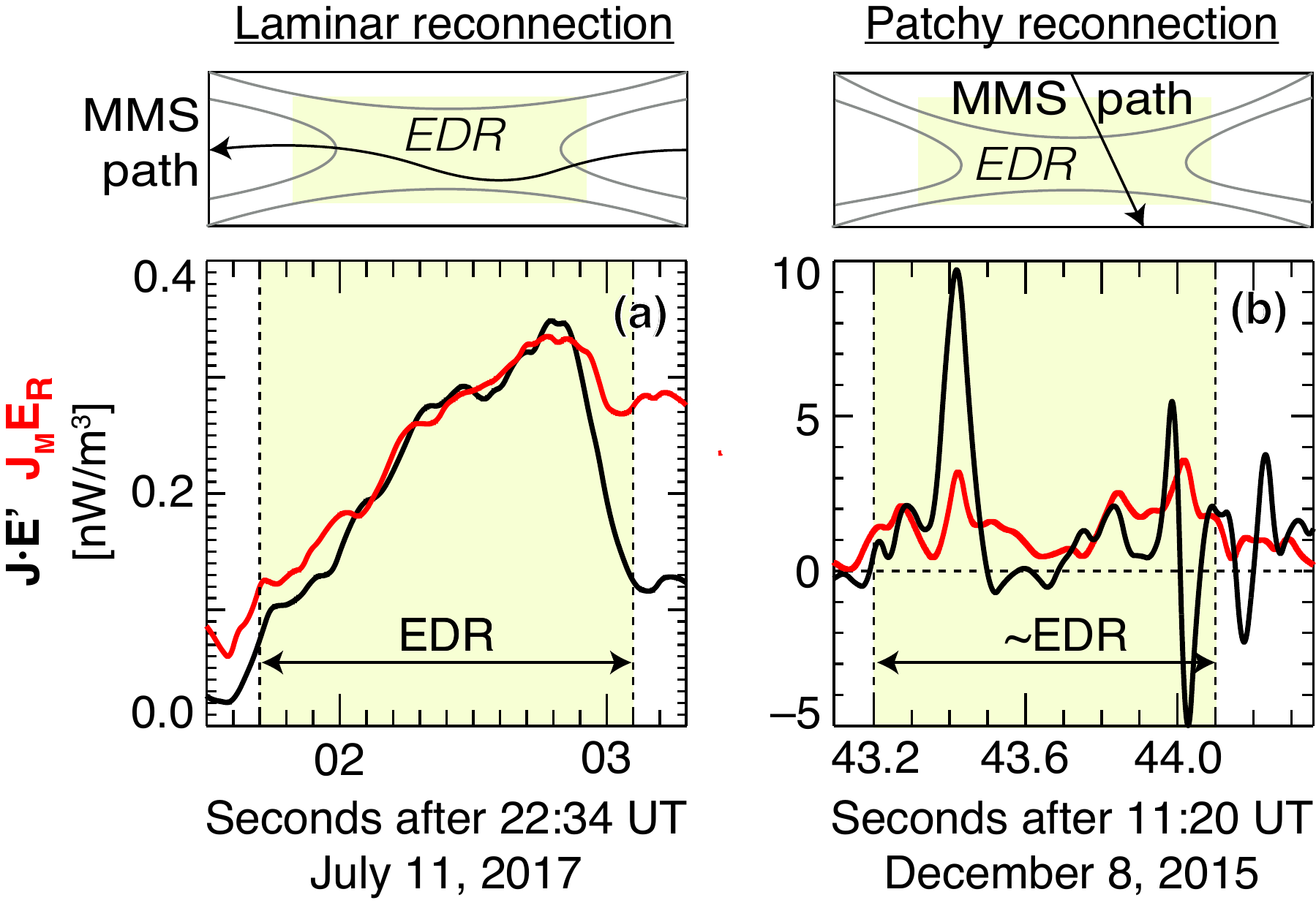}
\caption{A comparison of the observed non-ideal energy conversion rate $\jep$ and the rate expected based on a uniform and constant reconnection electric field $J_ME_R$. Two events are shown, which were identified in Earth's magnetotail \cite{Torbert.2017} (left) and at the magnetopause \cite{BurchandPhan.2016} (right).}
\label{example}
\end{figure}

The ``patchiness" of the EDR energy conversion rate $\sigma_{J\cdot E'}$ is quantified as the standard deviation of the red and black curves in Figure \ref{example}, normalized by the maximum value of the red curve, i.e.,

\begin{equation}
\sigma_{J\cdot E'}=\frac{\sqrt{\left<\left|\vec{J}\cdot\vec{E}'-J_ME_R\right|^2\right>-\left<\left|\vec{J}\cdot\vec{E}'-J_ME_R\right|\right>^2}}{\mathrm{max}\left(J_ME_R\right)},
\label{eq:sjep}
\end{equation}

\noindent where $\vec{E}'\equiv\vec{E}+\vec{v}_e\times\vec{B}$ is the electric field in the electron rest frame and the normalization quantity is the maximum value of $J_ME_R$ in the EDR. $E_R$ is a constant value determined as $E_R=R<V_{Ai0}B_0>$, where $R$ is the normalized reconnection rate and the theoretical maximum $R\simeq0.2$ value\cite{Liu.2017,Liu.2018a} is assumed, $V_{Ai0}$ is the hybrid asymptotic upstream ion Alfv\'en speed, and $B_0$ is the hybrid asymptotic upstream reconnecting magnetic field $B_L$. With the exception of $E_R$, all other parameters in equation (4) are evaluated in the EDR. Figure \ref{example}a shows an extremely laminar EDR energy conversion case while Figure \ref{example}b shows an extremely patchy event. 

If patchy EDR energy conversion results from charge accumulation at the $\Se$ line then large-amplitude and spatially-oscillatory $J_NE_N'$ should contribute predominantly to the overall product $\vec{J}\cdot\vec{E}'$. We also define and calculate separate ``patchiness'' terms for $J_LE_L'$, $J_ME_M'$, and $J_NE_N'$,

\begin{equation}
\sigma_{J_i\cdot E_i'}=\frac{\sqrt{\left<\left|J_iE_i'-\delta_{iM}J_ME_R\right|^2\right>-\left<\left|J_iE_i'-\delta_{iM}J_ME_R\right|\right>^2}}{\mathrm{max}\left(J_ME_R\right)},
\label{eq:sjeplmn}
\end{equation} 

\noindent where $i$ is $L$, $M$, or $N$ and $\delta_{iM}=1$ for $i=M$ and zero otherwise. 

\subsection{MMS dataset}

MMS consists of four identically-equipped spacecraft that, during the periods studied here, flew in an electron-scale tetrahedral formation \cite{Burch.2016a,Fuselier.2016}. MMS science data are available in two principal modes, burst and survey, which describe the resolution of the data returned to ground. High-resolution burst-mode data are typically only available during current sheet crossings and are required for analyzing EDRs. Lower-resolution survey mode data are used when analyzing the asymptotic inflow regions. 

The fast plasma investigation obtains 3-d velocity distribution functions and moments of ions and electrons once per 150-ms and 30-ms, respectively, in burst mode (4.5-second cadences for both species in survey mode)\cite{Pollock.2016}. For magnetopause EDRs, mass-per-charge-separated ion composition data from the hot plasma composition analyzer \cite{Young.2016} are used to help distinguish the magnetosheath, magnetosphere, and mixed boundary layer plasmas. Comparitively high He$^{++}$ and negligible O$^+$ concentrations are expected in the asymptotic upstream magnetosheath, while the opposite is expected in the magnetosphere inflow region. 3-d electric and magnetic field measurements are obtained by the electric field double probes\cite{Lindqvist.2016,Ergun.2016a} and fluxgate magnetometers\cite{Russell.2016}, respectively. Burst-mode electric field data are available at 8,192 Hz. Survey-mode magnetometer data are available at 8 Hz. The particle moments from the fast plasma investigation are used to calculate current densities uniquely at each of the four spacecraft\cite{Phan.2016a}. $\vec{J}\cdot\vec{E}'$ is also calculated uniquely at each spacecraft and is smoothed to remove sub-$d_e$-scale oscillations. 

\subsection{Analysis methods and event selection criteria}

First we identify EDR events. Throughout this paper EDR refers to the ``central EDR'', which is distinguished from the extended electron jet region often referred to as the ``outer EDR'' \cite{Phan.2007,Chen.2008}. Generally speaking, the central EDR is where field lines merge \cite{Hesse.2011,Zenitani.2011,Burch.2016a,Torbert.2018}. We started with 36 EDR events, 34 of which were identified at the dayside magnetopause\cite{Chen.2016,Chen.2017,Lavraud.2016,Burch.2016b,BurchandPhan.2016,Webster.2018,Ergun.2017,Torbert.2017,Genestreti.2018a,Pritchard.2019,Li.2020,Burch.2020} and 2 in the magnetotail\cite{Torbert.2018,Zhou.2019}.  

Next, we require that MMS observed both asymptotic inflow regions for several minutes. The trajectory MMS takes through an EDR depends almost entirely on the time-dependent motion of the EDR, which varies from event to event. In some cases, MMS does not fully cross the EDR into one inflow region; these events are discarded, leaving 27 EDRs. Three events for which plasma parameters during an inflow interval could not clearly be associated with the EDR interval (e.g., when large rotations in the upstream magnetic field were observed during the crossing) were discarded, leaving 24 EDRs.  

Average $LMN$ coordinates were determined for these 24 EDRs. Here, average specifies that a single coordinate system is used to define an EDR interval, whereas the axes may vary during the crossing \cite{Denton.2018}. The maximum directional derivative of $\vec{B}$ (MDD-B) technique\cite{Shi.2005} was used to identify the EDR current sheet normal $N$. For some events, MDD-B did not find a stable normal direction; in these cases, maximum variance of the electric field\cite{Paschmann.1986,Sonnerup.1987} (MVA-E) was used to identify $N$. Maximum variance of the magnetic field\cite{SonnerupandCahill.1967} (MVA-B) was then used to determine a direction $L^*$. $M$ was then evaluated as $N\times L^*/\left|N\times L^*\right|$ and $L=M\times N$. Similar hybrid techniques for finding LMN coordinates have been used previously \cite{Genestreti.2018c,Denton.2018}. Two events were discarded because EDR coordinates could not be confidently established, leaving 22 total EDR events for this study (20 magnetopause and 2 magnetotail events).

We use Spearman's $\rho$ coefficient to evaluate the strength of the correlations between the patchiness of the energy conversion in our 22 EDRs with the seven parameters enumerated in the list of section II.A. $\rho$ is a non-parametric measure of the strength with which two variables are associated \cite{myers2013research}. This approach was chosen because (1) the magnitude of $\rho$ is not strongly influenced by outlying data points and (2) we do not have to assume any particular functional form describing the relationships between the patchiness and the seven parameters; rather, only a monotonous relationship is assumed. We refer to correlations with $|\rho|\leq1/3$ as weak, $1/3\leq|\rho|\leq2/3$ as moderate, and $|\rho|\geq2/3$ as strong. We also evaluate a confidence interval for each correlation, i.e., the probability that a non-zero correlation is not the result of random chance, which is based on the sample size (22 EDRs) and the strength of the correlation ($=1-\rho\sqrt{2}$). We adopt a ``95$\%$ rule'', meaning that only correlations with $\geq$95$\%$ confidence (2$\sigma$) are deemed significant.

\section{Results}

\subsection{Multi-event study}

\begin{figure}
\noindent\includegraphics[width=30pc]{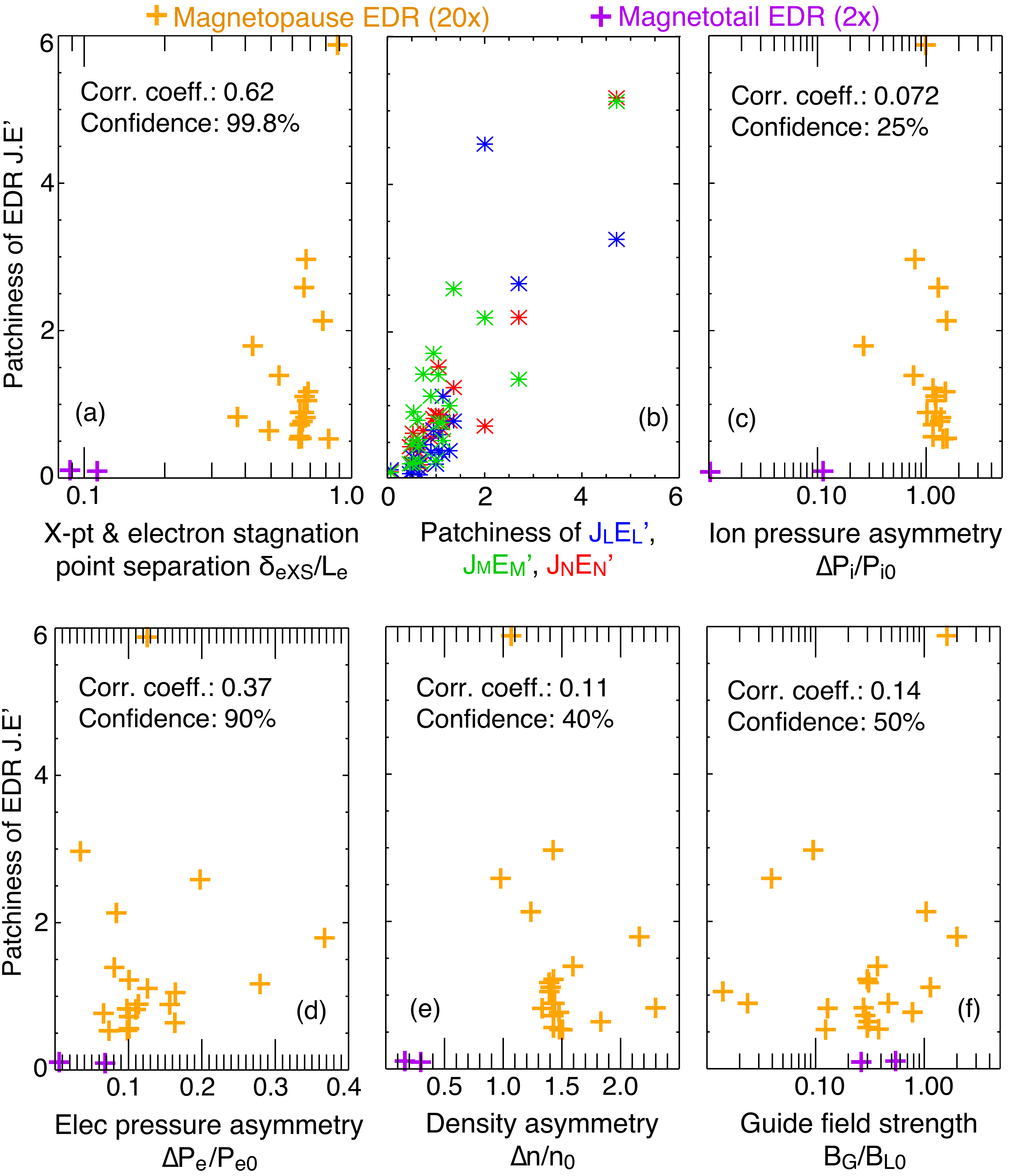}
\caption{Vertical axes are $\sigma_{J\cdot E'}$ defined in Eq. 1 for 22 EDRs. Horizontal axes are the normalized (a) separation between the X and electron stagnation ($\Se$) lines, a derived quantity based on the inflow magnetic field and density asymmetries, (b) the patchiness of $J_LE_L'$ (blue), $J_ME_M'$ (blue), and $J_NE_N'$ (red), as defined in equation \ref{eq:sjeplmn}, (c) scalar ion thermal pressure asymmetry, (d) scalar electron thermal pressure asymmetry, (e) density asymmetry, and (f) guide field strength, the definitions of which are found in the numbered list in section II.A. (a) and (c)-(f): Spearman correlation coefficients for and their confidence values are in the upper left of each panel, magnetopause EDRs are colored orange and magnetotail EDRs are purple.}
\label{scatter1}
\end{figure}

Figure \ref{scatter1}a-e show the first five parameters in the enumerated list in section II.A. Of these five parameters the separation between the X and $\Se$ lines (Fig. \ref{scatter1}a), as defined in equation \ref{eq:dxse} is the only parameter moderately and significantly correlated with the patchiness of the energy conversion. A note of caution is required, however, regarding the clustering of data points in the parameter space of figure \ref{scatter1}a. Since we do not have enough EDRs to control for all parameters simultaneously, it is not possible to discern whether the separation of magnetotail (two bottom/left-most data points in Fig. \ref{scatter1}a) and magnetosheath (twenty right-most data points in Fig. \ref{scatter1}a) EDRs are due to unique aspects of reconnection caused by X and $\Se$ line separations or due to other differences between the magnetopause and magnetosheath current sheets. However, when the two outlying magnetotail data points are excluded, the correlation coefficient and confidence drop only slightly to 0.5 and 97$\%$, respectively, meaning that the correlation is still moderate and significant.  Figure \ref{scatter1}f shows the component-specific patchiness parameter of equation \ref{eq:sjeplmn}. If charge accumulation at the $\Se$ line was the predominant cause of patchy energy conversion, then the energy conversion rates of patchier events is expected to be dominated by $J_NE_N$. However, there is no clear dominance of the patchiness of $J_LE_L'$ (blue), $J_ME_M'$ (green), and $J_NE_N'$ (red) to the overall patchiness of $\vec{J}\cdot\vec{E}'$. 

Weak correlations were found between the patchiness of the energy conversion and the ion (Fig. \ref{scatter1}b) and electron (Fig. \ref{scatter1}c) thermal pressure asymmetries, the density asymmetry (Fig. \ref{scatter1}d), and the guide field strength (Fig. \ref{scatter1}e); all correlations were all below our $95\%$ confidence threshold for significance. Observations and simulations suggest that these parameters may play a role in modulate the energy conversion rate at or very near the EDR, at least in some limiting circumstances. Since we are unable to control for all parameters simultaneously, the results of Fig. \ref{scatter1} may only be interpreted as evidence that these parameters do not exert a singular or predominant influence on the patchiness of the EDR energy conversion, over the parameters' ranges typically found in the magnetosphere. 

\begin{figure}
\noindent\includegraphics[width=25pc]{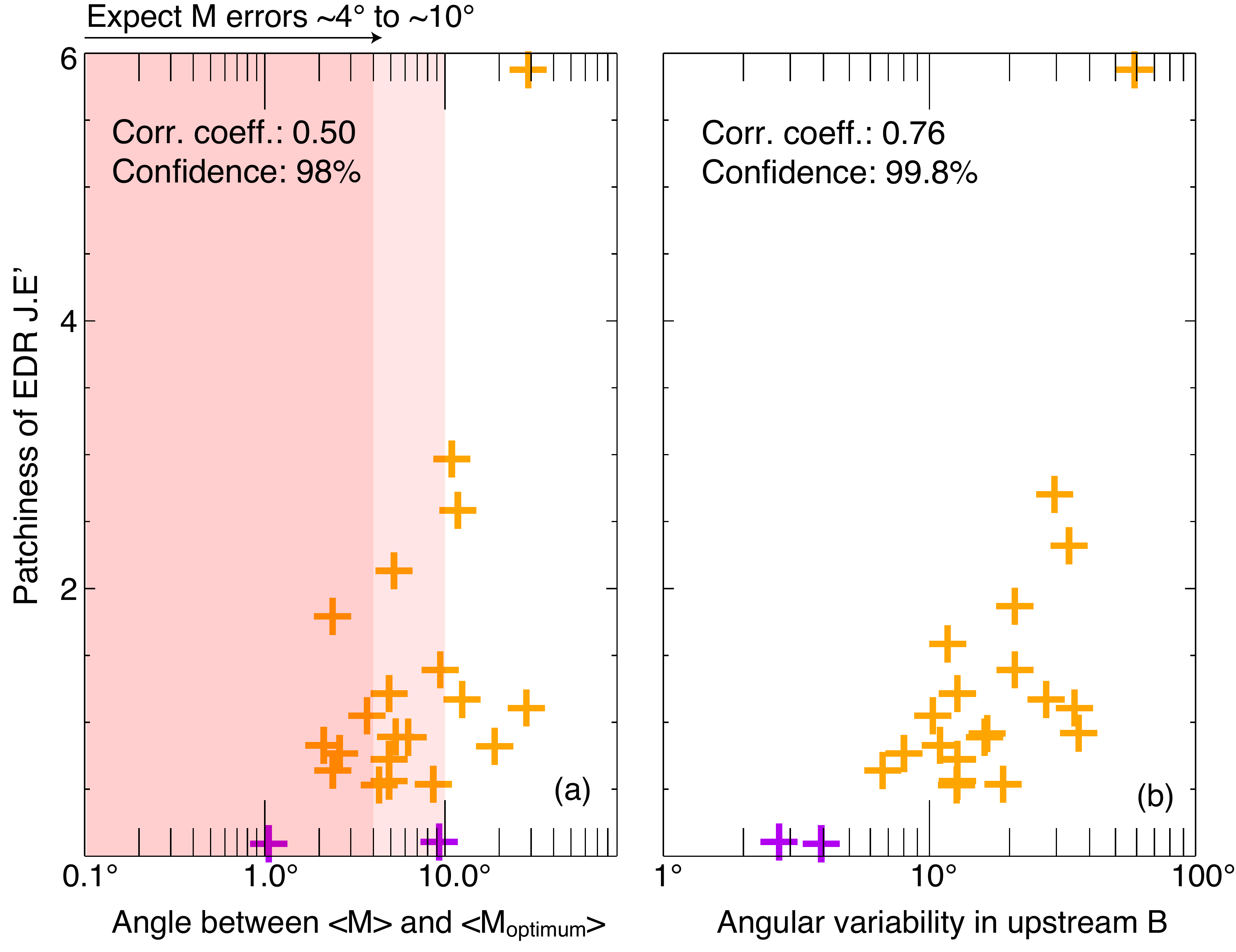}
\caption{Vertical axes are $\sigma_{J\cdot E'}$ defined in Eq. 1 for 22 EDRs. Horizontal axes are: (a) the angular difference in the $L-M$ plane between the EDR $M$ direction and the optimum $M$ direction, which bisects the time-averaged upstream magnetic fields and (b) the angular variability of the upstream magnetic field direction. Spearman correlation coefficients and confidence values are in the upper left of each panel. Magnetopause EDRs are colored orange and magnetotail EDRs are purple.}
\label{scatter2}
\end{figure}

The final two parameters from section II.A are shown in figure \ref{scatter2}a and \ref{scatter2}b: the angle between the actual EDR $M$ and optimum $M_{opt}$ directions and the angular variability of the upstream magnetic field, respectively. Errors in the EDR coordinate axes determined with the hybrid MDD-B/MVA technique may be $\sim$4$^\circ$-to-10$^\circ$ based on previous MMS case analyses \cite{Denton.2018,Genestreti.2018c}. We find that most of the EDRs are separated from the optimum $M_{opt}$ direction by angles less than our assumed $10^\circ$ of uncertainty. 

The parameter most strongly and significantly correlated with $\sigma_{J\cdot E'}$ is the time variability of the upstream magnetic field direction (Figure \ref{scatter2}b). This correlation may indicate that, while the EDR may be fairly well aligned with the time-averaged optimum $M_{opt}$ direction, time variations in $M_{opt}$ may also lead to secondary tearing growth. This result is in good agreement with recent 2-dimensional particle-in-cell simulations\cite{NakamuraTKM.2021} of reconnection with fluctuating magnetic fields. In the following section we investigate this result further by analyzing a 3-dimensional simulation of reconnection with a non-uniform inflow magnetic field.

\subsection{Simulation of reconnection with varying inflow conditions}

A three-dimensional fully-kinetic simulation was performed to investigate the behavior of reconnection under non-uniform inflow conditions. The simulation was run using the electromagnetic particle-in-cell code {\it VPIC} \cite{Bowers.2008}. The initial magnetic field profile of the primary asymmetric current sheet was taken from a previous work\cite{Liu.2018b}; however, a tangential discontinuity (TD) was added in the upstream magnetosheath (see Figure \ref{picsetup}b). The TD convects with the inflow toward the X-line, meaning the spatial variations in the inflow field translate to time-varying boundary conditions for the diffusion region. The upstream TD was an ion-scale rotation of the inflow magnetic field by $45^\circ$, which was chosen to loosely match the largest  variations in the upstream field direction for the event of Figure \ref{example}b. To reduce turbulence resulting from periodic conditions at the $M$ boundaries, the simulation box was oriented such that the optimal $M_{opt}$ direction of the primary reconnecting current sheet was aligned with the simulation $M$ coordinate\cite{Liu.2018b}. A full description of the simulation set-up is provided in Appendix A.

We limit our investigation to a single time of the simulation, $t=128/\Omega_{ci}$ (where $\Omega_{ci}$ is the ion cyclotron period), which is roughly 50$\Omega_{ci}$ after the TD convected into the diffusion region. At this time, strong $\vec{J}\cdot\vec{E}$ resulting from the initial conditions were no longer apparent. At $t=128/\Omega_{ci}$ some readily identifiable impacts of the time-varying inflow appear in the simulation.

\begin{figure}
\noindent\includegraphics[width=25pc]{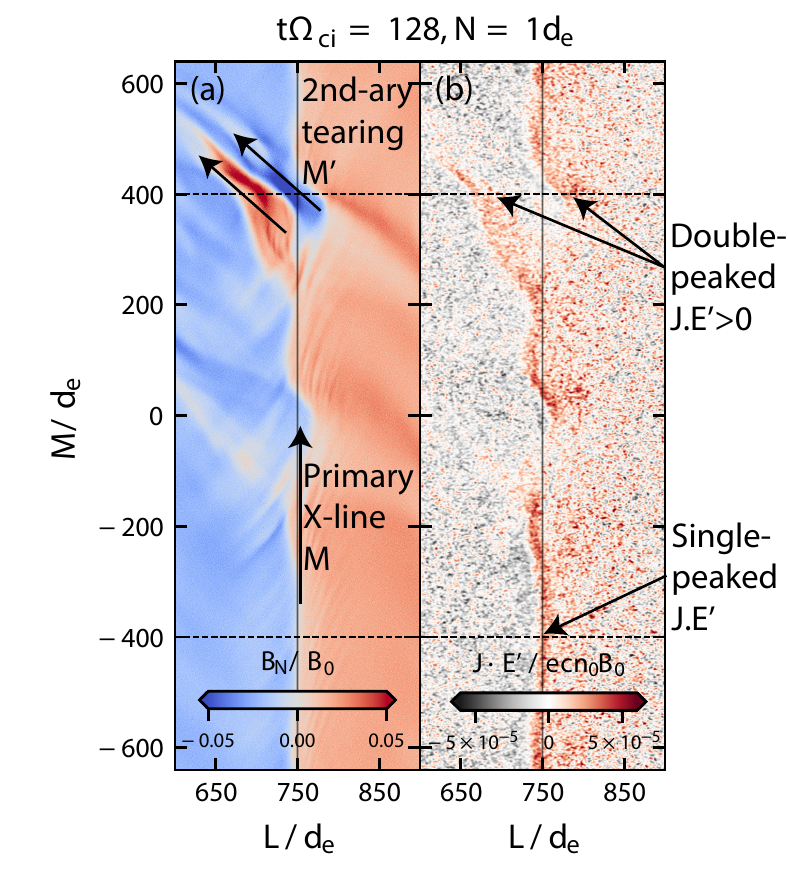}
\caption{The (a) reconnected component of the magnetic field and (b) electron-frame energy conversion rate in the simulation $L-M$ plane.}
\label{sim1}
\end{figure}

Figure \ref{sim1}a shows a cut through the $L-M$ plane at $N/d_e=1$, the approximate location of the primary X-line. In the $L-M$ plane, reconnection lines are identified as dividing lines that separate oppositely-directed $B_N$. Secondary tearing lines, shown in Figure \ref{sim1}a, are likely a result of the TD impact. The axes of the secondary tearing lines $M'$ are tilted by roughly $45^\circ$ relative to the primary X-line $M$, which is consistent with the expected optimal $M$ direction (along the line bisecting the upstream fields) after the $45^\circ$ rotation of the upstream $\vec{B}$ associated with the TD. Persistent features associated with the primary X-line, which was oriented in the optimal direction under the initial upstream conditions, appear simultaneously with the secondary tearing modes. 

\section{Summary, interpretation of results, and future work}

The overarching goal of this study was to determine the origin of patchy non-ideal energy conversion rates $\vec{J}\cdot\vec{E}'$ commonly found in MMS-observed electron diffusion regions (EDRs). We examined 36 EDRs, finding 22 that were suitable for a multi-event study. The patchiness of the energy conversion rate was quantified by $\sigma_{J\cdot E'}$, as defined in equation \ref{eq:sjep}, which is the difference between the MMS-observed energy conversion rate and the rate expected from a uniform, steady reconnection electric field with a normalized strength of 0.2. The patchiness of the energy conversion was then correlated with seven parameters describing the geometry of the diffusion region and its upstream conditions: the (1) separation between the X and electron stagnation ($\Se$) lines, a function of the magnetic field and density asymmetry, (2) ion scalar pressure asymmetry, (3) electron scalar pressure, (4) density asymmetry, (5) guide field strength, (6) the angle between the average EDR and optimum $M$ directions, the latter being the line bisecting the time-averaged upstream magnetic fields, and (7) the time variability of the upstream field. 

The principal findings from the multi-event study are:
\begin{enumerate}
\item The patchiness of the energy conversion rates in our EDR events is not correlated with the density asymmetry, ion and electron pressure asymmetries, nor the guide field strength. 
\item A moderate and significant correlation is observed between the patchiness of the EDR energy conversion and the separation between the X and $\Se$ lines, which is a function of the magnetic field and density asymmetry. There is no clear dominance of $J_LE_L'$, $J_ME_M'$ or $J_NE_N'$ in EDRs with patchy energy conversion.
\item The majority of EDRs have an average $M$ direction within (10$^\circ$) uncertainty bars of being aligned with the optimum direction, which bisects the time-averaged upstream magnetic fields and maximizes the reconnection rate.
\item The best correlation is observed between the patchiness of the EDR energy conversion and the time variability of the upstream magnetic field direction.
\end{enumerate}

A three-dimensional particle-in-cell (PIC) simulation was performed to investigate the behavior of reconnection with non-uniform inflow conditions. Reconnection began along a primary X-line, which had an optimum orientation that bisected the initial upstream field, thereby maximizing the initial reconnection rate. After a tangential discontinuity impacted the diffusion region and the immediately-upstream magnetic field rotated by 45$^\circ$ secondary tearing lines developed, which radiate from the primary X-line at an angle consistent with the change in the magnetic shear (by 45$^\circ$). Due to high noise levels, which may have been due in part to an initial state of disequilibrium, we were not able to quantify the patchiness of the energy conversion during the TD impact in the simulation studied here.

We interpret the findings in the following way: of the sources studied here, the predominant source of patchiness in the EDR energy conversion rate is the time variability of the inflowing magnetic field directions. The causal relationship may be due to the formation of secondary tearing lines, which develop from a primary tearing line in unsteady inflow conditions, as was seen in the simulation. Whereas the direction of the primary reconnection line seems to be (at least, most commonly) set by the direction that bisects the time-averaged upstream fields, the growth of secondary tearing lines may be the mechanism that maximizes the reconnection rate under time-varying inflow fields. This is just one possible interpretation, since no clear enhancement in $\vec{J}\cdot\vec{E}'$ was observed at the simulated secondary tearing lines. It is possible the single clean variation in the simulated magnetic field was not complex enough in its structure to lead to entangled flux rope formation\cite{Ergun.2016c,Daughton.2011,Lapenta.2015} and discernibly patchy $\vec{J}\cdot\vec{E}'$. Additionally, it is possible that the initially noisy $\vec{J}\cdot\vec{E}'$, found early in the simulation, did not provide an adequate benchmark with which patchy $\vec{J}\cdot\vec{E}$ could be identified.

This interpretation is comparable to findings from previous works, which studied in two-dimensional PIC simulations and found that the growth of secondary tearing lines and modulations in the reconnection rate result from time-varying inflow magnetic field configurations \cite{NakamuraTKM.2021,Spinnangr.2021}. In comparison to the aforementioned two-dimensional pictures, we suggest that the secondary tearing lines may form with oblique (3-d) geometries such that the reconnection rate is maximized for the time-varying field. Our interpretation and findings are also comparable with earlier MMS-based investigations. These studies suggested that patchiness in the reconnection rate may lead to the formation of tangled flux ropes, which, in turn, may reconnect with one another and generate patchy and large-amplitude electric fields \cite{Ergun.2016c}. 

Further simulation work is needed to develop a quantitive relationship between unsteady inflow magnetic fields and patchy reconnection. In addition to existing studies of two-dimensional simulations, three-dimensional simulations should be conducted to determine whether entanglement and reconnection of secondary flux ropes lead to enhanced energy conversion rates. 

One question that cannot be answered at present is whether or not patchy electron-scale reconnection has a discernible impact on reconnection at larger scales. It has recently been suggested that at/above ion scales, reconnection at Earth's magnetopause appears to have a continuous global-scale structure \cite{Fuselier.2021}. Reconciling the patchiness of reconnection at electron-scales with the apparent continuous and quasi-two-dimensional nature of reconnection at much larger scales may be possible in the near future as, in its current extended mission, the inter-spacecraft separations will be increased such that MMS will be able to resolve electron and ion-scales simultaneously.


%
%

%

\begin{acknowledgments}
We acknowledge the contributions made by many MMS team members to the success of the mission and the accessibility and high quality of the MMS data. This study has used several routines from the Space Physics Environment Data Analysis System \cite{spedas} and has benefited from conversations with Dr. Michael Hesse, Dr. Richard Denton, Dr. Paul Cassak, and Dr. Dominic Payne. The simulation is performed on Frontera at Texas Advanced Computer Center (TACC). KJG is supported by NASA grant 80NSSC20K0848. XL and YL are supported by MMS grant 80NSSC18K0289.
\end{acknowledgments}

\section{Data Availability}
MMS data are publicly available at https://lasp.colorado.edu/mms/sdc/public/. The simulation is available upon request to Xiaocan Li (Xiaocan.Li@dartmouth.edu)

\appendix

\section{Simulation set-up and analysis details}

\begin{figure}
\noindent\includegraphics[width=35pc]{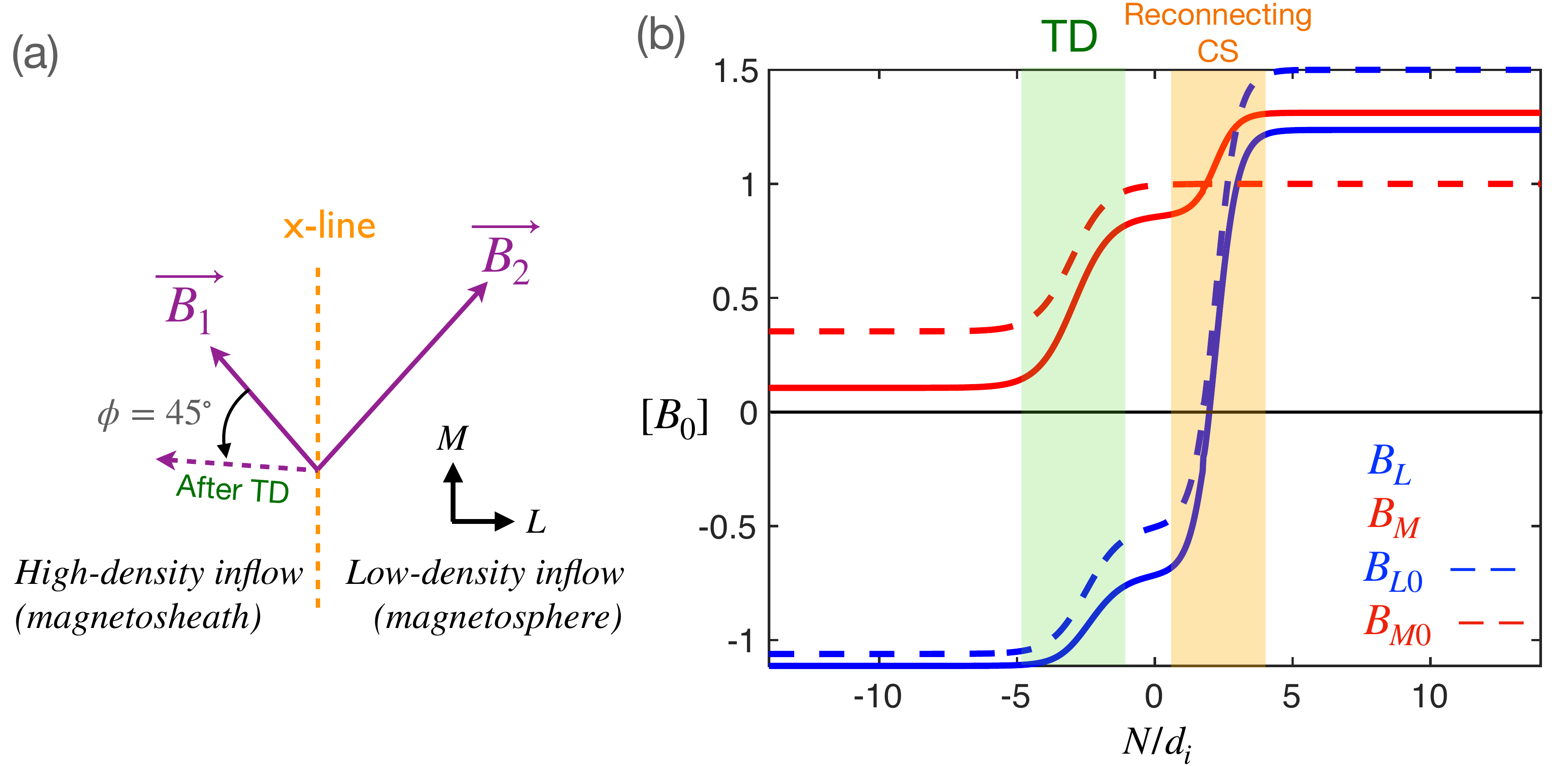}
\caption{Initial magnetic field condition used for the PIC simulation. (a) Illustration of the magnetic field rotation in the TD on the magnetosheath side. (b) Initial profiles of $B_L$, $B_M$, $B_{M0}$ and $B_{M0}$.}
\label{picsetup}
\end{figure}

The initial magnetic profile is

\begin{equation}
\begin{split}
B_{L0}(N)=B_0\left[(0.5+S)^2+1-\left(\frac{B_{M0}}{B_0}\right)^2\right]^{1/2}\mbox{sgn}(N-N_n), \\
B_{M0}(N)=B_0\left[\frac{1-b_{Ms}}{2}\mbox{tanh}\left(\frac{N-N_{TD}}{\lambda_{TD}}\right)+\frac{1+b_{Ms}}{2}\right], 
\end{split}
\label{profiles}
\end{equation}

\noindent where $S=\mbox{tanh}[(N-N_n)/\lambda]$. The primary reconnection current sheet is located at $N_n=2.2d_i$, while the TD is initially located at $N_{TD}=-3.0d_i$. Parameter $b_{Ms}=\mbox{cos}\phi-0.5\mbox{sin}\phi$ determines the net rotational angle $\phi$ of the TD and we used $\phi=45^\circ$, as illustrated in Fig.~\ref{picsetup}(a), which was chosen based on the largest angular deflections observed upstream of the EDR in Figure \ref{example}b. The initial half-thicknesses of the current sheet and TD are $\lambda= 0.8 d_i$ and $\lambda_{TD}=1.3 d_i$, respectively. These magnetic components are shown as dashed curves in Fig.~\ref{picsetup}(b).

The plasma has the same density profile $n=n_0[1-(S+S^2)/3]$ as in Liu et al. (2018), which is $n_2=n_0/3$ and $n_1=n_0$, where the subscripts ``1'' and ``2'' correspond to the magnetosheath and magnetosphere sides respectively.
The uniform total temperature is $T=3B_0^2/(8\pi n_0)$ that consists of contributions from ions and electrons with ratio $T_i/T_e=5$. The mass ratio is $m_i/m_e=25$. The ratio of the electron plasma to gyro-frequency is $\omega_{pe}/\Omega_{ce}=4$ where $\omega_{pe}\equiv(4\pi n_0 e^2/m_e)^{1/2}$ and $\Omega_{ce}\equiv eB_0/m_e c$. In the presentation, densities, time, velocities, spatial scales, magnetic fields and electric fields are normalized to $n_0$, the ion gyro-frequency $\Omega_{ci}$, the Alfv\'enic speed $V_A\equiv B_0/(4\pi n_0 m_i)^{1/2}$, the ion inertia length $d_i\equiv c/\omega_{pi}$, $B_0$ and $V_A B_0/c$, respectively.

From Liu et al. (2018), we determined the preferred orientation of the primary x-line of this asymmetric current sheet, which we like to align with the ${\bf y}$ axis of our simulation. Thus we rotate the simulation box by $\theta_{box}=-13^\circ$. 
The resulting magnetic field in the new coordinate will be
\begin{equation}
\begin{split}
B_L(N)=B_{L0}(N)\mbox{cos}\theta_{box}+B_{M0}(N)\mbox{sin}\theta_{box}, \\
B_M(N)=-B_{L0}(N)\mbox{sin}\theta_{box}+B_{M0}(N)\mbox{cos}\theta_{box}.
\end{split}
\label{oblique}
\end{equation}
and they are shown as solid curves in Fig.~\ref{picsetup}(b).
This large 3D run has a domain size $L_L \times L_M \times L_N=300d_i \times 256d_i \times 28d_i$ and $4800 \times 4096 \times 448$ cells. The origin of the coordinate locates at the center of this simulation domain. This run has 0.88 trillion macro particles. The boundary conditions are periodic both in the $L$- and $M$-directions, while in the $N$-direction they are conducting for fields and reflecting for particles.  We use the perturbation to uniformly initiate a reconnection x-line along the $M$-direction at $L$=0.
\\

\bibliography{patchy_jeprime.bib}

\begin{thebibliography}{76}%
\makeatletter
\providecommand \@ifxundefined [1]{%
 \@ifx{#1\undefined}
}%
\providecommand \@ifnum [1]{%
 \ifnum #1\expandafter \@firstoftwo
 \else \expandafter \@secondoftwo
 \fi
}%
\providecommand \@ifx [1]{%
 \ifx #1\expandafter \@firstoftwo
 \else \expandafter \@secondoftwo
 \fi
}%
\providecommand \natexlab [1]{#1}%
\providecommand \enquote  [1]{``#1''}%
\providecommand \bibnamefont  [1]{#1}%
\providecommand \bibfnamefont [1]{#1}%
\providecommand \citenamefont [1]{#1}%
\providecommand \href@noop [0]{\@secondoftwo}%
\providecommand \href [0]{\begingroup \@sanitize@url \@href}%
\providecommand \@href[1]{\@@startlink{#1}\@@href}%
\providecommand \@@href[1]{\endgroup#1\@@endlink}%
\providecommand \@sanitize@url [0]{\catcode `\\12\catcode `\$12\catcode
  `\&12\catcode `\#12\catcode `\^12\catcode `\_12\catcode `\%12\relax}%
\providecommand \@@startlink[1]{}%
\providecommand \@@endlink[0]{}%
\providecommand \url  [0]{\begingroup\@sanitize@url \@url }%
\providecommand \@url [1]{\endgroup\@href {#1}{\urlprefix }}%
\providecommand \urlprefix  [0]{URL }%
\providecommand \Eprint [0]{\href }%
\providecommand \doibase [0]{http://dx.doi.org/}%
\providecommand \selectlanguage [0]{\@gobble}%
\providecommand \bibinfo  [0]{\@secondoftwo}%
\providecommand \bibfield  [0]{\@secondoftwo}%
\providecommand \translation [1]{[#1]}%
\providecommand \BibitemOpen [0]{}%
\providecommand \bibitemStop [0]{}%
\providecommand \bibitemNoStop [0]{.\EOS\space}%
\providecommand \EOS [0]{\spacefactor3000\relax}%
\providecommand \BibitemShut  [1]{\csname bibitem#1\endcsname}%
\let\auto@bib@innerbib\@empty
\bibitem [{\citenamefont {{Sonnerup}}(1979)}]{Sonnerup.1979}%
  \BibitemOpen
  \bibfield  {author} {\bibinfo {author} {\bibfnamefont {B.~U.~{\"O}.}\
  \bibnamefont {{Sonnerup}}},\ }\enquote {\bibinfo {title} {{Magnetic field
  reconnection}},}\ in\ \href@noop {} {\emph {\bibinfo {booktitle} {In: Solar
  system plasma physics. Volume 3. (A79-53667 24-46) Amsterdam, North-Holland
  Publishing Co., 1979, p. 45-108.}}},\ Vol.~\bibinfo {volume} {3}\ (\bibinfo
  {year} {1979})\ pp.\ \bibinfo {pages} {45--108}\BibitemShut {NoStop}%
\bibitem [{\citenamefont {Burch}\ and\ \citenamefont
  {Drake}(2009)}]{BurchandDrake.2009}%
  \BibitemOpen
  \bibfield  {author} {\bibinfo {author} {\bibfnamefont {J.~L.}\ \bibnamefont
  {Burch}}\ and\ \bibinfo {author} {\bibfnamefont {J.~F.}\ \bibnamefont
  {Drake}},\ }\bibfield  {title} {\enquote {\bibinfo {title} {Reconnecting
  magnetic fields: The huge amounts of energy released from the relinking of
  magnetic fields in outer space are both mysterious and potentially
  destructive},}\ }\href@noop {} {\bibfield  {journal} {\bibinfo  {journal}
  {American Scientist}\ }\textbf {\bibinfo {volume} {97}},\ \bibinfo {pages}
  {392--399} (\bibinfo {year} {2009})}\BibitemShut {NoStop}%
\bibitem [{\citenamefont {Vasyliunas}(1975)}]{Vasilyunas.1975}%
  \BibitemOpen
  \bibfield  {author} {\bibinfo {author} {\bibfnamefont {V.~M.}\ \bibnamefont
  {Vasyliunas}},\ }\bibfield  {title} {\enquote {\bibinfo {title} {Theoretical
  models of magnetic field line merging},}\ }\href {\doibase
  10.1029/RG013i001p00303} {\bibfield  {journal} {\bibinfo  {journal} {Reviews
  of Geophysics}\ }\textbf {\bibinfo {volume} {13}},\ \bibinfo {pages}
  {303--336} (\bibinfo {year} {1975})}\BibitemShut {NoStop}%
\bibitem [{\citenamefont {{Hesse}}\ \emph {et~al.}(2011)\citenamefont
  {{Hesse}}, \citenamefont {{Neukirch}}, \citenamefont {{Schindler}},
  \citenamefont {{Kuznetsova}},\ and\ \citenamefont {{Zenitani}}}]{Hesse.2011}%
  \BibitemOpen
  \bibfield  {author} {\bibinfo {author} {\bibfnamefont {M.}~\bibnamefont
  {{Hesse}}}, \bibinfo {author} {\bibfnamefont {T.}~\bibnamefont {{Neukirch}}},
  \bibinfo {author} {\bibfnamefont {K.}~\bibnamefont {{Schindler}}}, \bibinfo
  {author} {\bibfnamefont {M.}~\bibnamefont {{Kuznetsova}}}, \ and\ \bibinfo
  {author} {\bibfnamefont {S.}~\bibnamefont {{Zenitani}}},\ }\bibfield  {title}
  {\enquote {\bibinfo {title} {{The Diffusion Region in Collisionless Magnetic
  Reconnection}},}\ }\href {\doibase 10.1007/s11214-010-9740-1} {\bibfield
  {journal} {\bibinfo  {journal} {Spa. Sci. Rev.}\ }\textbf {\bibinfo {volume}
  {160}},\ \bibinfo {pages} {3--23} (\bibinfo {year} {2011})}\BibitemShut
  {NoStop}%
\bibitem [{\citenamefont {{Hesse}}\ \emph {et~al.}(2018)\citenamefont
  {{Hesse}}, \citenamefont {{Liu}}, \citenamefont {{Chen}}, \citenamefont
  {{Bessho}}, \citenamefont {{Wang}}, \citenamefont {{Burch}}, \citenamefont
  {{Moretto}}, \citenamefont {{Norgren}}, \citenamefont {{Genestreti}},
  \citenamefont {{Phan}},\ and\ \citenamefont {{Tenfjord}}}]{Hesse.2018}%
  \BibitemOpen
  \bibfield  {author} {\bibinfo {author} {\bibfnamefont {M.}~\bibnamefont
  {{Hesse}}}, \bibinfo {author} {\bibfnamefont {Y.~H.}\ \bibnamefont {{Liu}}},
  \bibinfo {author} {\bibfnamefont {L.~J.}\ \bibnamefont {{Chen}}}, \bibinfo
  {author} {\bibfnamefont {N.}~\bibnamefont {{Bessho}}}, \bibinfo {author}
  {\bibfnamefont {S.}~\bibnamefont {{Wang}}}, \bibinfo {author} {\bibfnamefont
  {J.~L.}\ \bibnamefont {{Burch}}}, \bibinfo {author} {\bibfnamefont
  {T.}~\bibnamefont {{Moretto}}}, \bibinfo {author} {\bibfnamefont
  {C.}~\bibnamefont {{Norgren}}}, \bibinfo {author} {\bibfnamefont {K.~J.}\
  \bibnamefont {{Genestreti}}}, \bibinfo {author} {\bibfnamefont {T.~D.}\
  \bibnamefont {{Phan}}}, \ and\ \bibinfo {author} {\bibfnamefont
  {P.}~\bibnamefont {{Tenfjord}}},\ }\bibfield  {title} {\enquote {\bibinfo
  {title} {{The physical foundation of the reconnection electric field}},}\
  }\href {\doibase 10.1063/1.5021461} {\bibfield  {journal} {\bibinfo
  {journal} {Physics of Plasmas}\ }\textbf {\bibinfo {volume} {25}},\ \bibinfo
  {eid} {032901} (\bibinfo {year} {2018})},\ \Eprint
  {http://arxiv.org/abs/1801.01090} {arXiv:1801.01090 [physics.space-ph]}
  \BibitemShut {NoStop}%
\bibitem [{\citenamefont {{Burch}}\ \emph {et~al.}(2016)\citenamefont
  {{Burch}}, \citenamefont {{Moore}}, \citenamefont {{Torbert}},\ and\
  \citenamefont {{Giles}}}]{Burch.2016a}%
  \BibitemOpen
  \bibfield  {author} {\bibinfo {author} {\bibfnamefont {J.~L.}\ \bibnamefont
  {{Burch}}}, \bibinfo {author} {\bibfnamefont {T.~E.}\ \bibnamefont
  {{Moore}}}, \bibinfo {author} {\bibfnamefont {R.~B.}\ \bibnamefont
  {{Torbert}}}, \ and\ \bibinfo {author} {\bibfnamefont {B.~L.}\ \bibnamefont
  {{Giles}}},\ }\bibfield  {title} {\enquote {\bibinfo {title} {{Magnetospheric
  Multiscale Overview and Science Objectives}},}\ }\href {\doibase
  10.1007/s11214-015-0164-9} {\bibfield  {journal} {\bibinfo  {journal} {Spa.
  Sci. Rev.}\ }\textbf {\bibinfo {volume} {199}},\ \bibinfo {pages} {5--21}
  (\bibinfo {year} {2016})}\BibitemShut {NoStop}%
\bibitem [{\citenamefont {{Torbert}}\ \emph {et~al.}(2018)\citenamefont
  {{Torbert}}, \citenamefont {{Burch}}, \citenamefont {{Phan}}, \citenamefont
  {{Hesse}}, \citenamefont {{Argall}}, \citenamefont {{Shuster}}, \citenamefont
  {{Ergun}}, \citenamefont {{Alm}}, \citenamefont {{Nakamura}}, \citenamefont
  {{Genestreti}}, \citenamefont {{Gershman}}, \citenamefont {{Paterson}},
  \citenamefont {{Turner}}, \citenamefont {{Cohen}}, \citenamefont {{Giles}},
  \citenamefont {{Pollock}}, \citenamefont {{Wang}}, \citenamefont {{Chen}},
  \citenamefont {{Stawarz}}, \citenamefont {{Eastwood}}, \citenamefont
  {{Hwang}}, \citenamefont {{Farrugia}}, \citenamefont {{Dors}}, \citenamefont
  {{Vaith}}, \citenamefont {{Mouikis}}, \citenamefont {{Ardakani}},
  \citenamefont {{Mauk}}, \citenamefont {{Fuselier}}, \citenamefont
  {{Russell}}, \citenamefont {{Strangeway}}, \citenamefont {{Moore}},
  \citenamefont {{Drake}}, \citenamefont {{Shay}}, \citenamefont
  {{Khotyaintsev}}, \citenamefont {{Lindqvist}}, \citenamefont {{Baumjohann}},
  \citenamefont {{Wilder}}, \citenamefont {{Ahmadi}}, \citenamefont
  {{Dorelli}}, \citenamefont {{Avanov}}, \citenamefont {{Oka}}, \citenamefont
  {{Baker}}, \citenamefont {{Fennell}}, \citenamefont {{Blake}}, \citenamefont
  {{Jaynes}}, \citenamefont {{Le Contel}}, \citenamefont {{Petrinec}},
  \citenamefont {{Lavraud}},\ and\ \citenamefont {{Saito}}}]{Torbert.2018}%
  \BibitemOpen
  \bibfield  {author} {\bibinfo {author} {\bibfnamefont {R.~B.}\ \bibnamefont
  {{Torbert}}}, \bibinfo {author} {\bibfnamefont {J.~L.}\ \bibnamefont
  {{Burch}}}, \bibinfo {author} {\bibfnamefont {T.~D.}\ \bibnamefont {{Phan}}},
  \bibinfo {author} {\bibfnamefont {M.}~\bibnamefont {{Hesse}}}, \bibinfo
  {author} {\bibfnamefont {M.~R.}\ \bibnamefont {{Argall}}}, \bibinfo {author}
  {\bibfnamefont {J.}~\bibnamefont {{Shuster}}}, \bibinfo {author}
  {\bibfnamefont {R.~E.}\ \bibnamefont {{Ergun}}}, \bibinfo {author}
  {\bibfnamefont {L.}~\bibnamefont {{Alm}}}, \bibinfo {author} {\bibfnamefont
  {R.}~\bibnamefont {{Nakamura}}}, \bibinfo {author} {\bibfnamefont {K.~J.}\
  \bibnamefont {{Genestreti}}}, \bibinfo {author} {\bibfnamefont {D.~J.}\
  \bibnamefont {{Gershman}}}, \bibinfo {author} {\bibfnamefont {W.~R.}\
  \bibnamefont {{Paterson}}}, \bibinfo {author} {\bibfnamefont {D.~L.}\
  \bibnamefont {{Turner}}}, \bibinfo {author} {\bibfnamefont {I.}~\bibnamefont
  {{Cohen}}}, \bibinfo {author} {\bibfnamefont {B.~L.}\ \bibnamefont
  {{Giles}}}, \bibinfo {author} {\bibfnamefont {C.~J.}\ \bibnamefont
  {{Pollock}}}, \bibinfo {author} {\bibfnamefont {S.}~\bibnamefont {{Wang}}},
  \bibinfo {author} {\bibfnamefont {L.~J.}\ \bibnamefont {{Chen}}}, \bibinfo
  {author} {\bibfnamefont {J.~E.}\ \bibnamefont {{Stawarz}}}, \bibinfo {author}
  {\bibfnamefont {J.~P.}\ \bibnamefont {{Eastwood}}}, \bibinfo {author}
  {\bibfnamefont {K.~J.}\ \bibnamefont {{Hwang}}}, \bibinfo {author}
  {\bibfnamefont {C.}~\bibnamefont {{Farrugia}}}, \bibinfo {author}
  {\bibfnamefont {I.}~\bibnamefont {{Dors}}}, \bibinfo {author} {\bibfnamefont
  {H.}~\bibnamefont {{Vaith}}}, \bibinfo {author} {\bibfnamefont
  {C.}~\bibnamefont {{Mouikis}}}, \bibinfo {author} {\bibfnamefont
  {A.}~\bibnamefont {{Ardakani}}}, \bibinfo {author} {\bibfnamefont {B.~H.}\
  \bibnamefont {{Mauk}}}, \bibinfo {author} {\bibfnamefont {S.~A.}\
  \bibnamefont {{Fuselier}}}, \bibinfo {author} {\bibfnamefont {C.~T.}\
  \bibnamefont {{Russell}}}, \bibinfo {author} {\bibfnamefont {R.~J.}\
  \bibnamefont {{Strangeway}}}, \bibinfo {author} {\bibfnamefont {T.~E.}\
  \bibnamefont {{Moore}}}, \bibinfo {author} {\bibfnamefont {J.~F.}\
  \bibnamefont {{Drake}}}, \bibinfo {author} {\bibfnamefont {M.~A.}\
  \bibnamefont {{Shay}}}, \bibinfo {author} {\bibfnamefont {Y.~V.}\
  \bibnamefont {{Khotyaintsev}}}, \bibinfo {author} {\bibfnamefont {P.~A.}\
  \bibnamefont {{Lindqvist}}}, \bibinfo {author} {\bibfnamefont
  {W.}~\bibnamefont {{Baumjohann}}}, \bibinfo {author} {\bibfnamefont {F.~D.}\
  \bibnamefont {{Wilder}}}, \bibinfo {author} {\bibfnamefont {N.}~\bibnamefont
  {{Ahmadi}}}, \bibinfo {author} {\bibfnamefont {J.~C.}\ \bibnamefont
  {{Dorelli}}}, \bibinfo {author} {\bibfnamefont {L.~A.}\ \bibnamefont
  {{Avanov}}}, \bibinfo {author} {\bibfnamefont {M.}~\bibnamefont {{Oka}}},
  \bibinfo {author} {\bibfnamefont {D.~N.}\ \bibnamefont {{Baker}}}, \bibinfo
  {author} {\bibfnamefont {J.~F.}\ \bibnamefont {{Fennell}}}, \bibinfo {author}
  {\bibfnamefont {J.~B.}\ \bibnamefont {{Blake}}}, \bibinfo {author}
  {\bibfnamefont {A.~N.}\ \bibnamefont {{Jaynes}}}, \bibinfo {author}
  {\bibfnamefont {O.}~\bibnamefont {{Le Contel}}}, \bibinfo {author}
  {\bibfnamefont {S.~M.}\ \bibnamefont {{Petrinec}}}, \bibinfo {author}
  {\bibfnamefont {B.}~\bibnamefont {{Lavraud}}}, \ and\ \bibinfo {author}
  {\bibfnamefont {Y.}~\bibnamefont {{Saito}}},\ }\bibfield  {title} {\enquote
  {\bibinfo {title} {{Electron-scale dynamics of the diffusion region during
  symmetric magnetic reconnection in space}},}\ }\href {\doibase
  10.1126/science.aat2998} {\bibfield  {journal} {\bibinfo  {journal}
  {Science}\ }\textbf {\bibinfo {volume} {362}},\ \bibinfo {pages} {1391--1395}
  (\bibinfo {year} {2018})},\ \Eprint {http://arxiv.org/abs/1809.06932}
  {arXiv:1809.06932 [physics.space-ph]} \BibitemShut {NoStop}%
\bibitem [{\citenamefont {Genestreti}\ \emph {et~al.}(018b)\citenamefont
  {Genestreti}, \citenamefont {Nakamura}, \citenamefont {Nakamura},
  \citenamefont {Denton}, \citenamefont {Torbert}, \citenamefont {Burch},
  \citenamefont {Plaschke}, \citenamefont {Fuselier}, \citenamefont {Ergun},
  \citenamefont {Giles},\ and\ \citenamefont {Russell}}]{Genestreti.2018c}%
  \BibitemOpen
  \bibfield  {author} {\bibinfo {author} {\bibfnamefont {K.~J.}\ \bibnamefont
  {Genestreti}}, \bibinfo {author} {\bibfnamefont {T.~K.~M.}\ \bibnamefont
  {Nakamura}}, \bibinfo {author} {\bibfnamefont {R.}~\bibnamefont {Nakamura}},
  \bibinfo {author} {\bibfnamefont {R.~E.}\ \bibnamefont {Denton}}, \bibinfo
  {author} {\bibfnamefont {R.~B.}\ \bibnamefont {Torbert}}, \bibinfo {author}
  {\bibfnamefont {J.~L.}\ \bibnamefont {Burch}}, \bibinfo {author}
  {\bibfnamefont {F.}~\bibnamefont {Plaschke}}, \bibinfo {author}
  {\bibfnamefont {S.~A.}\ \bibnamefont {Fuselier}}, \bibinfo {author}
  {\bibfnamefont {R.~E.}\ \bibnamefont {Ergun}}, \bibinfo {author}
  {\bibfnamefont {B.~L.}\ \bibnamefont {Giles}}, \ and\ \bibinfo {author}
  {\bibfnamefont {C.~T.}\ \bibnamefont {Russell}},\ }\bibfield  {title}
  {\enquote {\bibinfo {title} {How accurately can we measure the reconnection
  rate em for the mms diffusion region event of 11 july 2017?}}\ }\href
  {\doibase 10.1029/2018JA025711} {\bibfield  {journal} {\bibinfo  {journal}
  {Journal of Geophysical Research: Space Physics}\ }\textbf {\bibinfo {volume}
  {123}},\ \bibinfo {pages} {9130--9149} (\bibinfo {year} {2018b})},\ \Eprint
  {http://arxiv.org/abs/https://agupubs.onlinelibrary.wiley.com/doi/pdf/10.1029/2018JA025711}
  {https://agupubs.onlinelibrary.wiley.com/doi/pdf/10.1029/2018JA025711}
  \BibitemShut {NoStop}%
\bibitem [{\citenamefont {Nakamura}\ \emph {et~al.}(2018)\citenamefont
  {Nakamura}, \citenamefont {Genestreti}, \citenamefont {Liu}, \citenamefont
  {Nakamura}, \citenamefont {Teh}, \citenamefont {Hasegawa}, \citenamefont
  {Daughton}, \citenamefont {Hesse}, \citenamefont {Torbert}, \citenamefont
  {Burch},\ and\ \citenamefont {Giles}}]{NakamuraTKM.2018}%
  \BibitemOpen
  \bibfield  {author} {\bibinfo {author} {\bibfnamefont {T.~K.~M.}\
  \bibnamefont {Nakamura}}, \bibinfo {author} {\bibfnamefont {K.~J.}\
  \bibnamefont {Genestreti}}, \bibinfo {author} {\bibfnamefont {Y.-H.}\
  \bibnamefont {Liu}}, \bibinfo {author} {\bibfnamefont {R.}~\bibnamefont
  {Nakamura}}, \bibinfo {author} {\bibfnamefont {W.-L.}\ \bibnamefont {Teh}},
  \bibinfo {author} {\bibfnamefont {H.}~\bibnamefont {Hasegawa}}, \bibinfo
  {author} {\bibfnamefont {W.}~\bibnamefont {Daughton}}, \bibinfo {author}
  {\bibfnamefont {M.}~\bibnamefont {Hesse}}, \bibinfo {author} {\bibfnamefont
  {R.~B.}\ \bibnamefont {Torbert}}, \bibinfo {author} {\bibfnamefont {J.~L.}\
  \bibnamefont {Burch}}, \ and\ \bibinfo {author} {\bibfnamefont {B.~L.}\
  \bibnamefont {Giles}},\ }\bibfield  {title} {\enquote {\bibinfo {title}
  {Measurement of the magnetic reconnection rate in the earth's magnetotail},}\
  }\href {\doibase 10.1029/2018JA025713} {\bibfield  {journal} {\bibinfo
  {journal} {Journal of Geophysical Research: Space Physics}\ }\textbf
  {\bibinfo {volume} {123}},\ \bibinfo {pages} {9150--9168} (\bibinfo {year}
  {2018})}\BibitemShut {NoStop}%
\bibitem [{\citenamefont {Bessho}\ \emph {et~al.}(2018)\citenamefont {Bessho},
  \citenamefont {Chen}, \citenamefont {Wang},\ and\ \citenamefont
  {Hesse}}]{Bessho.2018}%
  \BibitemOpen
  \bibfield  {author} {\bibinfo {author} {\bibfnamefont {N.}~\bibnamefont
  {Bessho}}, \bibinfo {author} {\bibfnamefont {L.-J.}\ \bibnamefont {Chen}},
  \bibinfo {author} {\bibfnamefont {S.}~\bibnamefont {Wang}}, \ and\ \bibinfo
  {author} {\bibfnamefont {M.}~\bibnamefont {Hesse}},\ }\bibfield  {title}
  {\enquote {\bibinfo {title} {Effect of the reconnection electric field on
  electron distribution functions in the diffusion region of magnetotail
  reconnection},}\ }\href {\doibase https://doi.org/10.1029/2018GL081216}
  {\bibfield  {journal} {\bibinfo  {journal} {Geophysical Research Letters}\
  }\textbf {\bibinfo {volume} {45}},\ \bibinfo {pages} {12,142--12,152}
  (\bibinfo {year} {2018})},\ \Eprint
  {http://arxiv.org/abs/https://agupubs.onlinelibrary.wiley.com/doi/pdf/10.1029/2018GL081216}
  {https://agupubs.onlinelibrary.wiley.com/doi/pdf/10.1029/2018GL081216}
  \BibitemShut {NoStop}%
\bibitem [{\citenamefont {Burch}\ \emph {et~al.}(2016)\citenamefont {Burch},
  \citenamefont {Torbert}, \citenamefont {Phan}, \citenamefont {Chen},
  \citenamefont {Moore}, \citenamefont {Ergun}, \citenamefont {Eastwood},
  \citenamefont {Gershman}, \citenamefont {Cassak}, \citenamefont {Argall},
  \citenamefont {Wang}, \citenamefont {Hesse}, \citenamefont {Pollock},
  \citenamefont {Giles}, \citenamefont {Nakamura}, \citenamefont {Mauk},
  \citenamefont {Fuselier}, \citenamefont {Russell}, \citenamefont
  {Strangeway}, \citenamefont {Drake}, \citenamefont {Shay}, \citenamefont
  {Khotyaintsev}, \citenamefont {Lindqvist}, \citenamefont {Marklund},
  \citenamefont {Wilder}, \citenamefont {Young}, \citenamefont {Torkar},
  \citenamefont {Goldstein}, \citenamefont {Dorelli}, \citenamefont {Avanov},
  \citenamefont {Oka}, \citenamefont {Baker}, \citenamefont {Jaynes},
  \citenamefont {Goodrich}, \citenamefont {Cohen}, \citenamefont {Turner},
  \citenamefont {Fennell}, \citenamefont {Blake}, \citenamefont {Clemmons},
  \citenamefont {Goldman}, \citenamefont {Newman}, \citenamefont {Petrinec},
  \citenamefont {Trattner}, \citenamefont {Lavraud}, \citenamefont {Reiff},
  \citenamefont {Baumjohann}, \citenamefont {Magnes}, \citenamefont {Steller},
  \citenamefont {Lewis}, \citenamefont {Saito}, \citenamefont {Coffey},\ and\
  \citenamefont {Chandler}}]{Burch.2016b}%
  \BibitemOpen
  \bibfield  {author} {\bibinfo {author} {\bibfnamefont {J.~L.}\ \bibnamefont
  {Burch}}, \bibinfo {author} {\bibfnamefont {R.~B.}\ \bibnamefont {Torbert}},
  \bibinfo {author} {\bibfnamefont {T.~D.}\ \bibnamefont {Phan}}, \bibinfo
  {author} {\bibfnamefont {L.-J.}\ \bibnamefont {Chen}}, \bibinfo {author}
  {\bibfnamefont {T.~E.}\ \bibnamefont {Moore}}, \bibinfo {author}
  {\bibfnamefont {R.~E.}\ \bibnamefont {Ergun}}, \bibinfo {author}
  {\bibfnamefont {J.~P.}\ \bibnamefont {Eastwood}}, \bibinfo {author}
  {\bibfnamefont {D.~J.}\ \bibnamefont {Gershman}}, \bibinfo {author}
  {\bibfnamefont {P.~A.}\ \bibnamefont {Cassak}}, \bibinfo {author}
  {\bibfnamefont {M.~R.}\ \bibnamefont {Argall}}, \bibinfo {author}
  {\bibfnamefont {S.}~\bibnamefont {Wang}}, \bibinfo {author} {\bibfnamefont
  {M.}~\bibnamefont {Hesse}}, \bibinfo {author} {\bibfnamefont {C.~J.}\
  \bibnamefont {Pollock}}, \bibinfo {author} {\bibfnamefont {B.~L.}\
  \bibnamefont {Giles}}, \bibinfo {author} {\bibfnamefont {R.}~\bibnamefont
  {Nakamura}}, \bibinfo {author} {\bibfnamefont {B.~H.}\ \bibnamefont {Mauk}},
  \bibinfo {author} {\bibfnamefont {S.~A.}\ \bibnamefont {Fuselier}}, \bibinfo
  {author} {\bibfnamefont {C.~T.}\ \bibnamefont {Russell}}, \bibinfo {author}
  {\bibfnamefont {R.~J.}\ \bibnamefont {Strangeway}}, \bibinfo {author}
  {\bibfnamefont {J.~F.}\ \bibnamefont {Drake}}, \bibinfo {author}
  {\bibfnamefont {M.~A.}\ \bibnamefont {Shay}}, \bibinfo {author}
  {\bibfnamefont {Y.~V.}\ \bibnamefont {Khotyaintsev}}, \bibinfo {author}
  {\bibfnamefont {P.-A.}\ \bibnamefont {Lindqvist}}, \bibinfo {author}
  {\bibfnamefont {G.}~\bibnamefont {Marklund}}, \bibinfo {author}
  {\bibfnamefont {F.~D.}\ \bibnamefont {Wilder}}, \bibinfo {author}
  {\bibfnamefont {D.~T.}\ \bibnamefont {Young}}, \bibinfo {author}
  {\bibfnamefont {K.}~\bibnamefont {Torkar}}, \bibinfo {author} {\bibfnamefont
  {J.}~\bibnamefont {Goldstein}}, \bibinfo {author} {\bibfnamefont {J.~C.}\
  \bibnamefont {Dorelli}}, \bibinfo {author} {\bibfnamefont {L.~A.}\
  \bibnamefont {Avanov}}, \bibinfo {author} {\bibfnamefont {M.}~\bibnamefont
  {Oka}}, \bibinfo {author} {\bibfnamefont {D.~N.}\ \bibnamefont {Baker}},
  \bibinfo {author} {\bibfnamefont {A.~N.}\ \bibnamefont {Jaynes}}, \bibinfo
  {author} {\bibfnamefont {K.~A.}\ \bibnamefont {Goodrich}}, \bibinfo {author}
  {\bibfnamefont {I.~J.}\ \bibnamefont {Cohen}}, \bibinfo {author}
  {\bibfnamefont {D.~L.}\ \bibnamefont {Turner}}, \bibinfo {author}
  {\bibfnamefont {J.~F.}\ \bibnamefont {Fennell}}, \bibinfo {author}
  {\bibfnamefont {J.~B.}\ \bibnamefont {Blake}}, \bibinfo {author}
  {\bibfnamefont {J.}~\bibnamefont {Clemmons}}, \bibinfo {author}
  {\bibfnamefont {M.}~\bibnamefont {Goldman}}, \bibinfo {author} {\bibfnamefont
  {D.}~\bibnamefont {Newman}}, \bibinfo {author} {\bibfnamefont {S.~M.}\
  \bibnamefont {Petrinec}}, \bibinfo {author} {\bibfnamefont {K.~J.}\
  \bibnamefont {Trattner}}, \bibinfo {author} {\bibfnamefont {B.}~\bibnamefont
  {Lavraud}}, \bibinfo {author} {\bibfnamefont {P.~H.}\ \bibnamefont {Reiff}},
  \bibinfo {author} {\bibfnamefont {W.}~\bibnamefont {Baumjohann}}, \bibinfo
  {author} {\bibfnamefont {W.}~\bibnamefont {Magnes}}, \bibinfo {author}
  {\bibfnamefont {M.}~\bibnamefont {Steller}}, \bibinfo {author} {\bibfnamefont
  {W.}~\bibnamefont {Lewis}}, \bibinfo {author} {\bibfnamefont
  {Y.}~\bibnamefont {Saito}}, \bibinfo {author} {\bibfnamefont
  {V.}~\bibnamefont {Coffey}}, \ and\ \bibinfo {author} {\bibfnamefont
  {M.}~\bibnamefont {Chandler}},\ }\bibfield  {title} {\enquote {\bibinfo
  {title} {Electron-scale measurements of magnetic reconnection in space},}\
  }\href {\doibase 10.1126/science.aaf2939} {\bibfield  {journal} {\bibinfo
  {journal} {Science}\ } (\bibinfo {year} {2016}),\ 10.1126/science.aaf2939},\
  \Eprint
  {http://arxiv.org/abs/http://science.sciencemag.org/content/early/2016/05/10/science.aaf2939.full.pdf}
  {http://science.sciencemag.org/content/early/2016/05/10/science.aaf2939.full.pdf}
  \BibitemShut {NoStop}%
\bibitem [{\citenamefont {Burch}\ \emph {et~al.}(2018)\citenamefont {Burch},
  \citenamefont {Ergun}, \citenamefont {Cassak}, \citenamefont {Webster},
  \citenamefont {Torbert}, \citenamefont {Giles}, \citenamefont {Dorelli},
  \citenamefont {Rager}, \citenamefont {Hwang}, \citenamefont {Phan},
  \citenamefont {Genestreti}, \citenamefont {Allen}, \citenamefont {Chen},
  \citenamefont {Wang}, \citenamefont {Gershman}, \citenamefont {Le~Contel},
  \citenamefont {Russell}, \citenamefont {Strangeway}, \citenamefont {Wilder},
  \citenamefont {Graham}, \citenamefont {Hesse}, \citenamefont {Drake},
  \citenamefont {Swisdak}, \citenamefont {Price}, \citenamefont {Shay},
  \citenamefont {Lindqvist}, \citenamefont {Pollock}, \citenamefont {Denton},\
  and\ \citenamefont {Newman}}]{Burch.2018a}%
  \BibitemOpen
  \bibfield  {author} {\bibinfo {author} {\bibfnamefont {J.~L.}\ \bibnamefont
  {Burch}}, \bibinfo {author} {\bibfnamefont {R.~E.}\ \bibnamefont {Ergun}},
  \bibinfo {author} {\bibfnamefont {P.~A.}\ \bibnamefont {Cassak}}, \bibinfo
  {author} {\bibfnamefont {J.~M.}\ \bibnamefont {Webster}}, \bibinfo {author}
  {\bibfnamefont {R.~B.}\ \bibnamefont {Torbert}}, \bibinfo {author}
  {\bibfnamefont {B.~L.}\ \bibnamefont {Giles}}, \bibinfo {author}
  {\bibfnamefont {J.~C.}\ \bibnamefont {Dorelli}}, \bibinfo {author}
  {\bibfnamefont {A.~C.}\ \bibnamefont {Rager}}, \bibinfo {author}
  {\bibfnamefont {K.-J.}\ \bibnamefont {Hwang}}, \bibinfo {author}
  {\bibfnamefont {T.~D.}\ \bibnamefont {Phan}}, \bibinfo {author}
  {\bibfnamefont {K.~J.}\ \bibnamefont {Genestreti}}, \bibinfo {author}
  {\bibfnamefont {R.~C.}\ \bibnamefont {Allen}}, \bibinfo {author}
  {\bibfnamefont {L.-J.}\ \bibnamefont {Chen}}, \bibinfo {author}
  {\bibfnamefont {S.}~\bibnamefont {Wang}}, \bibinfo {author} {\bibfnamefont
  {D.}~\bibnamefont {Gershman}}, \bibinfo {author} {\bibfnamefont
  {O.}~\bibnamefont {Le~Contel}}, \bibinfo {author} {\bibfnamefont {C.~T.}\
  \bibnamefont {Russell}}, \bibinfo {author} {\bibfnamefont {R.~J.}\
  \bibnamefont {Strangeway}}, \bibinfo {author} {\bibfnamefont {F.~D.}\
  \bibnamefont {Wilder}}, \bibinfo {author} {\bibfnamefont {D.~B.}\
  \bibnamefont {Graham}}, \bibinfo {author} {\bibfnamefont {M.}~\bibnamefont
  {Hesse}}, \bibinfo {author} {\bibfnamefont {J.~F.}\ \bibnamefont {Drake}},
  \bibinfo {author} {\bibfnamefont {M.}~\bibnamefont {Swisdak}}, \bibinfo
  {author} {\bibfnamefont {L.~M.}\ \bibnamefont {Price}}, \bibinfo {author}
  {\bibfnamefont {M.~A.}\ \bibnamefont {Shay}}, \bibinfo {author}
  {\bibfnamefont {P.-A.}\ \bibnamefont {Lindqvist}}, \bibinfo {author}
  {\bibfnamefont {C.~J.}\ \bibnamefont {Pollock}}, \bibinfo {author}
  {\bibfnamefont {R.~E.}\ \bibnamefont {Denton}}, \ and\ \bibinfo {author}
  {\bibfnamefont {D.~L.}\ \bibnamefont {Newman}},\ }\bibfield  {title}
  {\enquote {\bibinfo {title} {Localized oscillatory energy conversion in
  magnetopause reconnection},}\ }\href {\doibase 10.1002/2017GL076809}
  {\bibfield  {journal} {\bibinfo  {journal} {Geophysical Research Letters}\
  }\textbf {\bibinfo {volume} {45}},\ \bibinfo {pages} {1237--1245} (\bibinfo
  {year} {2018})}\BibitemShut {NoStop}%
\bibitem [{\citenamefont {{Burch}}\ \emph {et~al.}(2018)\citenamefont
  {{Burch}}, \citenamefont {{Webster}}, \citenamefont {{Genestreti}},
  \citenamefont {{Torbert}}, \citenamefont {{Giles}}, \citenamefont
  {{Fuselier}}, \citenamefont {{Dorelli}}, \citenamefont {{Rager}},
  \citenamefont {{Phan}}, \citenamefont {{Allen}}, \citenamefont {{Chen}},
  \citenamefont {{Wang}}, \citenamefont {{Le Contel}}, \citenamefont
  {{Russell}}, \citenamefont {{Strangeway}}, \citenamefont {{Ergun}},
  \citenamefont {{Jaynes}}, \citenamefont {{Lindqvist}}, \citenamefont
  {{Graham}}, \citenamefont {{Wilder}}, \citenamefont {{Hwang}},\ and\
  \citenamefont {{Goldstein}}}]{Burch.2018b}%
  \BibitemOpen
  \bibfield  {author} {\bibinfo {author} {\bibfnamefont {J.~L.}\ \bibnamefont
  {{Burch}}}, \bibinfo {author} {\bibfnamefont {J.~M.}\ \bibnamefont
  {{Webster}}}, \bibinfo {author} {\bibfnamefont {K.~J.}\ \bibnamefont
  {{Genestreti}}}, \bibinfo {author} {\bibfnamefont {R.~B.}\ \bibnamefont
  {{Torbert}}}, \bibinfo {author} {\bibfnamefont {B.~L.}\ \bibnamefont
  {{Giles}}}, \bibinfo {author} {\bibfnamefont {S.~A.}\ \bibnamefont
  {{Fuselier}}}, \bibinfo {author} {\bibfnamefont {J.~C.}\ \bibnamefont
  {{Dorelli}}}, \bibinfo {author} {\bibfnamefont {A.~C.}\ \bibnamefont
  {{Rager}}}, \bibinfo {author} {\bibfnamefont {T.~D.}\ \bibnamefont {{Phan}}},
  \bibinfo {author} {\bibfnamefont {R.~C.}\ \bibnamefont {{Allen}}}, \bibinfo
  {author} {\bibfnamefont {L.~J.}\ \bibnamefont {{Chen}}}, \bibinfo {author}
  {\bibfnamefont {S.}~\bibnamefont {{Wang}}}, \bibinfo {author} {\bibfnamefont
  {O.}~\bibnamefont {{Le Contel}}}, \bibinfo {author} {\bibfnamefont {C.~T.}\
  \bibnamefont {{Russell}}}, \bibinfo {author} {\bibfnamefont {R.~J.}\
  \bibnamefont {{Strangeway}}}, \bibinfo {author} {\bibfnamefont {R.~E.}\
  \bibnamefont {{Ergun}}}, \bibinfo {author} {\bibfnamefont {A.~N.}\
  \bibnamefont {{Jaynes}}}, \bibinfo {author} {\bibfnamefont {P.~A.}\
  \bibnamefont {{Lindqvist}}}, \bibinfo {author} {\bibfnamefont {D.~B.}\
  \bibnamefont {{Graham}}}, \bibinfo {author} {\bibfnamefont {F.~D.}\
  \bibnamefont {{Wilder}}}, \bibinfo {author} {\bibfnamefont {K.~J.}\
  \bibnamefont {{Hwang}}}, \ and\ \bibinfo {author} {\bibfnamefont
  {J.}~\bibnamefont {{Goldstein}}},\ }\bibfield  {title} {\enquote {\bibinfo
  {title} {{Wave Phenomena and Beam-Plasma Interactions at the Magnetopause
  Reconnection Region}},}\ }\href {\doibase 10.1002/2017JA024789} {\bibfield
  {journal} {\bibinfo  {journal} {Journal of Geophysical Research (Space
  Physics)}\ }\textbf {\bibinfo {volume} {123}},\ \bibinfo {pages} {1118--1133}
  (\bibinfo {year} {2018})}\BibitemShut {NoStop}%
\bibitem [{\citenamefont {Burch}\ \emph {et~al.}(2020)\citenamefont {Burch},
  \citenamefont {Webster}, \citenamefont {Hesse}, \citenamefont {Genestreti},
  \citenamefont {Denton}, \citenamefont {Phan}, \citenamefont {Hasegawa},
  \citenamefont {Cassak}, \citenamefont {Torbert}, \citenamefont {Giles},
  \citenamefont {Gershman}, \citenamefont {Ergun}, \citenamefont {Russell},
  \citenamefont {Strangeway}, \citenamefont {Le~Contel}, \citenamefont
  {Pritchard}, \citenamefont {Marshall}, \citenamefont {Hwang}, \citenamefont
  {Dokgo}, \citenamefont {Fuselier}, \citenamefont {Chen}, \citenamefont
  {Wang}, \citenamefont {Swisdak}, \citenamefont {Drake}, \citenamefont
  {Argall}, \citenamefont {Trattner}, \citenamefont {Yamada},\ and\
  \citenamefont {Paschmann}}]{Burch.2020}%
  \BibitemOpen
  \bibfield  {author} {\bibinfo {author} {\bibfnamefont {J.~L.}\ \bibnamefont
  {Burch}}, \bibinfo {author} {\bibfnamefont {J.~M.}\ \bibnamefont {Webster}},
  \bibinfo {author} {\bibfnamefont {M.}~\bibnamefont {Hesse}}, \bibinfo
  {author} {\bibfnamefont {K.~J.}\ \bibnamefont {Genestreti}}, \bibinfo
  {author} {\bibfnamefont {R.~E.}\ \bibnamefont {Denton}}, \bibinfo {author}
  {\bibfnamefont {T.~D.}\ \bibnamefont {Phan}}, \bibinfo {author}
  {\bibfnamefont {H.}~\bibnamefont {Hasegawa}}, \bibinfo {author}
  {\bibfnamefont {P.~A.}\ \bibnamefont {Cassak}}, \bibinfo {author}
  {\bibfnamefont {R.~B.}\ \bibnamefont {Torbert}}, \bibinfo {author}
  {\bibfnamefont {B.~L.}\ \bibnamefont {Giles}}, \bibinfo {author}
  {\bibfnamefont {D.~J.}\ \bibnamefont {Gershman}}, \bibinfo {author}
  {\bibfnamefont {R.~E.}\ \bibnamefont {Ergun}}, \bibinfo {author}
  {\bibfnamefont {C.~T.}\ \bibnamefont {Russell}}, \bibinfo {author}
  {\bibfnamefont {R.~J.}\ \bibnamefont {Strangeway}}, \bibinfo {author}
  {\bibfnamefont {O.}~\bibnamefont {Le~Contel}}, \bibinfo {author}
  {\bibfnamefont {K.~R.}\ \bibnamefont {Pritchard}}, \bibinfo {author}
  {\bibfnamefont {A.~T.}\ \bibnamefont {Marshall}}, \bibinfo {author}
  {\bibfnamefont {K.-J.}\ \bibnamefont {Hwang}}, \bibinfo {author}
  {\bibfnamefont {K.}~\bibnamefont {Dokgo}}, \bibinfo {author} {\bibfnamefont
  {S.~A.}\ \bibnamefont {Fuselier}}, \bibinfo {author} {\bibfnamefont {L.-J.}\
  \bibnamefont {Chen}}, \bibinfo {author} {\bibfnamefont {S.}~\bibnamefont
  {Wang}}, \bibinfo {author} {\bibfnamefont {M.}~\bibnamefont {Swisdak}},
  \bibinfo {author} {\bibfnamefont {J.~F.}\ \bibnamefont {Drake}}, \bibinfo
  {author} {\bibfnamefont {M.~R.}\ \bibnamefont {Argall}}, \bibinfo {author}
  {\bibfnamefont {K.~J.}\ \bibnamefont {Trattner}}, \bibinfo {author}
  {\bibfnamefont {M.}~\bibnamefont {Yamada}}, \ and\ \bibinfo {author}
  {\bibfnamefont {G.}~\bibnamefont {Paschmann}},\ }\bibfield  {title} {\enquote
  {\bibinfo {title} {Electron inflow velocities and reconnection rates at
  earth's magnetopause and magnetosheath},}\ }\href {\doibase
  https://doi.org/10.1029/2020GL089082} {\bibfield  {journal} {\bibinfo
  {journal} {Geophysical Research Letters}\ }\textbf {\bibinfo {volume} {47}},\
  \bibinfo {pages} {e2020GL089082} (\bibinfo {year} {2020})},\ \bibinfo {note}
  {e2020GL089082 2020GL089082},\ \Eprint
  {http://arxiv.org/abs/https://agupubs.onlinelibrary.wiley.com/doi/pdf/10.1029/2020GL089082}
  {https://agupubs.onlinelibrary.wiley.com/doi/pdf/10.1029/2020GL089082}
  \BibitemShut {NoStop}%
\bibitem [{\citenamefont {Cassak}\ \emph {et~al.}(017a)\citenamefont {Cassak},
  \citenamefont {Genestreti}, \citenamefont {Burch}, \citenamefont {Phan},
  \citenamefont {Shay}, \citenamefont {Swisdak}, \citenamefont {Drake},
  \citenamefont {Price}, \citenamefont {Eriksson}, \citenamefont {Ergun},
  \citenamefont {Anderson}, \citenamefont {Merkin},\ and\ \citenamefont
  {Komar}}]{Cassak.2017a}%
  \BibitemOpen
  \bibfield  {author} {\bibinfo {author} {\bibfnamefont {P.~A.}\ \bibnamefont
  {Cassak}}, \bibinfo {author} {\bibfnamefont {K.~J.}\ \bibnamefont
  {Genestreti}}, \bibinfo {author} {\bibfnamefont {J.~L.}\ \bibnamefont
  {Burch}}, \bibinfo {author} {\bibfnamefont {T.-D.}\ \bibnamefont {Phan}},
  \bibinfo {author} {\bibfnamefont {M.~A.}\ \bibnamefont {Shay}}, \bibinfo
  {author} {\bibfnamefont {M.}~\bibnamefont {Swisdak}}, \bibinfo {author}
  {\bibfnamefont {J.~F.}\ \bibnamefont {Drake}}, \bibinfo {author}
  {\bibfnamefont {L.}~\bibnamefont {Price}}, \bibinfo {author} {\bibfnamefont
  {S.}~\bibnamefont {Eriksson}}, \bibinfo {author} {\bibfnamefont {R.~E.}\
  \bibnamefont {Ergun}}, \bibinfo {author} {\bibfnamefont {B.~J.}\ \bibnamefont
  {Anderson}}, \bibinfo {author} {\bibfnamefont {V.~G.}\ \bibnamefont
  {Merkin}}, \ and\ \bibinfo {author} {\bibfnamefont {C.~M.}\ \bibnamefont
  {Komar}},\ }\bibfield  {title} {\enquote {\bibinfo {title} {The effect of a
  guide field on local energy conversion during asymmetric magnetic
  reconnection: Particle-in-cell simulations},}\ }\href {\doibase
  10.1002/2017JA024555} {\bibfield  {journal} {\bibinfo  {journal} {Journal of
  Geophysical Research: Space Physics}\ }\textbf {\bibinfo {volume} {122}},\
  \bibinfo {pages} {11,523--11,542} (\bibinfo {year} {2017a})},\ \Eprint
  {http://arxiv.org/abs/https://agupubs.onlinelibrary.wiley.com/doi/pdf/10.1002/2017JA024555}
  {https://agupubs.onlinelibrary.wiley.com/doi/pdf/10.1002/2017JA024555}
  \BibitemShut {NoStop}%
\bibitem [{\citenamefont {Genestreti}\ \emph {et~al.}(2017)\citenamefont
  {Genestreti}, \citenamefont {Burch}, \citenamefont {Cassak}, \citenamefont
  {Torbert}, \citenamefont {Ergun}, \citenamefont {Varsani}, \citenamefont
  {Phan}, \citenamefont {Giles}, \citenamefont {Russell}, \citenamefont {Wang},
  \citenamefont {Akhavan-Tafti},\ and\ \citenamefont
  {Allen}}]{Genestreti.2017}%
  \BibitemOpen
  \bibfield  {author} {\bibinfo {author} {\bibfnamefont {K.~J.}\ \bibnamefont
  {Genestreti}}, \bibinfo {author} {\bibfnamefont {J.~L.}\ \bibnamefont
  {Burch}}, \bibinfo {author} {\bibfnamefont {P.~A.}\ \bibnamefont {Cassak}},
  \bibinfo {author} {\bibfnamefont {R.~B.}\ \bibnamefont {Torbert}}, \bibinfo
  {author} {\bibfnamefont {R.~E.}\ \bibnamefont {Ergun}}, \bibinfo {author}
  {\bibfnamefont {A.}~\bibnamefont {Varsani}}, \bibinfo {author} {\bibfnamefont
  {T.~D.}\ \bibnamefont {Phan}}, \bibinfo {author} {\bibfnamefont {B.~L.}\
  \bibnamefont {Giles}}, \bibinfo {author} {\bibfnamefont {C.~T.}\ \bibnamefont
  {Russell}}, \bibinfo {author} {\bibfnamefont {S.}~\bibnamefont {Wang}},
  \bibinfo {author} {\bibfnamefont {M.}~\bibnamefont {Akhavan-Tafti}}, \ and\
  \bibinfo {author} {\bibfnamefont {R.~C.}\ \bibnamefont {Allen}},\ }\bibfield
  {title} {\enquote {\bibinfo {title} {The effect of a guide field on local
  energy conversion during asymmetric magnetic reconnection: Mms
  observations},}\ }\href {\doibase 10.1002/2017JA024247} {\bibfield  {journal}
  {\bibinfo  {journal} {Journal of Geophysical Research: Space Physics}\
  }\textbf {\bibinfo {volume} {122}},\ \bibinfo {pages} {11,342--11,353}
  (\bibinfo {year} {2017})},\ \Eprint
  {http://arxiv.org/abs/https://agupubs.onlinelibrary.wiley.com/doi/pdf/10.1002/2017JA024247}
  {https://agupubs.onlinelibrary.wiley.com/doi/pdf/10.1002/2017JA024247}
  \BibitemShut {NoStop}%
\bibitem [{\citenamefont {Genestreti}\ \emph {et~al.}(018a)\citenamefont
  {Genestreti}, \citenamefont {Varsani}, \citenamefont {Burch}, \citenamefont
  {Cassak}, \citenamefont {Torbert}, \citenamefont {Nakamura}, \citenamefont
  {Ergun}, \citenamefont {Phan}, \citenamefont {Toledo-Redondo}, \citenamefont
  {Hesse}, \citenamefont {Wang}, \citenamefont {Giles}, \citenamefont
  {Russell}, \citenamefont {Vörös}, \citenamefont {Hwang}, \citenamefont
  {Eastwood}, \citenamefont {Lavraud}, \citenamefont {Escoubet}, \citenamefont
  {Fear}, \citenamefont {Khotyaintsev}, \citenamefont {Nakamura}, \citenamefont
  {Webster},\ and\ \citenamefont {Baumjohann}}]{Genestreti.2018a}%
  \BibitemOpen
  \bibfield  {author} {\bibinfo {author} {\bibfnamefont {K.~J.}\ \bibnamefont
  {Genestreti}}, \bibinfo {author} {\bibfnamefont {A.}~\bibnamefont {Varsani}},
  \bibinfo {author} {\bibfnamefont {J.~L.}\ \bibnamefont {Burch}}, \bibinfo
  {author} {\bibfnamefont {P.~A.}\ \bibnamefont {Cassak}}, \bibinfo {author}
  {\bibfnamefont {R.~B.}\ \bibnamefont {Torbert}}, \bibinfo {author}
  {\bibfnamefont {R.}~\bibnamefont {Nakamura}}, \bibinfo {author}
  {\bibfnamefont {R.~E.}\ \bibnamefont {Ergun}}, \bibinfo {author}
  {\bibfnamefont {T.-D.}\ \bibnamefont {Phan}}, \bibinfo {author}
  {\bibfnamefont {S.}~\bibnamefont {Toledo-Redondo}}, \bibinfo {author}
  {\bibfnamefont {M.}~\bibnamefont {Hesse}}, \bibinfo {author} {\bibfnamefont
  {S.}~\bibnamefont {Wang}}, \bibinfo {author} {\bibfnamefont {B.~L.}\
  \bibnamefont {Giles}}, \bibinfo {author} {\bibfnamefont {C.~T.}\ \bibnamefont
  {Russell}}, \bibinfo {author} {\bibfnamefont {Z.}~\bibnamefont {Vörös}},
  \bibinfo {author} {\bibfnamefont {K.-J.}\ \bibnamefont {Hwang}}, \bibinfo
  {author} {\bibfnamefont {J.~P.}\ \bibnamefont {Eastwood}}, \bibinfo {author}
  {\bibfnamefont {B.}~\bibnamefont {Lavraud}}, \bibinfo {author} {\bibfnamefont
  {C.~P.}\ \bibnamefont {Escoubet}}, \bibinfo {author} {\bibfnamefont {R.~C.}\
  \bibnamefont {Fear}}, \bibinfo {author} {\bibfnamefont {Y.}~\bibnamefont
  {Khotyaintsev}}, \bibinfo {author} {\bibfnamefont {T.~K.~M.}\ \bibnamefont
  {Nakamura}}, \bibinfo {author} {\bibfnamefont {J.~M.}\ \bibnamefont
  {Webster}}, \ and\ \bibinfo {author} {\bibfnamefont {W.}~\bibnamefont
  {Baumjohann}},\ }\bibfield  {title} {\enquote {\bibinfo {title} {Mms
  observation of asymmetric reconnection supported by 3-d electron pressure
  divergence},}\ }\href {\doibase 10.1002/2017JA025019} {\bibfield  {journal}
  {\bibinfo  {journal} {Journal of Geophysical Research: Space Physics}\
  }\textbf {\bibinfo {volume} {123}},\ \bibinfo {pages} {1806--1821} (\bibinfo
  {year} {2018a})},\ \Eprint
  {http://arxiv.org/abs/https://agupubs.onlinelibrary.wiley.com/doi/pdf/10.1002/2017JA025019}
  {https://agupubs.onlinelibrary.wiley.com/doi/pdf/10.1002/2017JA025019}
  \BibitemShut {NoStop}%
\bibitem [{\citenamefont {{Fuselier}}\ \emph {et~al.}(2017)\citenamefont
  {{Fuselier}}, \citenamefont {{Vines}}, \citenamefont {{Burch}}, \citenamefont
  {{Petrinec}}, \citenamefont {{Trattner}}, \citenamefont {{Cassak}},
  \citenamefont {{Chen}}, \citenamefont {{Ergun}}, \citenamefont {{Eriksson}},
  \citenamefont {{Giles}}, \citenamefont {{Graham}}, \citenamefont
  {{Khotyaintsev}}, \citenamefont {{Lavraud}}, \citenamefont {{Lewis}},
  \citenamefont {{Mukherjee}}, \citenamefont {{Norgren}}, \citenamefont
  {{Phan}}, \citenamefont {{Russell}}, \citenamefont {{Strangeway}},
  \citenamefont {{Torbert}},\ and\ \citenamefont {{Webster}}}]{Fuselier.2017}%
  \BibitemOpen
  \bibfield  {author} {\bibinfo {author} {\bibfnamefont {S.~A.}\ \bibnamefont
  {{Fuselier}}}, \bibinfo {author} {\bibfnamefont {S.~K.}\ \bibnamefont
  {{Vines}}}, \bibinfo {author} {\bibfnamefont {J.~L.}\ \bibnamefont
  {{Burch}}}, \bibinfo {author} {\bibfnamefont {S.~M.}\ \bibnamefont
  {{Petrinec}}}, \bibinfo {author} {\bibfnamefont {K.~J.}\ \bibnamefont
  {{Trattner}}}, \bibinfo {author} {\bibfnamefont {P.~A.}\ \bibnamefont
  {{Cassak}}}, \bibinfo {author} {\bibfnamefont {L.-J.}\ \bibnamefont
  {{Chen}}}, \bibinfo {author} {\bibfnamefont {R.~E.}\ \bibnamefont {{Ergun}}},
  \bibinfo {author} {\bibfnamefont {S.}~\bibnamefont {{Eriksson}}}, \bibinfo
  {author} {\bibfnamefont {B.~L.}\ \bibnamefont {{Giles}}}, \bibinfo {author}
  {\bibfnamefont {D.~B.}\ \bibnamefont {{Graham}}}, \bibinfo {author}
  {\bibfnamefont {Y.~V.}\ \bibnamefont {{Khotyaintsev}}}, \bibinfo {author}
  {\bibfnamefont {B.}~\bibnamefont {{Lavraud}}}, \bibinfo {author}
  {\bibfnamefont {W.~S.}\ \bibnamefont {{Lewis}}}, \bibinfo {author}
  {\bibfnamefont {J.}~\bibnamefont {{Mukherjee}}}, \bibinfo {author}
  {\bibfnamefont {C.}~\bibnamefont {{Norgren}}}, \bibinfo {author}
  {\bibfnamefont {T.-D.}\ \bibnamefont {{Phan}}}, \bibinfo {author}
  {\bibfnamefont {C.~T.}\ \bibnamefont {{Russell}}}, \bibinfo {author}
  {\bibfnamefont {R.~J.}\ \bibnamefont {{Strangeway}}}, \bibinfo {author}
  {\bibfnamefont {R.~B.}\ \bibnamefont {{Torbert}}}, \ and\ \bibinfo {author}
  {\bibfnamefont {J.~M.}\ \bibnamefont {{Webster}}},\ }\bibfield  {title}
  {\enquote {\bibinfo {title} {{Large-scale characteristics of reconnection
  diffusion regions and associated magnetopause crossings observed by MMS}},}\
  }\href {\doibase 10.1002/2017JA024024} {\bibfield  {journal} {\bibinfo
  {journal} {Journal of Geophysical Research (Space Physics)}\ }\textbf
  {\bibinfo {volume} {122}},\ \bibinfo {pages} {5466--5486} (\bibinfo {year}
  {2017})}\BibitemShut {NoStop}%
\bibitem [{\citenamefont {Eastwood}\ \emph {et~al.}(2010)\citenamefont
  {Eastwood}, \citenamefont {Phan}, \citenamefont {Øieroset},\ and\
  \citenamefont {Shay}}]{Eastwood.2010}%
  \BibitemOpen
  \bibfield  {author} {\bibinfo {author} {\bibfnamefont {J.~P.}\ \bibnamefont
  {Eastwood}}, \bibinfo {author} {\bibfnamefont {T.~D.}\ \bibnamefont {Phan}},
  \bibinfo {author} {\bibfnamefont {M.}~\bibnamefont {Øieroset}}, \ and\
  \bibinfo {author} {\bibfnamefont {M.~A.}\ \bibnamefont {Shay}},\ }\bibfield
  {title} {\enquote {\bibinfo {title} {Average properties of the magnetic
  reconnection ion diffusion region in the earth's magnetotail: The 2001–2005
  cluster observations and comparison with simulations},}\ }\href {\doibase
  10.1029/2009JA014962} {\bibfield  {journal} {\bibinfo  {journal} {Journal of
  Geophysical Research: Space Physics}\ }\textbf {\bibinfo {volume} {115}}
  (\bibinfo {year} {2010}),\ 10.1029/2009JA014962}\BibitemShut {NoStop}%
\bibitem [{\citenamefont {{Cassak}}\ and\ \citenamefont
  {{Shay}}(2007)}]{CassakandShay.2007}%
  \BibitemOpen
  \bibfield  {author} {\bibinfo {author} {\bibfnamefont {P.~A.}\ \bibnamefont
  {{Cassak}}}\ and\ \bibinfo {author} {\bibfnamefont {M.~A.}\ \bibnamefont
  {{Shay}}},\ }\bibfield  {title} {\enquote {\bibinfo {title} {{Scaling of
  asymmetric magnetic reconnection: General theory and collisional
  simulations}},}\ }\href {\doibase 10.1063/1.2795630} {\bibfield  {journal}
  {\bibinfo  {journal} {Physics of Plasmas}\ }\textbf {\bibinfo {volume}
  {14}},\ \bibinfo {eid} {102114} (\bibinfo {year} {2007})}\BibitemShut
  {NoStop}%
\bibitem [{\citenamefont {{Cassak}}\ and\ \citenamefont
  {{Shay}}(2008)}]{CassakandShay.2008}%
  \BibitemOpen
  \bibfield  {author} {\bibinfo {author} {\bibfnamefont {P.~A.}\ \bibnamefont
  {{Cassak}}}\ and\ \bibinfo {author} {\bibfnamefont {M.~A.}\ \bibnamefont
  {{Shay}}},\ }\bibfield  {title} {\enquote {\bibinfo {title} {{Scaling of
  asymmetric Hall magnetic reconnection}},}\ }\href {\doibase
  10.1029/2008GL035268} {\bibfield  {journal} {\bibinfo  {journal} {Geophys.
  Res. Lett.}\ }\textbf {\bibinfo {volume} {35}},\ \bibinfo {eid} {L19102}
  (\bibinfo {year} {2008})}\BibitemShut {NoStop}%
\bibitem [{\citenamefont {{Pritchett}}\ and\ \citenamefont
  {{Mozer}}(2009)}]{PritchettandMozer.2009}%
  \BibitemOpen
  \bibfield  {author} {\bibinfo {author} {\bibfnamefont {P.~L.}\ \bibnamefont
  {{Pritchett}}}\ and\ \bibinfo {author} {\bibfnamefont {F.~S.}\ \bibnamefont
  {{Mozer}}},\ }\bibfield  {title} {\enquote {\bibinfo {title} {{The magnetic
  field reconnection site and dissipation region}},}\ }\href {\doibase
  10.1063/1.3206947} {\bibfield  {journal} {\bibinfo  {journal} {Physics of
  Plasmas}\ }\textbf {\bibinfo {volume} {16}},\ \bibinfo {eid} {080702}
  (\bibinfo {year} {2009})}\BibitemShut {NoStop}%
\bibitem [{\citenamefont {Swisdak}\ \emph {et~al.}(2018)\citenamefont
  {Swisdak}, \citenamefont {Drake}, \citenamefont {Price}, \citenamefont
  {Burch}, \citenamefont {Cassak},\ and\ \citenamefont {Phan}}]{Swisdak.2018}%
  \BibitemOpen
  \bibfield  {author} {\bibinfo {author} {\bibfnamefont {M.}~\bibnamefont
  {Swisdak}}, \bibinfo {author} {\bibfnamefont {J.~F.}\ \bibnamefont {Drake}},
  \bibinfo {author} {\bibfnamefont {L.}~\bibnamefont {Price}}, \bibinfo
  {author} {\bibfnamefont {J.~L.}\ \bibnamefont {Burch}}, \bibinfo {author}
  {\bibfnamefont {P.~A.}\ \bibnamefont {Cassak}}, \ and\ \bibinfo {author}
  {\bibfnamefont {T.-D.}\ \bibnamefont {Phan}},\ }\bibfield  {title} {\enquote
  {\bibinfo {title} {Localized and intense energy conversion in the diffusion
  region of asymmetric magnetic reconnection},}\ }\href {\doibase
  https://doi.org/10.1029/2017GL076862} {\bibfield  {journal} {\bibinfo
  {journal} {Geophysical Research Letters}\ }\textbf {\bibinfo {volume} {45}},\
  \bibinfo {pages} {5260--5267} (\bibinfo {year} {2018})},\ \Eprint
  {http://arxiv.org/abs/https://agupubs.onlinelibrary.wiley.com/doi/pdf/10.1029/2017GL076862}
  {https://agupubs.onlinelibrary.wiley.com/doi/pdf/10.1029/2017GL076862}
  \BibitemShut {NoStop}%
\bibitem [{\citenamefont {Pritchard}\ \emph {et~al.}(2019)\citenamefont
  {Pritchard}, \citenamefont {Burch}, \citenamefont {Fuselier}, \citenamefont
  {Webster}, \citenamefont {Torbert}, \citenamefont {Argall}, \citenamefont
  {Broll}, \citenamefont {Genestreti}, \citenamefont {Giles}, \citenamefont
  {Le~Contel}, \citenamefont {Mukherjee}, \citenamefont {Phan}, \citenamefont
  {Rager}, \citenamefont {Russell},\ and\ \citenamefont
  {Strangeway}}]{Pritchard.2019}%
  \BibitemOpen
  \bibfield  {author} {\bibinfo {author} {\bibfnamefont {K.~R.}\ \bibnamefont
  {Pritchard}}, \bibinfo {author} {\bibfnamefont {J.~L.}\ \bibnamefont
  {Burch}}, \bibinfo {author} {\bibfnamefont {S.~A.}\ \bibnamefont {Fuselier}},
  \bibinfo {author} {\bibfnamefont {J.~M.}\ \bibnamefont {Webster}}, \bibinfo
  {author} {\bibfnamefont {R.~B.}\ \bibnamefont {Torbert}}, \bibinfo {author}
  {\bibfnamefont {M.~R.}\ \bibnamefont {Argall}}, \bibinfo {author}
  {\bibfnamefont {J.}~\bibnamefont {Broll}}, \bibinfo {author} {\bibfnamefont
  {K.~J.}\ \bibnamefont {Genestreti}}, \bibinfo {author} {\bibfnamefont
  {B.~L.}\ \bibnamefont {Giles}}, \bibinfo {author} {\bibfnamefont
  {O.}~\bibnamefont {Le~Contel}}, \bibinfo {author} {\bibfnamefont
  {J.}~\bibnamefont {Mukherjee}}, \bibinfo {author} {\bibfnamefont {T.~D.}\
  \bibnamefont {Phan}}, \bibinfo {author} {\bibfnamefont {A.~C.}\ \bibnamefont
  {Rager}}, \bibinfo {author} {\bibfnamefont {C.~T.}\ \bibnamefont {Russell}},
  \ and\ \bibinfo {author} {\bibfnamefont {R.~J.}\ \bibnamefont {Strangeway}},\
  }\bibfield  {title} {\enquote {\bibinfo {title} {Energy conversion and
  electron acceleration in the magnetopause reconnection diffusion region},}\
  }\href {\doibase https://doi.org/10.1029/2019GL084636} {\bibfield  {journal}
  {\bibinfo  {journal} {Geophysical Research Letters}\ }\textbf {\bibinfo
  {volume} {46}},\ \bibinfo {pages} {10274--10282} (\bibinfo {year} {2019})},\
  \Eprint
  {http://arxiv.org/abs/https://agupubs.onlinelibrary.wiley.com/doi/pdf/10.1029/2019GL084636}
  {https://agupubs.onlinelibrary.wiley.com/doi/pdf/10.1029/2019GL084636}
  \BibitemShut {NoStop}%
\bibitem [{\citenamefont {Swisdak}\ \emph {et~al.}(2003)\citenamefont
  {Swisdak}, \citenamefont {Rogers}, \citenamefont {Drake},\ and\ \citenamefont
  {Shay}}]{Swisdak.2003}%
  \BibitemOpen
  \bibfield  {author} {\bibinfo {author} {\bibfnamefont {M.}~\bibnamefont
  {Swisdak}}, \bibinfo {author} {\bibfnamefont {B.~N.}\ \bibnamefont {Rogers}},
  \bibinfo {author} {\bibfnamefont {J.~F.}\ \bibnamefont {Drake}}, \ and\
  \bibinfo {author} {\bibfnamefont {M.~A.}\ \bibnamefont {Shay}},\ }\bibfield
  {title} {\enquote {\bibinfo {title} {Diamagnetic suppression of component
  magnetic reconnection at the magnetopause},}\ }\href {\doibase
  10.1029/2002JA009726} {\bibfield  {journal} {\bibinfo  {journal} {Journal of
  Geophysical Research: Space Physics}\ }\textbf {\bibinfo {volume} {108}}
  (\bibinfo {year} {2003}),\ 10.1029/2002JA009726}\BibitemShut {NoStop}%
\bibitem [{\citenamefont {{Price}}\ \emph {et~al.}(2016)\citenamefont
  {{Price}}, \citenamefont {{Swisdak}}, \citenamefont {{Drake}}, \citenamefont
  {{Cassak}}, \citenamefont {{Dahlin}},\ and\ \citenamefont
  {{Ergun}}}]{Price.2016}%
  \BibitemOpen
  \bibfield  {author} {\bibinfo {author} {\bibfnamefont {L.}~\bibnamefont
  {{Price}}}, \bibinfo {author} {\bibfnamefont {M.}~\bibnamefont {{Swisdak}}},
  \bibinfo {author} {\bibfnamefont {J.~F.}\ \bibnamefont {{Drake}}}, \bibinfo
  {author} {\bibfnamefont {P.~A.}\ \bibnamefont {{Cassak}}}, \bibinfo {author}
  {\bibfnamefont {J.~T.}\ \bibnamefont {{Dahlin}}}, \ and\ \bibinfo {author}
  {\bibfnamefont {R.~E.}\ \bibnamefont {{Ergun}}},\ }\bibfield  {title}
  {\enquote {\bibinfo {title} {{The effects of turbulence on three-dimensional
  magnetic reconnection at the magnetopause}},}\ }\href {\doibase
  10.1002/2016GL069578} {\bibfield  {journal} {\bibinfo  {journal} {Geophys.
  Res. Lett.}\ }\textbf {\bibinfo {volume} {43}},\ \bibinfo {pages}
  {6020--6027} (\bibinfo {year} {2016})},\ \Eprint
  {http://arxiv.org/abs/1604.08172} {arXiv:1604.08172 [physics.space-ph]}
  \BibitemShut {NoStop}%
\bibitem [{\citenamefont {{Ergun}}\ \emph {et~al.}(2017)\citenamefont
  {{Ergun}}, \citenamefont {{Chen}}, \citenamefont {{Wilder}}, \citenamefont
  {{Ahmadi}}, \citenamefont {{Eriksson}}, \citenamefont {{Usanova}},
  \citenamefont {{Goodrich}}, \citenamefont {{Holmes}}, \citenamefont
  {{Sturner}}, \citenamefont {{Malaspina}}, \citenamefont {{Newman}},
  \citenamefont {{Torbert}}, \citenamefont {{Argall}}, \citenamefont
  {{Lindqvist}}, \citenamefont {{Burch}}, \citenamefont {{Webster}},
  \citenamefont {{Drake}}, \citenamefont {{Price}}, \citenamefont {{Cassak}},
  \citenamefont {{Swisdak}}, \citenamefont {{Shay}}, \citenamefont {{Graham}},
  \citenamefont {{Strangeway}}, \citenamefont {{Russell}}, \citenamefont
  {{Giles}}, \citenamefont {{Dorelli}}, \citenamefont {{Gershman}},
  \citenamefont {{Avanov}}, \citenamefont {{Hesse}}, \citenamefont {{Lavraud}},
  \citenamefont {{Le Contel}}, \citenamefont {{Retino}}, \citenamefont
  {{Phan}}, \citenamefont {{Goldman}}, \citenamefont {{Stawarz}}, \citenamefont
  {{Schwartz}}, \citenamefont {{Eastwood}}, \citenamefont {{Hwang}},
  \citenamefont {{Nakamura}},\ and\ \citenamefont {{Wang}}}]{Ergun.2017}%
  \BibitemOpen
  \bibfield  {author} {\bibinfo {author} {\bibfnamefont {R.~E.}\ \bibnamefont
  {{Ergun}}}, \bibinfo {author} {\bibfnamefont {L.-J.}\ \bibnamefont {{Chen}}},
  \bibinfo {author} {\bibfnamefont {F.~D.}\ \bibnamefont {{Wilder}}}, \bibinfo
  {author} {\bibfnamefont {N.}~\bibnamefont {{Ahmadi}}}, \bibinfo {author}
  {\bibfnamefont {S.}~\bibnamefont {{Eriksson}}}, \bibinfo {author}
  {\bibfnamefont {M.~E.}\ \bibnamefont {{Usanova}}}, \bibinfo {author}
  {\bibfnamefont {K.~A.}\ \bibnamefont {{Goodrich}}}, \bibinfo {author}
  {\bibfnamefont {J.~C.}\ \bibnamefont {{Holmes}}}, \bibinfo {author}
  {\bibfnamefont {A.~P.}\ \bibnamefont {{Sturner}}}, \bibinfo {author}
  {\bibfnamefont {D.~M.}\ \bibnamefont {{Malaspina}}}, \bibinfo {author}
  {\bibfnamefont {D.~L.}\ \bibnamefont {{Newman}}}, \bibinfo {author}
  {\bibfnamefont {R.~B.}\ \bibnamefont {{Torbert}}}, \bibinfo {author}
  {\bibfnamefont {M.~R.}\ \bibnamefont {{Argall}}}, \bibinfo {author}
  {\bibfnamefont {P.-A.}\ \bibnamefont {{Lindqvist}}}, \bibinfo {author}
  {\bibfnamefont {J.~L.}\ \bibnamefont {{Burch}}}, \bibinfo {author}
  {\bibfnamefont {J.~M.}\ \bibnamefont {{Webster}}}, \bibinfo {author}
  {\bibfnamefont {J.~F.}\ \bibnamefont {{Drake}}}, \bibinfo {author}
  {\bibfnamefont {L.}~\bibnamefont {{Price}}}, \bibinfo {author} {\bibfnamefont
  {P.~A.}\ \bibnamefont {{Cassak}}}, \bibinfo {author} {\bibfnamefont
  {M.}~\bibnamefont {{Swisdak}}}, \bibinfo {author} {\bibfnamefont {M.~A.}\
  \bibnamefont {{Shay}}}, \bibinfo {author} {\bibfnamefont {D.~B.}\
  \bibnamefont {{Graham}}}, \bibinfo {author} {\bibfnamefont {R.~J.}\
  \bibnamefont {{Strangeway}}}, \bibinfo {author} {\bibfnamefont {C.~T.}\
  \bibnamefont {{Russell}}}, \bibinfo {author} {\bibfnamefont {B.~L.}\
  \bibnamefont {{Giles}}}, \bibinfo {author} {\bibfnamefont {J.~C.}\
  \bibnamefont {{Dorelli}}}, \bibinfo {author} {\bibfnamefont {D.}~\bibnamefont
  {{Gershman}}}, \bibinfo {author} {\bibfnamefont {L.}~\bibnamefont
  {{Avanov}}}, \bibinfo {author} {\bibfnamefont {M.}~\bibnamefont {{Hesse}}},
  \bibinfo {author} {\bibfnamefont {B.}~\bibnamefont {{Lavraud}}}, \bibinfo
  {author} {\bibfnamefont {O.}~\bibnamefont {{Le Contel}}}, \bibinfo {author}
  {\bibfnamefont {A.}~\bibnamefont {{Retino}}}, \bibinfo {author}
  {\bibfnamefont {T.~D.}\ \bibnamefont {{Phan}}}, \bibinfo {author}
  {\bibfnamefont {M.~V.}\ \bibnamefont {{Goldman}}}, \bibinfo {author}
  {\bibfnamefont {J.~E.}\ \bibnamefont {{Stawarz}}}, \bibinfo {author}
  {\bibfnamefont {S.~J.}\ \bibnamefont {{Schwartz}}}, \bibinfo {author}
  {\bibfnamefont {J.~P.}\ \bibnamefont {{Eastwood}}}, \bibinfo {author}
  {\bibfnamefont {K.-J.}\ \bibnamefont {{Hwang}}}, \bibinfo {author}
  {\bibfnamefont {R.}~\bibnamefont {{Nakamura}}}, \ and\ \bibinfo {author}
  {\bibfnamefont {S.}~\bibnamefont {{Wang}}},\ }\bibfield  {title} {\enquote
  {\bibinfo {title} {{Drift waves, intense parallel electric fields, and
  turbulence associated with asymmetric magnetic reconnection at the
  magnetopause}},}\ }\href {\doibase 10.1002/2016GL072493} {\bibfield
  {journal} {\bibinfo  {journal} {Geophys. Rev. Lett.}\ }\textbf {\bibinfo
  {volume} {44}},\ \bibinfo {pages} {2978--2986} (\bibinfo {year}
  {2017})}\BibitemShut {NoStop}%
\bibitem [{\citenamefont {Ergun}\ \emph
  {et~al.}(2019{\natexlab{a}})\citenamefont {Ergun}, \citenamefont {Hoilijoki},
  \citenamefont {Ahmadi}, \citenamefont {Schwartz}, \citenamefont {Wilder},
  \citenamefont {Burch}, \citenamefont {Torbert}, \citenamefont {Lindqvist},
  \citenamefont {Graham}, \citenamefont {Strangeway}, \citenamefont
  {Le~Contel}, \citenamefont {Holmes}, \citenamefont {Stawarz}, \citenamefont
  {Goodrich}, \citenamefont {Eriksson}, \citenamefont {Giles}, \citenamefont
  {Gershman},\ and\ \citenamefont {Chen}}]{Ergun.2019a}%
  \BibitemOpen
  \bibfield  {author} {\bibinfo {author} {\bibfnamefont {R.~E.}\ \bibnamefont
  {Ergun}}, \bibinfo {author} {\bibfnamefont {S.}~\bibnamefont {Hoilijoki}},
  \bibinfo {author} {\bibfnamefont {N.}~\bibnamefont {Ahmadi}}, \bibinfo
  {author} {\bibfnamefont {S.~J.}\ \bibnamefont {Schwartz}}, \bibinfo {author}
  {\bibfnamefont {F.~D.}\ \bibnamefont {Wilder}}, \bibinfo {author}
  {\bibfnamefont {J.~L.}\ \bibnamefont {Burch}}, \bibinfo {author}
  {\bibfnamefont {R.~B.}\ \bibnamefont {Torbert}}, \bibinfo {author}
  {\bibfnamefont {P.-A.}\ \bibnamefont {Lindqvist}}, \bibinfo {author}
  {\bibfnamefont {D.~B.}\ \bibnamefont {Graham}}, \bibinfo {author}
  {\bibfnamefont {R.~J.}\ \bibnamefont {Strangeway}}, \bibinfo {author}
  {\bibfnamefont {O.}~\bibnamefont {Le~Contel}}, \bibinfo {author}
  {\bibfnamefont {J.~C.}\ \bibnamefont {Holmes}}, \bibinfo {author}
  {\bibfnamefont {J.~E.}\ \bibnamefont {Stawarz}}, \bibinfo {author}
  {\bibfnamefont {K.~A.}\ \bibnamefont {Goodrich}}, \bibinfo {author}
  {\bibfnamefont {S.}~\bibnamefont {Eriksson}}, \bibinfo {author}
  {\bibfnamefont {B.~L.}\ \bibnamefont {Giles}}, \bibinfo {author}
  {\bibfnamefont {D.}~\bibnamefont {Gershman}}, \ and\ \bibinfo {author}
  {\bibfnamefont {L.~J.}\ \bibnamefont {Chen}},\ }\bibfield  {title} {\enquote
  {\bibinfo {title} {Magnetic reconnection in three dimensions: Observations of
  electromagnetic drift waves in the adjacent current sheet},}\ }\href
  {\doibase 10.1029/2019JA027228} {\bibfield  {journal} {\bibinfo  {journal}
  {Journal of Geophysical Research: Space Physics}\ }\textbf {\bibinfo {volume}
  {n/a}} (\bibinfo {year} {2019}{\natexlab{a}}),\ 10.1029/2019JA027228},\
  \Eprint
  {http://arxiv.org/abs/https://agupubs.onlinelibrary.wiley.com/doi/pdf/10.1029/2019JA027228}
  {https://agupubs.onlinelibrary.wiley.com/doi/pdf/10.1029/2019JA027228}
  \BibitemShut {NoStop}%
\bibitem [{\citenamefont {Ergun}\ \emph
  {et~al.}(2019{\natexlab{b}})\citenamefont {Ergun}, \citenamefont {Hoilijoki},
  \citenamefont {Ahmadi}, \citenamefont {Schwartz}, \citenamefont {Wilder},
  \citenamefont {Drake}, \citenamefont {Hesse}, \citenamefont {Shay},
  \citenamefont {Ji}, \citenamefont {Yamada}, \citenamefont {Graham},
  \citenamefont {Cassak}, \citenamefont {Swisdak}, \citenamefont {Burch},
  \citenamefont {Torbert}, \citenamefont {Holmes}, \citenamefont {Stawarz},
  \citenamefont {Goodrich}, \citenamefont {Eriksson}, \citenamefont
  {Strangeway},\ and\ \citenamefont {LeContel}}]{Ergun.2019b}%
  \BibitemOpen
  \bibfield  {author} {\bibinfo {author} {\bibfnamefont {R.~E.}\ \bibnamefont
  {Ergun}}, \bibinfo {author} {\bibfnamefont {S.}~\bibnamefont {Hoilijoki}},
  \bibinfo {author} {\bibfnamefont {N.}~\bibnamefont {Ahmadi}}, \bibinfo
  {author} {\bibfnamefont {S.~J.}\ \bibnamefont {Schwartz}}, \bibinfo {author}
  {\bibfnamefont {F.~D.}\ \bibnamefont {Wilder}}, \bibinfo {author}
  {\bibfnamefont {J.~F.}\ \bibnamefont {Drake}}, \bibinfo {author}
  {\bibfnamefont {M.}~\bibnamefont {Hesse}}, \bibinfo {author} {\bibfnamefont
  {M.~A.}\ \bibnamefont {Shay}}, \bibinfo {author} {\bibfnamefont
  {H.}~\bibnamefont {Ji}}, \bibinfo {author} {\bibfnamefont {M.}~\bibnamefont
  {Yamada}}, \bibinfo {author} {\bibfnamefont {D.~B.}\ \bibnamefont {Graham}},
  \bibinfo {author} {\bibfnamefont {P.~A.}\ \bibnamefont {Cassak}}, \bibinfo
  {author} {\bibfnamefont {M.}~\bibnamefont {Swisdak}}, \bibinfo {author}
  {\bibfnamefont {J.~L.}\ \bibnamefont {Burch}}, \bibinfo {author}
  {\bibfnamefont {R.~B.}\ \bibnamefont {Torbert}}, \bibinfo {author}
  {\bibfnamefont {J.~C.}\ \bibnamefont {Holmes}}, \bibinfo {author}
  {\bibfnamefont {J.~E.}\ \bibnamefont {Stawarz}}, \bibinfo {author}
  {\bibfnamefont {K.~A.}\ \bibnamefont {Goodrich}}, \bibinfo {author}
  {\bibfnamefont {S.}~\bibnamefont {Eriksson}}, \bibinfo {author}
  {\bibfnamefont {R.~J.}\ \bibnamefont {Strangeway}}, \ and\ \bibinfo {author}
  {\bibfnamefont {O.}~\bibnamefont {LeContel}},\ }\bibfield  {title} {\enquote
  {\bibinfo {title} {Magnetic reconnection in three dimensions: Modeling and
  analysis of electromagnetic drift waves in the adjacent current sheet},}\
  }\href {\doibase 10.1029/2019JA027275} {\bibfield  {journal} {\bibinfo
  {journal} {Journal of Geophysical Research: Space Physics}\ }\textbf
  {\bibinfo {volume} {n/a}} (\bibinfo {year} {2019}{\natexlab{b}}),\
  10.1029/2019JA027275},\ \Eprint
  {http://arxiv.org/abs/https://agupubs.onlinelibrary.wiley.com/doi/pdf/10.1029/2019JA027275}
  {https://agupubs.onlinelibrary.wiley.com/doi/pdf/10.1029/2019JA027275}
  \BibitemShut {NoStop}%
\bibitem [{\citenamefont {Graham}\ \emph {et~al.}(2017)\citenamefont {Graham},
  \citenamefont {Khotyaintsev}, \citenamefont {Norgren}, \citenamefont
  {Vaivads}, \citenamefont {André}, \citenamefont {Toledo-Redondo},
  \citenamefont {Lindqvist}, \citenamefont {Marklund}, \citenamefont {Ergun},
  \citenamefont {Paterson}, \citenamefont {Gershman}, \citenamefont {Giles},
  \citenamefont {Pollock}, \citenamefont {Dorelli}, \citenamefont {Avanov},
  \citenamefont {Lavraud}, \citenamefont {Saito}, \citenamefont {Magnes},
  \citenamefont {Russell}, \citenamefont {Strangeway}, \citenamefont
  {Torbert},\ and\ \citenamefont {Burch}}]{Graham.2017}%
  \BibitemOpen
  \bibfield  {author} {\bibinfo {author} {\bibfnamefont {D.~B.}\ \bibnamefont
  {Graham}}, \bibinfo {author} {\bibfnamefont {Y.~V.}\ \bibnamefont
  {Khotyaintsev}}, \bibinfo {author} {\bibfnamefont {C.}~\bibnamefont
  {Norgren}}, \bibinfo {author} {\bibfnamefont {A.}~\bibnamefont {Vaivads}},
  \bibinfo {author} {\bibfnamefont {M.}~\bibnamefont {André}}, \bibinfo
  {author} {\bibfnamefont {S.}~\bibnamefont {Toledo-Redondo}}, \bibinfo
  {author} {\bibfnamefont {P.-A.}\ \bibnamefont {Lindqvist}}, \bibinfo {author}
  {\bibfnamefont {G.~T.}\ \bibnamefont {Marklund}}, \bibinfo {author}
  {\bibfnamefont {R.~E.}\ \bibnamefont {Ergun}}, \bibinfo {author}
  {\bibfnamefont {W.~R.}\ \bibnamefont {Paterson}}, \bibinfo {author}
  {\bibfnamefont {D.~J.}\ \bibnamefont {Gershman}}, \bibinfo {author}
  {\bibfnamefont {B.~L.}\ \bibnamefont {Giles}}, \bibinfo {author}
  {\bibfnamefont {C.~J.}\ \bibnamefont {Pollock}}, \bibinfo {author}
  {\bibfnamefont {J.~C.}\ \bibnamefont {Dorelli}}, \bibinfo {author}
  {\bibfnamefont {L.~A.}\ \bibnamefont {Avanov}}, \bibinfo {author}
  {\bibfnamefont {B.}~\bibnamefont {Lavraud}}, \bibinfo {author} {\bibfnamefont
  {Y.}~\bibnamefont {Saito}}, \bibinfo {author} {\bibfnamefont
  {W.}~\bibnamefont {Magnes}}, \bibinfo {author} {\bibfnamefont {C.~T.}\
  \bibnamefont {Russell}}, \bibinfo {author} {\bibfnamefont {R.~J.}\
  \bibnamefont {Strangeway}}, \bibinfo {author} {\bibfnamefont {R.~B.}\
  \bibnamefont {Torbert}}, \ and\ \bibinfo {author} {\bibfnamefont {J.~L.}\
  \bibnamefont {Burch}},\ }\bibfield  {title} {\enquote {\bibinfo {title}
  {Lower hybrid waves in the ion diffusion and magnetospheric inflow
  regions},}\ }\href {\doibase 10.1002/2016JA023572} {\bibfield  {journal}
  {\bibinfo  {journal} {Journal of Geophysical Research: Space Physics}\
  }\textbf {\bibinfo {volume} {122}},\ \bibinfo {pages} {517--533} (\bibinfo
  {year} {2017})},\ \Eprint
  {http://arxiv.org/abs/https://agupubs.onlinelibrary.wiley.com/doi/pdf/10.1002/2016JA023572}
  {https://agupubs.onlinelibrary.wiley.com/doi/pdf/10.1002/2016JA023572}
  \BibitemShut {NoStop}%
\bibitem [{\citenamefont {Le}\ \emph {et~al.}(2017)\citenamefont {Le},
  \citenamefont {Daughton}, \citenamefont {Chen},\ and\ \citenamefont
  {Egedal}}]{Le.2017}%
  \BibitemOpen
  \bibfield  {author} {\bibinfo {author} {\bibfnamefont {A.}~\bibnamefont
  {Le}}, \bibinfo {author} {\bibfnamefont {W.}~\bibnamefont {Daughton}},
  \bibinfo {author} {\bibfnamefont {L.-J.}\ \bibnamefont {Chen}}, \ and\
  \bibinfo {author} {\bibfnamefont {J.}~\bibnamefont {Egedal}},\ }\bibfield
  {title} {\enquote {\bibinfo {title} {Enhanced electron mixing and heating in
  3-d asymmetric reconnection at the earth's magnetopause},}\ }\href {\doibase
  10.1002/2017GL072522} {\bibfield  {journal} {\bibinfo  {journal} {Geophysical
  Research Letters}\ }\textbf {\bibinfo {volume} {44}},\ \bibinfo {pages}
  {2096--2104} (\bibinfo {year} {2017})}\BibitemShut {NoStop}%
\bibitem [{\citenamefont {Wilder}\ \emph {et~al.}(2019)\citenamefont {Wilder},
  \citenamefont {Ergun}, \citenamefont {Hoilijoki}, \citenamefont {Webster},
  \citenamefont {Argall}, \citenamefont {Ahmadi}, \citenamefont {Eriksson},
  \citenamefont {Burch}, \citenamefont {Torbert}, \citenamefont {Le~Contel},
  \citenamefont {Strangeway},\ and\ \citenamefont {Giles}}]{Wilder.2019}%
  \BibitemOpen
  \bibfield  {author} {\bibinfo {author} {\bibfnamefont {F.~D.}\ \bibnamefont
  {Wilder}}, \bibinfo {author} {\bibfnamefont {R.~E.}\ \bibnamefont {Ergun}},
  \bibinfo {author} {\bibfnamefont {S.}~\bibnamefont {Hoilijoki}}, \bibinfo
  {author} {\bibfnamefont {J.}~\bibnamefont {Webster}}, \bibinfo {author}
  {\bibfnamefont {M.~R.}\ \bibnamefont {Argall}}, \bibinfo {author}
  {\bibfnamefont {N.}~\bibnamefont {Ahmadi}}, \bibinfo {author} {\bibfnamefont
  {S.}~\bibnamefont {Eriksson}}, \bibinfo {author} {\bibfnamefont {J.~L.}\
  \bibnamefont {Burch}}, \bibinfo {author} {\bibfnamefont {R.~B.}\ \bibnamefont
  {Torbert}}, \bibinfo {author} {\bibfnamefont {O.}~\bibnamefont {Le~Contel}},
  \bibinfo {author} {\bibfnamefont {R.~J.}\ \bibnamefont {Strangeway}}, \ and\
  \bibinfo {author} {\bibfnamefont {B.~L.}\ \bibnamefont {Giles}},\ }\bibfield
  {title} {\enquote {\bibinfo {title} {A survey of plasma waves appearing near
  dayside magnetopause electron diffusion region events},}\ }\href {\doibase
  https://doi.org/10.1029/2019JA027060} {\bibfield  {journal} {\bibinfo
  {journal} {Journal of Geophysical Research: Space Physics}\ }\textbf
  {\bibinfo {volume} {124}},\ \bibinfo {pages} {7837--7849} (\bibinfo {year}
  {2019})},\ \Eprint
  {http://arxiv.org/abs/https://agupubs.onlinelibrary.wiley.com/doi/pdf/10.1029/2019JA027060}
  {https://agupubs.onlinelibrary.wiley.com/doi/pdf/10.1029/2019JA027060}
  \BibitemShut {NoStop}%
\bibitem [{\citenamefont {{Hesse}}\ \emph {et~al.}(2014)\citenamefont
  {{Hesse}}, \citenamefont {{Aunai}}, \citenamefont {{Sibeck}},\ and\
  \citenamefont {{Birn}}}]{Hesse.2014}%
  \BibitemOpen
  \bibfield  {author} {\bibinfo {author} {\bibfnamefont {M.}~\bibnamefont
  {{Hesse}}}, \bibinfo {author} {\bibfnamefont {N.}~\bibnamefont {{Aunai}}},
  \bibinfo {author} {\bibfnamefont {D.}~\bibnamefont {{Sibeck}}}, \ and\
  \bibinfo {author} {\bibfnamefont {J.}~\bibnamefont {{Birn}}},\ }\bibfield
  {title} {\enquote {\bibinfo {title} {{On the electron diffusion region in
  planar, asymmetric, systems}},}\ }\href {\doibase 10.1002/2014GL061586}
  {\bibfield  {journal} {\bibinfo  {journal} {Geophys. Res. Lett.}\ }\textbf
  {\bibinfo {volume} {41}},\ \bibinfo {pages} {8673--8680} (\bibinfo {year}
  {2014})}\BibitemShut {NoStop}%
\bibitem [{\citenamefont {Wilder}\ \emph {et~al.}(2017)\citenamefont {Wilder},
  \citenamefont {Ergun}, \citenamefont {Eriksson}, \citenamefont {Phan},
  \citenamefont {Burch}, \citenamefont {Ahmadi}, \citenamefont {Goodrich},
  \citenamefont {Newman}, \citenamefont {Trattner}, \citenamefont {Torbert},
  \citenamefont {Giles}, \citenamefont {Strangeway}, \citenamefont {Magnes},
  \citenamefont {Lindqvist},\ and\ \citenamefont {Khotyaintsev}}]{Wilder.2017}%
  \BibitemOpen
  \bibfield  {author} {\bibinfo {author} {\bibfnamefont {F.~D.}\ \bibnamefont
  {Wilder}}, \bibinfo {author} {\bibfnamefont {R.~E.}\ \bibnamefont {Ergun}},
  \bibinfo {author} {\bibfnamefont {S.}~\bibnamefont {Eriksson}}, \bibinfo
  {author} {\bibfnamefont {T.~D.}\ \bibnamefont {Phan}}, \bibinfo {author}
  {\bibfnamefont {J.~L.}\ \bibnamefont {Burch}}, \bibinfo {author}
  {\bibfnamefont {N.}~\bibnamefont {Ahmadi}}, \bibinfo {author} {\bibfnamefont
  {K.~A.}\ \bibnamefont {Goodrich}}, \bibinfo {author} {\bibfnamefont {D.~L.}\
  \bibnamefont {Newman}}, \bibinfo {author} {\bibfnamefont {K.~J.}\
  \bibnamefont {Trattner}}, \bibinfo {author} {\bibfnamefont {R.~B.}\
  \bibnamefont {Torbert}}, \bibinfo {author} {\bibfnamefont {B.~L.}\
  \bibnamefont {Giles}}, \bibinfo {author} {\bibfnamefont {R.~J.}\ \bibnamefont
  {Strangeway}}, \bibinfo {author} {\bibfnamefont {W.}~\bibnamefont {Magnes}},
  \bibinfo {author} {\bibfnamefont {P.-A.}\ \bibnamefont {Lindqvist}}, \ and\
  \bibinfo {author} {\bibfnamefont {Y.-V.}\ \bibnamefont {Khotyaintsev}},\
  }\bibfield  {title} {\enquote {\bibinfo {title} {Multipoint measurements of
  the electron jet of symmetric magnetic reconnection with a moderate guide
  field},}\ }\href {\doibase 10.1103/PhysRevLett.118.265101} {\bibfield
  {journal} {\bibinfo  {journal} {Phys. Rev. Lett.}\ }\textbf {\bibinfo
  {volume} {118}},\ \bibinfo {pages} {265101} (\bibinfo {year}
  {2017})}\BibitemShut {NoStop}%
\bibitem [{\citenamefont {{Eriksson}}\ \emph {et~al.}(2016)\citenamefont
  {{Eriksson}}, \citenamefont {{Wilder}}, \citenamefont {{Ergun}},
  \citenamefont {{Schwartz}}, \citenamefont {{Cassak}}, \citenamefont
  {{Burch}}, \citenamefont {{Chen}}, \citenamefont {{Torbert}}, \citenamefont
  {{Phan}}, \citenamefont {{Lavraud}}, \citenamefont {{Goodrich}},
  \citenamefont {{Holmes}}, \citenamefont {{Stawarz}}, \citenamefont
  {{Sturner}}, \citenamefont {{Malaspina}}, \citenamefont {{Usanova}},
  \citenamefont {{Trattner}}, \citenamefont {{Strangeway}}, \citenamefont
  {{Russell}}, \citenamefont {{Pollock}}, \citenamefont {{Giles}},
  \citenamefont {{Hesse}}, \citenamefont {{Lindqvist}}, \citenamefont
  {{Drake}}, \citenamefont {{Shay}}, \citenamefont {{Nakamura}},\ and\
  \citenamefont {{Marklund}}}]{Eriksson.2016}%
  \BibitemOpen
  \bibfield  {author} {\bibinfo {author} {\bibfnamefont {S.}~\bibnamefont
  {{Eriksson}}}, \bibinfo {author} {\bibfnamefont {F.~D.}\ \bibnamefont
  {{Wilder}}}, \bibinfo {author} {\bibfnamefont {R.~E.}\ \bibnamefont
  {{Ergun}}}, \bibinfo {author} {\bibfnamefont {S.~J.}\ \bibnamefont
  {{Schwartz}}}, \bibinfo {author} {\bibfnamefont {P.~A.}\ \bibnamefont
  {{Cassak}}}, \bibinfo {author} {\bibfnamefont {J.~L.}\ \bibnamefont
  {{Burch}}}, \bibinfo {author} {\bibfnamefont {L.-J.}\ \bibnamefont {{Chen}}},
  \bibinfo {author} {\bibfnamefont {R.~B.}\ \bibnamefont {{Torbert}}}, \bibinfo
  {author} {\bibfnamefont {T.~D.}\ \bibnamefont {{Phan}}}, \bibinfo {author}
  {\bibfnamefont {B.}~\bibnamefont {{Lavraud}}}, \bibinfo {author}
  {\bibfnamefont {K.~A.}\ \bibnamefont {{Goodrich}}}, \bibinfo {author}
  {\bibfnamefont {J.~C.}\ \bibnamefont {{Holmes}}}, \bibinfo {author}
  {\bibfnamefont {J.~E.}\ \bibnamefont {{Stawarz}}}, \bibinfo {author}
  {\bibfnamefont {A.~P.}\ \bibnamefont {{Sturner}}}, \bibinfo {author}
  {\bibfnamefont {D.~M.}\ \bibnamefont {{Malaspina}}}, \bibinfo {author}
  {\bibfnamefont {M.~E.}\ \bibnamefont {{Usanova}}}, \bibinfo {author}
  {\bibfnamefont {K.~J.}\ \bibnamefont {{Trattner}}}, \bibinfo {author}
  {\bibfnamefont {R.~J.}\ \bibnamefont {{Strangeway}}}, \bibinfo {author}
  {\bibfnamefont {C.~T.}\ \bibnamefont {{Russell}}}, \bibinfo {author}
  {\bibfnamefont {C.~J.}\ \bibnamefont {{Pollock}}}, \bibinfo {author}
  {\bibfnamefont {B.~L.}\ \bibnamefont {{Giles}}}, \bibinfo {author}
  {\bibfnamefont {M.}~\bibnamefont {{Hesse}}}, \bibinfo {author} {\bibfnamefont
  {P.-A.}\ \bibnamefont {{Lindqvist}}}, \bibinfo {author} {\bibfnamefont
  {J.~F.}\ \bibnamefont {{Drake}}}, \bibinfo {author} {\bibfnamefont {M.~A.}\
  \bibnamefont {{Shay}}}, \bibinfo {author} {\bibfnamefont {R.}~\bibnamefont
  {{Nakamura}}}, \ and\ \bibinfo {author} {\bibfnamefont {G.~T.}\ \bibnamefont
  {{Marklund}}},\ }\bibfield  {title} {\enquote {\bibinfo {title}
  {{Magnetospheric Multiscale Observations of the Electron Diffusion Region of
  Large Guide Field Magnetic Reconnection}},}\ }\href {\doibase
  10.1103/PhysRevLett.117.015001} {\bibfield  {journal} {\bibinfo  {journal}
  {Phys. Rev. Lett.}\ }\textbf {\bibinfo {volume} {117}},\ \bibinfo {eid}
  {015001} (\bibinfo {year} {2016})}\BibitemShut {NoStop}%
\bibitem [{\citenamefont {{Burch}}\ and\ \citenamefont
  {{Phan}}(2016)}]{BurchandPhan.2016}%
  \BibitemOpen
  \bibfield  {author} {\bibinfo {author} {\bibfnamefont {J.~L.}\ \bibnamefont
  {{Burch}}}\ and\ \bibinfo {author} {\bibfnamefont {T.~D.}\ \bibnamefont
  {{Phan}}},\ }\bibfield  {title} {\enquote {\bibinfo {title} {{Magnetic
  reconnection at the dayside magnetopause: Advances with MMS}},}\ }\href
  {\doibase 10.1002/2016GL069787} {\bibfield  {journal} {\bibinfo  {journal}
  {Geophys. Res. Lett.}\ }\textbf {\bibinfo {volume} {43}},\ \bibinfo {pages}
  {8327--8338} (\bibinfo {year} {2016})}\BibitemShut {NoStop}%
\bibitem [{\citenamefont {Burch}\ \emph {et~al.}(2019)\citenamefont {Burch},
  \citenamefont {Dokgo}, \citenamefont {Hwang}, \citenamefont {Torbert},
  \citenamefont {Graham}, \citenamefont {Webster}, \citenamefont {Ergun},
  \citenamefont {Giles}, \citenamefont {Allen}, \citenamefont {Chen},
  \citenamefont {Wang}, \citenamefont {Genestreti}, \citenamefont {Russell},
  \citenamefont {Strangeway},\ and\ \citenamefont {Le~Contel}}]{Burch.2019}%
  \BibitemOpen
  \bibfield  {author} {\bibinfo {author} {\bibfnamefont {J.}~\bibnamefont
  {Burch}}, \bibinfo {author} {\bibfnamefont {K.}~\bibnamefont {Dokgo}},
  \bibinfo {author} {\bibfnamefont {K.}~\bibnamefont {Hwang}}, \bibinfo
  {author} {\bibfnamefont {R.}~\bibnamefont {Torbert}}, \bibinfo {author}
  {\bibfnamefont {D.}~\bibnamefont {Graham}}, \bibinfo {author} {\bibfnamefont
  {J.}~\bibnamefont {Webster}}, \bibinfo {author} {\bibfnamefont
  {R.}~\bibnamefont {Ergun}}, \bibinfo {author} {\bibfnamefont
  {B.}~\bibnamefont {Giles}}, \bibinfo {author} {\bibfnamefont
  {R.}~\bibnamefont {Allen}}, \bibinfo {author} {\bibfnamefont {L.-J.}\
  \bibnamefont {Chen}}, \bibinfo {author} {\bibfnamefont {S.}~\bibnamefont
  {Wang}}, \bibinfo {author} {\bibfnamefont {K.}~\bibnamefont {Genestreti}},
  \bibinfo {author} {\bibfnamefont {C.}~\bibnamefont {Russell}}, \bibinfo
  {author} {\bibfnamefont {R.}~\bibnamefont {Strangeway}}, \ and\ \bibinfo
  {author} {\bibfnamefont {O.}~\bibnamefont {Le~Contel}},\ }\bibfield  {title}
  {\enquote {\bibinfo {title} {High-frequency wave generation in magnetotail
  reconnection: Linear dispersion analysis},}\ }\href {\doibase
  https://doi.org/10.1029/2019GL082471} {\bibfield  {journal} {\bibinfo
  {journal} {Geophysical Research Letters}\ }\textbf {\bibinfo {volume} {46}},\
  \bibinfo {pages} {4089--4097} (\bibinfo {year} {2019})},\ \Eprint
  {http://arxiv.org/abs/https://agupubs.onlinelibrary.wiley.com/doi/pdf/10.1029/2019GL082471}
  {https://agupubs.onlinelibrary.wiley.com/doi/pdf/10.1029/2019GL082471}
  \BibitemShut {NoStop}%
\bibitem [{\citenamefont {Dokgo}\ \emph {et~al.}(2019)\citenamefont {Dokgo},
  \citenamefont {Hwang}, \citenamefont {Burch}, \citenamefont {Choi},
  \citenamefont {Yoon}, \citenamefont {Sibeck},\ and\ \citenamefont
  {Graham}}]{Dokgo.2019}%
  \BibitemOpen
  \bibfield  {author} {\bibinfo {author} {\bibfnamefont {K.}~\bibnamefont
  {Dokgo}}, \bibinfo {author} {\bibfnamefont {K.-J.}\ \bibnamefont {Hwang}},
  \bibinfo {author} {\bibfnamefont {J.~L.}\ \bibnamefont {Burch}}, \bibinfo
  {author} {\bibfnamefont {E.}~\bibnamefont {Choi}}, \bibinfo {author}
  {\bibfnamefont {P.~H.}\ \bibnamefont {Yoon}}, \bibinfo {author}
  {\bibfnamefont {D.~G.}\ \bibnamefont {Sibeck}}, \ and\ \bibinfo {author}
  {\bibfnamefont {D.~B.}\ \bibnamefont {Graham}},\ }\bibfield  {title}
  {\enquote {\bibinfo {title} {High-frequency wave generation in magnetotail
  reconnection: Nonlinear harmonics of upper hybrid waves},}\ }\href {\doibase
  https://doi.org/10.1029/2019GL083361} {\bibfield  {journal} {\bibinfo
  {journal} {Geophysical Research Letters}\ }\textbf {\bibinfo {volume} {46}},\
  \bibinfo {pages} {7873--7882} (\bibinfo {year} {2019})},\ \Eprint
  {http://arxiv.org/abs/https://agupubs.onlinelibrary.wiley.com/doi/pdf/10.1029/2019GL083361}
  {https://agupubs.onlinelibrary.wiley.com/doi/pdf/10.1029/2019GL083361}
  \BibitemShut {NoStop}%
\bibitem [{\citenamefont {Khotyaintsev}\ \emph {et~al.}(2020)\citenamefont
  {Khotyaintsev}, \citenamefont {Graham}, \citenamefont {Steinvall},
  \citenamefont {Alm}, \citenamefont {Vaivads}, \citenamefont {Johlander},
  \citenamefont {Norgren}, \citenamefont {Li}, \citenamefont {Divin},
  \citenamefont {Fu}, \citenamefont {Hwang}, \citenamefont {Burch},
  \citenamefont {Ahmadi}, \citenamefont {Le~Contel}, \citenamefont {Gershman},
  \citenamefont {Russell},\ and\ \citenamefont {Torbert}}]{Khotyaintsev.2019}%
  \BibitemOpen
  \bibfield  {author} {\bibinfo {author} {\bibfnamefont {Y.~V.}\ \bibnamefont
  {Khotyaintsev}}, \bibinfo {author} {\bibfnamefont {D.~B.}\ \bibnamefont
  {Graham}}, \bibinfo {author} {\bibfnamefont {K.}~\bibnamefont {Steinvall}},
  \bibinfo {author} {\bibfnamefont {L.}~\bibnamefont {Alm}}, \bibinfo {author}
  {\bibfnamefont {A.}~\bibnamefont {Vaivads}}, \bibinfo {author} {\bibfnamefont
  {A.}~\bibnamefont {Johlander}}, \bibinfo {author} {\bibfnamefont
  {C.}~\bibnamefont {Norgren}}, \bibinfo {author} {\bibfnamefont
  {W.}~\bibnamefont {Li}}, \bibinfo {author} {\bibfnamefont {A.}~\bibnamefont
  {Divin}}, \bibinfo {author} {\bibfnamefont {H.~S.}\ \bibnamefont {Fu}},
  \bibinfo {author} {\bibfnamefont {K.-J.}\ \bibnamefont {Hwang}}, \bibinfo
  {author} {\bibfnamefont {J.~L.}\ \bibnamefont {Burch}}, \bibinfo {author}
  {\bibfnamefont {N.}~\bibnamefont {Ahmadi}}, \bibinfo {author} {\bibfnamefont
  {O.}~\bibnamefont {Le~Contel}}, \bibinfo {author} {\bibfnamefont {D.~J.}\
  \bibnamefont {Gershman}}, \bibinfo {author} {\bibfnamefont {C.~T.}\
  \bibnamefont {Russell}}, \ and\ \bibinfo {author} {\bibfnamefont {R.~B.}\
  \bibnamefont {Torbert}},\ }\bibfield  {title} {\enquote {\bibinfo {title}
  {Electron heating by debye-scale turbulence in guide-field reconnection},}\
  }\href {\doibase 10.1103/PhysRevLett.124.045101} {\bibfield  {journal}
  {\bibinfo  {journal} {Phys. Rev. Lett.}\ }\textbf {\bibinfo {volume} {124}},\
  \bibinfo {pages} {045101} (\bibinfo {year} {2020})}\BibitemShut {NoStop}%
\bibitem [{\citenamefont {Huba}, \citenamefont {Gladd},\ and\ \citenamefont
  {Drake}(1982)}]{Huba.1982}%
  \BibitemOpen
  \bibfield  {author} {\bibinfo {author} {\bibfnamefont {J.~D.}\ \bibnamefont
  {Huba}}, \bibinfo {author} {\bibfnamefont {N.~T.}\ \bibnamefont {Gladd}}, \
  and\ \bibinfo {author} {\bibfnamefont {J.~F.}\ \bibnamefont {Drake}},\
  }\bibfield  {title} {\enquote {\bibinfo {title} {The lower hybrid drift
  instability in nonantiparallel reversed field plasmas},}\ }\href {\doibase
  https://doi.org/10.1029/JA087iA03p01697} {\bibfield  {journal} {\bibinfo
  {journal} {Journal of Geophysical Research: Space Physics}\ }\textbf
  {\bibinfo {volume} {87}},\ \bibinfo {pages} {1697--1701} (\bibinfo {year}
  {1982})},\ \Eprint
  {http://arxiv.org/abs/https://agupubs.onlinelibrary.wiley.com/doi/pdf/10.1029/JA087iA03p01697}
  {https://agupubs.onlinelibrary.wiley.com/doi/pdf/10.1029/JA087iA03p01697}
  \BibitemShut {NoStop}%
\bibitem [{\citenamefont {Price}\ \emph {et~al.}(2020)\citenamefont {Price},
  \citenamefont {Swisdak}, \citenamefont {Drake},\ and\ \citenamefont
  {Graham}}]{Price.2020}%
  \BibitemOpen
  \bibfield  {author} {\bibinfo {author} {\bibfnamefont {L.}~\bibnamefont
  {Price}}, \bibinfo {author} {\bibfnamefont {M.}~\bibnamefont {Swisdak}},
  \bibinfo {author} {\bibfnamefont {J.~F.}\ \bibnamefont {Drake}}, \ and\
  \bibinfo {author} {\bibfnamefont {D.~B.}\ \bibnamefont {Graham}},\ }\bibfield
   {title} {\enquote {\bibinfo {title} {Turbulence and transport during guide
  field reconnection at the magnetopause},}\ }\href {\doibase
  https://doi.org/10.1029/2019JA027498} {\bibfield  {journal} {\bibinfo
  {journal} {Journal of Geophysical Research: Space Physics}\ }\textbf
  {\bibinfo {volume} {125}},\ \bibinfo {pages} {e2019JA027498} (\bibinfo {year}
  {2020})},\ \bibinfo {note} {e2019JA027498 10.1029/2019JA027498},\ \Eprint
  {http://arxiv.org/abs/https://agupubs.onlinelibrary.wiley.com/doi/pdf/10.1029/2019JA027498}
  {https://agupubs.onlinelibrary.wiley.com/doi/pdf/10.1029/2019JA027498}
  \BibitemShut {NoStop}%
\bibitem [{\citenamefont {{Hesse}}\ \emph {et~al.}(2013)\citenamefont
  {{Hesse}}, \citenamefont {{Aunai}}, \citenamefont {{Zenitani}}, \citenamefont
  {{Kuznetsova}},\ and\ \citenamefont {{Birn}}}]{Hesse.2013}%
  \BibitemOpen
  \bibfield  {author} {\bibinfo {author} {\bibfnamefont {M.}~\bibnamefont
  {{Hesse}}}, \bibinfo {author} {\bibfnamefont {N.}~\bibnamefont {{Aunai}}},
  \bibinfo {author} {\bibfnamefont {S.}~\bibnamefont {{Zenitani}}}, \bibinfo
  {author} {\bibfnamefont {M.}~\bibnamefont {{Kuznetsova}}}, \ and\ \bibinfo
  {author} {\bibfnamefont {J.}~\bibnamefont {{Birn}}},\ }\bibfield  {title}
  {\enquote {\bibinfo {title} {{Aspects of collisionless magnetic reconnection
  in asymmetric systems}},}\ }\href {\doibase 10.1063/1.4811467} {\bibfield
  {journal} {\bibinfo  {journal} {Physics of Plasmas}\ }\textbf {\bibinfo
  {volume} {20}},\ \bibinfo {eid} {061210} (\bibinfo {year}
  {2013})}\BibitemShut {NoStop}%
\bibitem [{\citenamefont {{Liu}}\ \emph {et~al.}(018b)\citenamefont {{Liu}},
  \citenamefont {{Hesse}}, \citenamefont {{Li}}, \citenamefont {{Kuznetsova}},\
  and\ \citenamefont {{Le}}}]{Liu.2018b}%
  \BibitemOpen
  \bibfield  {author} {\bibinfo {author} {\bibfnamefont {Y.-H.}\ \bibnamefont
  {{Liu}}}, \bibinfo {author} {\bibfnamefont {M.}~\bibnamefont {{Hesse}}},
  \bibinfo {author} {\bibfnamefont {T.~C.}\ \bibnamefont {{Li}}}, \bibinfo
  {author} {\bibfnamefont {M.}~\bibnamefont {{Kuznetsova}}}, \ and\ \bibinfo
  {author} {\bibfnamefont {A.}~\bibnamefont {{Le}}},\ }\bibfield  {title}
  {\enquote {\bibinfo {title} {{Orientation and Stability of Asymmetric
  Magnetic Reconnection X Line}},}\ }\href {\doibase 10.1029/2018JA025410}
  {\bibfield  {journal} {\bibinfo  {journal} {Journal of Geophysical Research
  (Space Physics)}\ }\textbf {\bibinfo {volume} {123}},\ \bibinfo {pages}
  {4908--4920} (\bibinfo {year} {2018b})},\ \Eprint
  {http://arxiv.org/abs/1805.07774} {arXiv:1805.07774} \BibitemShut {NoStop}%
\bibitem [{\citenamefont {Nakamura}\ \emph {et~al.}(2021)\citenamefont
  {Nakamura}, \citenamefont {Hasegawa}, \citenamefont {Genestreti},
  \citenamefont {Denton}, \citenamefont {Phan}, \citenamefont {Stawarz},
  \citenamefont {Nakamura},\ and\ \citenamefont {Nystrom}}]{NakamuraTKM.2021}%
  \BibitemOpen
  \bibfield  {author} {\bibinfo {author} {\bibfnamefont {T.~K.~M.}\
  \bibnamefont {Nakamura}}, \bibinfo {author} {\bibfnamefont {H.}~\bibnamefont
  {Hasegawa}}, \bibinfo {author} {\bibfnamefont {K.~J.}\ \bibnamefont
  {Genestreti}}, \bibinfo {author} {\bibfnamefont {R.~E.}\ \bibnamefont
  {Denton}}, \bibinfo {author} {\bibfnamefont {T.~D.}\ \bibnamefont {Phan}},
  \bibinfo {author} {\bibfnamefont {J.~E.}\ \bibnamefont {Stawarz}}, \bibinfo
  {author} {\bibfnamefont {R.}~\bibnamefont {Nakamura}}, \ and\ \bibinfo
  {author} {\bibfnamefont {W.~D.}\ \bibnamefont {Nystrom}},\ }\bibfield
  {title} {\enquote {\bibinfo {title} {Fast cross-scale energy transfer during
  turbulent magnetic reconnection},}\ }\href {\doibase
  https://doi.org/10.1029/2021GL093524} {\bibfield  {journal} {\bibinfo
  {journal} {Geophysical Research Letters}\ }\textbf {\bibinfo {volume} {48}},\
  \bibinfo {pages} {e2021GL093524} (\bibinfo {year} {2021})},\ \bibinfo {note}
  {e2021GL093524 2021GL093524},\ \Eprint
  {http://arxiv.org/abs/https://agupubs.onlinelibrary.wiley.com/doi/pdf/10.1029/2021GL093524}
  {https://agupubs.onlinelibrary.wiley.com/doi/pdf/10.1029/2021GL093524}
  \BibitemShut {NoStop}%
\bibitem [{\citenamefont {Spinnangr}\ \emph {et~al.}(2021)\citenamefont
  {Spinnangr}, \citenamefont {Tenfjord}, \citenamefont {Hesse}, \citenamefont
  {Norgren}, \citenamefont {Kolstø}, \citenamefont {Kwagala}, \citenamefont
  {Jørgensen},\ and\ \citenamefont {Pérez-Coll~Jiménez}}]{Spinnangr.2021}%
  \BibitemOpen
  \bibfield  {author} {\bibinfo {author} {\bibfnamefont {S.~F.}\ \bibnamefont
  {Spinnangr}}, \bibinfo {author} {\bibfnamefont {P.}~\bibnamefont {Tenfjord}},
  \bibinfo {author} {\bibfnamefont {M.}~\bibnamefont {Hesse}}, \bibinfo
  {author} {\bibfnamefont {C.}~\bibnamefont {Norgren}}, \bibinfo {author}
  {\bibfnamefont {H.~M.}\ \bibnamefont {Kolstø}}, \bibinfo {author}
  {\bibfnamefont {N.~K.}\ \bibnamefont {Kwagala}}, \bibinfo {author}
  {\bibfnamefont {T.~M.}\ \bibnamefont {Jørgensen}}, \ and\ \bibinfo {author}
  {\bibfnamefont {J.}~\bibnamefont {Pérez-Coll~Jiménez}},\ }\bibfield
  {title} {\enquote {\bibinfo {title} {Asymmetrically varying guide field
  during magnetic reconnection: Particle-in-cell simulations},}\ }\href
  {\doibase 10.1002/essoar.10508041.1} {\bibfield  {journal} {\bibinfo
  {journal} {Earth and Space Science Open Archive}\ ,\ \bibinfo {pages} {19}}
  (\bibinfo {year} {2021})}\BibitemShut {NoStop}%
\bibitem [{\citenamefont {{Daughton}}\ \emph {et~al.}(2011)\citenamefont
  {{Daughton}}, \citenamefont {{Roytershteyn}}, \citenamefont {{Karimabadi}},
  \citenamefont {{Yin}}, \citenamefont {{Albright}}, \citenamefont {{Bergen}},\
  and\ \citenamefont {{Bowers}}}]{Daughton.2011}%
  \BibitemOpen
  \bibfield  {author} {\bibinfo {author} {\bibfnamefont {W.}~\bibnamefont
  {{Daughton}}}, \bibinfo {author} {\bibfnamefont {V.}~\bibnamefont
  {{Roytershteyn}}}, \bibinfo {author} {\bibfnamefont {H.}~\bibnamefont
  {{Karimabadi}}}, \bibinfo {author} {\bibfnamefont {L.}~\bibnamefont {{Yin}}},
  \bibinfo {author} {\bibfnamefont {B.~J.}\ \bibnamefont {{Albright}}},
  \bibinfo {author} {\bibfnamefont {B.}~\bibnamefont {{Bergen}}}, \ and\
  \bibinfo {author} {\bibfnamefont {K.~J.}\ \bibnamefont {{Bowers}}},\
  }\bibfield  {title} {\enquote {\bibinfo {title} {{Role of electron physics in
  the development of turbulent magnetic reconnection in collisionless
  plasmas}},}\ }\href {\doibase 10.1038/nphys1965} {\bibfield  {journal}
  {\bibinfo  {journal} {Nature Physics}\ }\textbf {\bibinfo {volume} {7}},\
  \bibinfo {pages} {539--542} (\bibinfo {year} {2011})}\BibitemShut {NoStop}%
\bibitem [{\citenamefont {{Lapenta}}\ \emph {et~al.}(2015)\citenamefont
  {{Lapenta}}, \citenamefont {{Markidis}}, \citenamefont {{Goldman}},\ and\
  \citenamefont {{Newman}}}]{Lapenta.2015}%
  \BibitemOpen
  \bibfield  {author} {\bibinfo {author} {\bibfnamefont {G.}~\bibnamefont
  {{Lapenta}}}, \bibinfo {author} {\bibfnamefont {S.}~\bibnamefont
  {{Markidis}}}, \bibinfo {author} {\bibfnamefont {M.~V.}\ \bibnamefont
  {{Goldman}}}, \ and\ \bibinfo {author} {\bibfnamefont {D.~L.}\ \bibnamefont
  {{Newman}}},\ }\bibfield  {title} {\enquote {\bibinfo {title} {{Secondary
  reconnection sites in reconnection-generated flux ropes and reconnection
  fronts}},}\ }\href {\doibase 10.1038/nphys3406} {\bibfield  {journal}
  {\bibinfo  {journal} {Nature Physics}\ }\textbf {\bibinfo {volume} {11}},\
  \bibinfo {pages} {690--695} (\bibinfo {year} {2015})}\BibitemShut {NoStop}%
\bibitem [{\citenamefont {{Ergun}}\ \emph
  {et~al.}(2016{\natexlab{a}})\citenamefont {{Ergun}}, \citenamefont
  {{Goodrich}}, \citenamefont {{Wilder}}, \citenamefont {{Holmes}},
  \citenamefont {{Stawarz}}, \citenamefont {{Eriksson}}, \citenamefont
  {{Sturner}}, \citenamefont {{Malaspina}}, \citenamefont {{Usanova}},
  \citenamefont {{Torbert}}, \citenamefont {{Lindqvist}}, \citenamefont
  {{Khotyaintsev}}, \citenamefont {{Burch}}, \citenamefont {{Strangeway}},
  \citenamefont {{Russell}}, \citenamefont {{Pollock}}, \citenamefont
  {{Giles}}, \citenamefont {{Hesse}}, \citenamefont {{Chen}}, \citenamefont
  {{Lapenta}}, \citenamefont {{Goldman}}, \citenamefont {{Newman}},
  \citenamefont {{Schwartz}}, \citenamefont {{Eastwood}}, \citenamefont
  {{Phan}}, \citenamefont {{Mozer}}, \citenamefont {{Drake}}, \citenamefont
  {{Shay}}, \citenamefont {{Cassak}}, \citenamefont {{Nakamura}},\ and\
  \citenamefont {{Marklund}}}]{Ergun.2016c}%
  \BibitemOpen
  \bibfield  {author} {\bibinfo {author} {\bibfnamefont {R.~E.}\ \bibnamefont
  {{Ergun}}}, \bibinfo {author} {\bibfnamefont {K.~A.}\ \bibnamefont
  {{Goodrich}}}, \bibinfo {author} {\bibfnamefont {F.~D.}\ \bibnamefont
  {{Wilder}}}, \bibinfo {author} {\bibfnamefont {J.~C.}\ \bibnamefont
  {{Holmes}}}, \bibinfo {author} {\bibfnamefont {J.~E.}\ \bibnamefont
  {{Stawarz}}}, \bibinfo {author} {\bibfnamefont {S.}~\bibnamefont
  {{Eriksson}}}, \bibinfo {author} {\bibfnamefont {A.~P.}\ \bibnamefont
  {{Sturner}}}, \bibinfo {author} {\bibfnamefont {D.~M.}\ \bibnamefont
  {{Malaspina}}}, \bibinfo {author} {\bibfnamefont {M.~E.}\ \bibnamefont
  {{Usanova}}}, \bibinfo {author} {\bibfnamefont {R.~B.}\ \bibnamefont
  {{Torbert}}}, \bibinfo {author} {\bibfnamefont {P.~A.}\ \bibnamefont
  {{Lindqvist}}}, \bibinfo {author} {\bibfnamefont {Y.}~\bibnamefont
  {{Khotyaintsev}}}, \bibinfo {author} {\bibfnamefont {J.~L.}\ \bibnamefont
  {{Burch}}}, \bibinfo {author} {\bibfnamefont {R.~J.}\ \bibnamefont
  {{Strangeway}}}, \bibinfo {author} {\bibfnamefont {C.~T.}\ \bibnamefont
  {{Russell}}}, \bibinfo {author} {\bibfnamefont {C.~J.}\ \bibnamefont
  {{Pollock}}}, \bibinfo {author} {\bibfnamefont {B.~L.}\ \bibnamefont
  {{Giles}}}, \bibinfo {author} {\bibfnamefont {M.}~\bibnamefont {{Hesse}}},
  \bibinfo {author} {\bibfnamefont {L.~J.}\ \bibnamefont {{Chen}}}, \bibinfo
  {author} {\bibfnamefont {G.}~\bibnamefont {{Lapenta}}}, \bibinfo {author}
  {\bibfnamefont {M.~V.}\ \bibnamefont {{Goldman}}}, \bibinfo {author}
  {\bibfnamefont {D.~L.}\ \bibnamefont {{Newman}}}, \bibinfo {author}
  {\bibfnamefont {S.~J.}\ \bibnamefont {{Schwartz}}}, \bibinfo {author}
  {\bibfnamefont {J.~P.}\ \bibnamefont {{Eastwood}}}, \bibinfo {author}
  {\bibfnamefont {T.~D.}\ \bibnamefont {{Phan}}}, \bibinfo {author}
  {\bibfnamefont {F.~S.}\ \bibnamefont {{Mozer}}}, \bibinfo {author}
  {\bibfnamefont {J.}~\bibnamefont {{Drake}}}, \bibinfo {author} {\bibfnamefont
  {M.~A.}\ \bibnamefont {{Shay}}}, \bibinfo {author} {\bibfnamefont {P.~A.}\
  \bibnamefont {{Cassak}}}, \bibinfo {author} {\bibfnamefont {R.}~\bibnamefont
  {{Nakamura}}}, \ and\ \bibinfo {author} {\bibfnamefont {G.}~\bibnamefont
  {{Marklund}}},\ }\bibfield  {title} {\enquote {\bibinfo {title}
  {{Magnetospheric Multiscale Satellites Observations of Parallel Electric
  Fields Associated with Magnetic Reconnection}},}\ }\href {\doibase
  10.1103/PhysRevLett.116.235102} {\bibfield  {journal} {\bibinfo  {journal}
  {Phys. Rev. Lett.}\ }\textbf {\bibinfo {volume} {116}},\ \bibinfo {eid}
  {235102} (\bibinfo {year} {2016}{\natexlab{a}})}\BibitemShut {NoStop}%
\bibitem [{\citenamefont {{Liu}}\ \emph {et~al.}(2017)\citenamefont {{Liu}},
  \citenamefont {{Hesse}}, \citenamefont {{Guo}}, \citenamefont {{Daughton}},
  \citenamefont {{Li}}, \citenamefont {{Cassak}},\ and\ \citenamefont
  {{Shay}}}]{Liu.2017}%
  \BibitemOpen
  \bibfield  {author} {\bibinfo {author} {\bibfnamefont {Y.-H.}\ \bibnamefont
  {{Liu}}}, \bibinfo {author} {\bibfnamefont {M.}~\bibnamefont {{Hesse}}},
  \bibinfo {author} {\bibfnamefont {F.}~\bibnamefont {{Guo}}}, \bibinfo
  {author} {\bibfnamefont {W.}~\bibnamefont {{Daughton}}}, \bibinfo {author}
  {\bibfnamefont {H.}~\bibnamefont {{Li}}}, \bibinfo {author} {\bibfnamefont
  {P.~A.}\ \bibnamefont {{Cassak}}}, \ and\ \bibinfo {author} {\bibfnamefont
  {M.~A.}\ \bibnamefont {{Shay}}},\ }\bibfield  {title} {\enquote {\bibinfo
  {title} {{Why does Steady-State Magnetic Reconnection have a Maximum Local
  Rate of Order 0.1?}}}\ }\href {\doibase 10.1103/PhysRevLett.118.085101}
  {\bibfield  {journal} {\bibinfo  {journal} {Phys. Rev. Lett}\ }\textbf
  {\bibinfo {volume} {118}},\ \bibinfo {eid} {085101} (\bibinfo {year}
  {2017})},\ \Eprint {http://arxiv.org/abs/1611.07859} {arXiv:1611.07859
  [physics.plasm-ph]} \BibitemShut {NoStop}%
\bibitem [{\citenamefont {{Liu}}\ \emph {et~al.}(018a)\citenamefont {{Liu}},
  \citenamefont {{Hesse}}, \citenamefont {{Cassak}}, \citenamefont {{Shay}},
  \citenamefont {{Wang}},\ and\ \citenamefont {{Chen}}}]{Liu.2018a}%
  \BibitemOpen
  \bibfield  {author} {\bibinfo {author} {\bibfnamefont {Y.-H.}\ \bibnamefont
  {{Liu}}}, \bibinfo {author} {\bibfnamefont {M.}~\bibnamefont {{Hesse}}},
  \bibinfo {author} {\bibfnamefont {P.~A.}\ \bibnamefont {{Cassak}}}, \bibinfo
  {author} {\bibfnamefont {M.~A.}\ \bibnamefont {{Shay}}}, \bibinfo {author}
  {\bibfnamefont {S.}~\bibnamefont {{Wang}}}, \ and\ \bibinfo {author}
  {\bibfnamefont {L.~J.}\ \bibnamefont {{Chen}}},\ }\bibfield  {title}
  {\enquote {\bibinfo {title} {{On the Collisionless Asymmetric Magnetic
  Reconnection Rate}},}\ }\href {\doibase 10.1002/2017GL076460} {\bibfield
  {journal} {\bibinfo  {journal} {Geophys. Res. Lett.}\ }\textbf {\bibinfo
  {volume} {45}},\ \bibinfo {pages} {3311--3318} (\bibinfo {year} {2018a})},\
  \Eprint {http://arxiv.org/abs/1711.06708} {arXiv:1711.06708
  [physics.plasm-ph]} \BibitemShut {NoStop}%
\bibitem [{\citenamefont {{Torbert}}\ \emph {et~al.}(2017)\citenamefont
  {{Torbert}}, \citenamefont {{Burch}}, \citenamefont {{Argall}}, \citenamefont
  {{Alm}}, \citenamefont {{Farrugia}}, \citenamefont {{Forbes}}, \citenamefont
  {{Giles}}, \citenamefont {{Rager}}, \citenamefont {{Dorelli}}, \citenamefont
  {{Strangeway}}, \citenamefont {{Ergun}}, \citenamefont {{Wilder}},
  \citenamefont {{Ahmadi}}, \citenamefont {{Lindqvist}},\ and\ \citenamefont
  {{Khotyaintsev}}}]{Torbert.2017}%
  \BibitemOpen
  \bibfield  {author} {\bibinfo {author} {\bibfnamefont {R.~B.}\ \bibnamefont
  {{Torbert}}}, \bibinfo {author} {\bibfnamefont {J.~L.}\ \bibnamefont
  {{Burch}}}, \bibinfo {author} {\bibfnamefont {M.~R.}\ \bibnamefont
  {{Argall}}}, \bibinfo {author} {\bibfnamefont {L.}~\bibnamefont {{Alm}}},
  \bibinfo {author} {\bibfnamefont {C.~J.}\ \bibnamefont {{Farrugia}}},
  \bibinfo {author} {\bibfnamefont {T.~G.}\ \bibnamefont {{Forbes}}}, \bibinfo
  {author} {\bibfnamefont {B.~L.}\ \bibnamefont {{Giles}}}, \bibinfo {author}
  {\bibfnamefont {A.}~\bibnamefont {{Rager}}}, \bibinfo {author} {\bibfnamefont
  {J.}~\bibnamefont {{Dorelli}}}, \bibinfo {author} {\bibfnamefont {R.~J.}\
  \bibnamefont {{Strangeway}}}, \bibinfo {author} {\bibfnamefont {R.~E.}\
  \bibnamefont {{Ergun}}}, \bibinfo {author} {\bibfnamefont {F.~D.}\
  \bibnamefont {{Wilder}}}, \bibinfo {author} {\bibfnamefont {N.}~\bibnamefont
  {{Ahmadi}}}, \bibinfo {author} {\bibfnamefont {P.~A.}\ \bibnamefont
  {{Lindqvist}}}, \ and\ \bibinfo {author} {\bibfnamefont {Y.}~\bibnamefont
  {{Khotyaintsev}}},\ }\bibfield  {title} {\enquote {\bibinfo {title}
  {{Structure and Dissipation Characteristics of an Electron Diffusion Region
  Observed by MMS During a Rapid, Normal-Incidence Magnetopause Crossing}},}\
  }\href {\doibase 10.1002/2017JA024579} {\bibfield  {journal} {\bibinfo
  {journal} {Journal of Geophysical Research (Space Physics)}\ }\textbf
  {\bibinfo {volume} {122}},\ \bibinfo {pages} {11,901--11,916} (\bibinfo
  {year} {2017})}\BibitemShut {NoStop}%
\bibitem [{\citenamefont {{Fuselier}}\ \emph {et~al.}(2016)\citenamefont
  {{Fuselier}}, \citenamefont {{Lewis}}, \citenamefont {{Schiff}},
  \citenamefont {{Ergun}}, \citenamefont {{Burch}}, \citenamefont
  {{Petrinec}},\ and\ \citenamefont {{Trattner}}}]{Fuselier.2016}%
  \BibitemOpen
  \bibfield  {author} {\bibinfo {author} {\bibfnamefont {S.~A.}\ \bibnamefont
  {{Fuselier}}}, \bibinfo {author} {\bibfnamefont {W.~S.}\ \bibnamefont
  {{Lewis}}}, \bibinfo {author} {\bibfnamefont {C.}~\bibnamefont {{Schiff}}},
  \bibinfo {author} {\bibfnamefont {R.}~\bibnamefont {{Ergun}}}, \bibinfo
  {author} {\bibfnamefont {J.~L.}\ \bibnamefont {{Burch}}}, \bibinfo {author}
  {\bibfnamefont {S.~M.}\ \bibnamefont {{Petrinec}}}, \ and\ \bibinfo {author}
  {\bibfnamefont {K.~J.}\ \bibnamefont {{Trattner}}},\ }\bibfield  {title}
  {\enquote {\bibinfo {title} {{Magnetospheric Multiscale Science Mission
  Profile and Operations}},}\ }\href {\doibase 10.1007/s11214-014-0087-x}
  {\bibfield  {journal} {\bibinfo  {journal} {Space Science Reviews}\ }\textbf
  {\bibinfo {volume} {199}},\ \bibinfo {pages} {77--103} (\bibinfo {year}
  {2016})}\BibitemShut {NoStop}%
\bibitem [{\citenamefont {{Pollock}}\ \emph {et~al.}(2016)\citenamefont
  {{Pollock}}, \citenamefont {{Moore}}, \citenamefont {{Jacques}},
  \citenamefont {{Burch}}, \citenamefont {{Gliese}}, \citenamefont {{Saito}},
  \citenamefont {{Omoto}}, \citenamefont {{Avanov}}, \citenamefont {{Barrie}},
  \citenamefont {{Coffey}}, \citenamefont {{Dorelli}}, \citenamefont
  {{Gershman}}, \citenamefont {{Giles}}, \citenamefont {{Rosnack}},
  \citenamefont {{Salo}}, \citenamefont {{Yokota}}, \citenamefont {{Adrian}},
  \citenamefont {{Aoustin}}, \citenamefont {{Auletti}}, \citenamefont {{Aung}},
  \citenamefont {{Bigio}}, \citenamefont {{Cao}}, \citenamefont {{Chandler}},
  \citenamefont {{Chornay}}, \citenamefont {{Christian}}, \citenamefont
  {{Clark}}, \citenamefont {{Collinson}}, \citenamefont {{Corris}},
  \citenamefont {{De Los Santos}}, \citenamefont {{Devlin}}, \citenamefont
  {{Diaz}}, \citenamefont {{Dickerson}}, \citenamefont {{Dickson}},
  \citenamefont {{Diekmann}}, \citenamefont {{Diggs}}, \citenamefont
  {{Duncan}}, \citenamefont {{Figueroa-Vinas}}, \citenamefont {{Firman}},
  \citenamefont {{Freeman}}, \citenamefont {{Galassi}}, \citenamefont
  {{Garcia}}, \citenamefont {{Goodhart}}, \citenamefont {{Guererro}},
  \citenamefont {{Hageman}}, \citenamefont {{Hanley}}, \citenamefont
  {{Hemminger}}, \citenamefont {{Holland}}, \citenamefont {{Hutchins}},
  \citenamefont {{James}}, \citenamefont {{Jones}}, \citenamefont {{Kreisler}},
  \citenamefont {{Kujawski}}, \citenamefont {{Lavu}}, \citenamefont {{Lobell}},
  \citenamefont {{LeCompte}}, \citenamefont {{Lukemire}}, \citenamefont
  {{MacDonald}}, \citenamefont {{Mariano}}, \citenamefont {{Mukai}},
  \citenamefont {{Narayanan}}, \citenamefont {{Nguyan}}, \citenamefont
  {{Onizuka}}, \citenamefont {{Paterson}}, \citenamefont {{Persyn}},
  \citenamefont {{Piepgrass}}, \citenamefont {{Cheney}}, \citenamefont
  {{Rager}}, \citenamefont {{Raghuram}}, \citenamefont {{Ramil}}, \citenamefont
  {{Reichenthal}}, \citenamefont {{Rodriguez}}, \citenamefont {{Rouzaud}},
  \citenamefont {{Rucker}}, \citenamefont {{Saito}}, \citenamefont {{Samara}},
  \citenamefont {{Sauvaud}}, \citenamefont {{Schuster}}, \citenamefont
  {{Shappirio}}, \citenamefont {{Shelton}}, \citenamefont {{Sher}},
  \citenamefont {{Smith}}, \citenamefont {{Smith}}, \citenamefont {{Smith}},
  \citenamefont {{Steinfeld}}, \citenamefont {{Szymkiewicz}}, \citenamefont
  {{Tanimoto}}, \citenamefont {{Taylor}}, \citenamefont {{Tucker}},
  \citenamefont {{Tull}}, \citenamefont {{Uhl}}, \citenamefont {{Vloet}},
  \citenamefont {{Walpole}}, \citenamefont {{Weidner}}, \citenamefont
  {{White}}, \citenamefont {{Winkert}}, \citenamefont {{Yeh}},\ and\
  \citenamefont {{Zeuch}}}]{Pollock.2016}%
  \BibitemOpen
  \bibfield  {author} {\bibinfo {author} {\bibfnamefont {C.}~\bibnamefont
  {{Pollock}}}, \bibinfo {author} {\bibfnamefont {T.}~\bibnamefont {{Moore}}},
  \bibinfo {author} {\bibfnamefont {A.}~\bibnamefont {{Jacques}}}, \bibinfo
  {author} {\bibfnamefont {J.}~\bibnamefont {{Burch}}}, \bibinfo {author}
  {\bibfnamefont {U.}~\bibnamefont {{Gliese}}}, \bibinfo {author}
  {\bibfnamefont {Y.}~\bibnamefont {{Saito}}}, \bibinfo {author} {\bibfnamefont
  {T.}~\bibnamefont {{Omoto}}}, \bibinfo {author} {\bibfnamefont
  {L.}~\bibnamefont {{Avanov}}}, \bibinfo {author} {\bibfnamefont
  {A.}~\bibnamefont {{Barrie}}}, \bibinfo {author} {\bibfnamefont
  {V.}~\bibnamefont {{Coffey}}}, \bibinfo {author} {\bibfnamefont
  {J.}~\bibnamefont {{Dorelli}}}, \bibinfo {author} {\bibfnamefont
  {D.}~\bibnamefont {{Gershman}}}, \bibinfo {author} {\bibfnamefont
  {B.}~\bibnamefont {{Giles}}}, \bibinfo {author} {\bibfnamefont
  {T.}~\bibnamefont {{Rosnack}}}, \bibinfo {author} {\bibfnamefont
  {C.}~\bibnamefont {{Salo}}}, \bibinfo {author} {\bibfnamefont
  {S.}~\bibnamefont {{Yokota}}}, \bibinfo {author} {\bibfnamefont
  {M.}~\bibnamefont {{Adrian}}}, \bibinfo {author} {\bibfnamefont
  {C.}~\bibnamefont {{Aoustin}}}, \bibinfo {author} {\bibfnamefont
  {C.}~\bibnamefont {{Auletti}}}, \bibinfo {author} {\bibfnamefont
  {S.}~\bibnamefont {{Aung}}}, \bibinfo {author} {\bibfnamefont
  {V.}~\bibnamefont {{Bigio}}}, \bibinfo {author} {\bibfnamefont
  {N.}~\bibnamefont {{Cao}}}, \bibinfo {author} {\bibfnamefont
  {M.}~\bibnamefont {{Chandler}}}, \bibinfo {author} {\bibfnamefont
  {D.}~\bibnamefont {{Chornay}}}, \bibinfo {author} {\bibfnamefont
  {K.}~\bibnamefont {{Christian}}}, \bibinfo {author} {\bibfnamefont
  {G.}~\bibnamefont {{Clark}}}, \bibinfo {author} {\bibfnamefont
  {G.}~\bibnamefont {{Collinson}}}, \bibinfo {author} {\bibfnamefont
  {T.}~\bibnamefont {{Corris}}}, \bibinfo {author} {\bibfnamefont
  {A.}~\bibnamefont {{De Los Santos}}}, \bibinfo {author} {\bibfnamefont
  {R.}~\bibnamefont {{Devlin}}}, \bibinfo {author} {\bibfnamefont
  {T.}~\bibnamefont {{Diaz}}}, \bibinfo {author} {\bibfnamefont
  {T.}~\bibnamefont {{Dickerson}}}, \bibinfo {author} {\bibfnamefont
  {C.}~\bibnamefont {{Dickson}}}, \bibinfo {author} {\bibfnamefont
  {A.}~\bibnamefont {{Diekmann}}}, \bibinfo {author} {\bibfnamefont
  {F.}~\bibnamefont {{Diggs}}}, \bibinfo {author} {\bibfnamefont
  {C.}~\bibnamefont {{Duncan}}}, \bibinfo {author} {\bibfnamefont
  {A.}~\bibnamefont {{Figueroa-Vinas}}}, \bibinfo {author} {\bibfnamefont
  {C.}~\bibnamefont {{Firman}}}, \bibinfo {author} {\bibfnamefont
  {M.}~\bibnamefont {{Freeman}}}, \bibinfo {author} {\bibfnamefont
  {N.}~\bibnamefont {{Galassi}}}, \bibinfo {author} {\bibfnamefont
  {K.}~\bibnamefont {{Garcia}}}, \bibinfo {author} {\bibfnamefont
  {G.}~\bibnamefont {{Goodhart}}}, \bibinfo {author} {\bibfnamefont
  {D.}~\bibnamefont {{Guererro}}}, \bibinfo {author} {\bibfnamefont
  {J.}~\bibnamefont {{Hageman}}}, \bibinfo {author} {\bibfnamefont
  {J.}~\bibnamefont {{Hanley}}}, \bibinfo {author} {\bibfnamefont
  {E.}~\bibnamefont {{Hemminger}}}, \bibinfo {author} {\bibfnamefont
  {M.}~\bibnamefont {{Holland}}}, \bibinfo {author} {\bibfnamefont
  {M.}~\bibnamefont {{Hutchins}}}, \bibinfo {author} {\bibfnamefont
  {T.}~\bibnamefont {{James}}}, \bibinfo {author} {\bibfnamefont
  {W.}~\bibnamefont {{Jones}}}, \bibinfo {author} {\bibfnamefont
  {S.}~\bibnamefont {{Kreisler}}}, \bibinfo {author} {\bibfnamefont
  {J.}~\bibnamefont {{Kujawski}}}, \bibinfo {author} {\bibfnamefont
  {V.}~\bibnamefont {{Lavu}}}, \bibinfo {author} {\bibfnamefont
  {J.}~\bibnamefont {{Lobell}}}, \bibinfo {author} {\bibfnamefont
  {E.}~\bibnamefont {{LeCompte}}}, \bibinfo {author} {\bibfnamefont
  {A.}~\bibnamefont {{Lukemire}}}, \bibinfo {author} {\bibfnamefont
  {E.}~\bibnamefont {{MacDonald}}}, \bibinfo {author} {\bibfnamefont
  {A.}~\bibnamefont {{Mariano}}}, \bibinfo {author} {\bibfnamefont
  {T.}~\bibnamefont {{Mukai}}}, \bibinfo {author} {\bibfnamefont
  {K.}~\bibnamefont {{Narayanan}}}, \bibinfo {author} {\bibfnamefont
  {Q.}~\bibnamefont {{Nguyan}}}, \bibinfo {author} {\bibfnamefont
  {M.}~\bibnamefont {{Onizuka}}}, \bibinfo {author} {\bibfnamefont
  {W.}~\bibnamefont {{Paterson}}}, \bibinfo {author} {\bibfnamefont
  {S.}~\bibnamefont {{Persyn}}}, \bibinfo {author} {\bibfnamefont
  {B.}~\bibnamefont {{Piepgrass}}}, \bibinfo {author} {\bibfnamefont
  {F.}~\bibnamefont {{Cheney}}}, \bibinfo {author} {\bibfnamefont
  {A.}~\bibnamefont {{Rager}}}, \bibinfo {author} {\bibfnamefont
  {T.}~\bibnamefont {{Raghuram}}}, \bibinfo {author} {\bibfnamefont
  {A.}~\bibnamefont {{Ramil}}}, \bibinfo {author} {\bibfnamefont
  {L.}~\bibnamefont {{Reichenthal}}}, \bibinfo {author} {\bibfnamefont
  {H.}~\bibnamefont {{Rodriguez}}}, \bibinfo {author} {\bibfnamefont
  {J.}~\bibnamefont {{Rouzaud}}}, \bibinfo {author} {\bibfnamefont
  {A.}~\bibnamefont {{Rucker}}}, \bibinfo {author} {\bibfnamefont
  {Y.}~\bibnamefont {{Saito}}}, \bibinfo {author} {\bibfnamefont
  {M.}~\bibnamefont {{Samara}}}, \bibinfo {author} {\bibfnamefont {J.-A.}\
  \bibnamefont {{Sauvaud}}}, \bibinfo {author} {\bibfnamefont {D.}~\bibnamefont
  {{Schuster}}}, \bibinfo {author} {\bibfnamefont {M.}~\bibnamefont
  {{Shappirio}}}, \bibinfo {author} {\bibfnamefont {K.}~\bibnamefont
  {{Shelton}}}, \bibinfo {author} {\bibfnamefont {D.}~\bibnamefont {{Sher}}},
  \bibinfo {author} {\bibfnamefont {D.}~\bibnamefont {{Smith}}}, \bibinfo
  {author} {\bibfnamefont {K.}~\bibnamefont {{Smith}}}, \bibinfo {author}
  {\bibfnamefont {S.}~\bibnamefont {{Smith}}}, \bibinfo {author} {\bibfnamefont
  {D.}~\bibnamefont {{Steinfeld}}}, \bibinfo {author} {\bibfnamefont
  {R.}~\bibnamefont {{Szymkiewicz}}}, \bibinfo {author} {\bibfnamefont
  {K.}~\bibnamefont {{Tanimoto}}}, \bibinfo {author} {\bibfnamefont
  {J.}~\bibnamefont {{Taylor}}}, \bibinfo {author} {\bibfnamefont
  {C.}~\bibnamefont {{Tucker}}}, \bibinfo {author} {\bibfnamefont
  {K.}~\bibnamefont {{Tull}}}, \bibinfo {author} {\bibfnamefont
  {A.}~\bibnamefont {{Uhl}}}, \bibinfo {author} {\bibfnamefont
  {J.}~\bibnamefont {{Vloet}}}, \bibinfo {author} {\bibfnamefont
  {P.}~\bibnamefont {{Walpole}}}, \bibinfo {author} {\bibfnamefont
  {S.}~\bibnamefont {{Weidner}}}, \bibinfo {author} {\bibfnamefont
  {D.}~\bibnamefont {{White}}}, \bibinfo {author} {\bibfnamefont
  {G.}~\bibnamefont {{Winkert}}}, \bibinfo {author} {\bibfnamefont {P.-S.}\
  \bibnamefont {{Yeh}}}, \ and\ \bibinfo {author} {\bibfnamefont
  {M.}~\bibnamefont {{Zeuch}}},\ }\bibfield  {title} {\enquote {\bibinfo
  {title} {{Fast Plasma Investigation for Magnetospheric Multiscale}},}\ }\href
  {\doibase 10.1007/s11214-016-0245-4} {\bibfield  {journal} {\bibinfo
  {journal} {Spa. Sci. Rev.}\ }\textbf {\bibinfo {volume} {199}},\ \bibinfo
  {pages} {331--406} (\bibinfo {year} {2016})}\BibitemShut {NoStop}%
\bibitem [{\citenamefont {{Young}}\ \emph {et~al.}(2016)\citenamefont
  {{Young}}, \citenamefont {{Burch}}, \citenamefont {{Gomez}}, \citenamefont
  {{De Los Santos}}, \citenamefont {{Miller}}, \citenamefont {{Wilson}},
  \citenamefont {{Paschalidis}}, \citenamefont {{Fuselier}}, \citenamefont
  {{Pickens}}, \citenamefont {{Hertzberg}}, \citenamefont {{Pollock}},
  \citenamefont {{Scherrer}}, \citenamefont {{Wood}}, \citenamefont {{Donald}},
  \citenamefont {{Aaron}}, \citenamefont {{Furman}}, \citenamefont {{George}},
  \citenamefont {{Gurnee}}, \citenamefont {{Hourani}}, \citenamefont
  {{Jacques}}, \citenamefont {{Johnson}}, \citenamefont {{Orr}}, \citenamefont
  {{Pan}}, \citenamefont {{Persyn}}, \citenamefont {{Pope}}, \citenamefont
  {{Roberts}}, \citenamefont {{Stokes}}, \citenamefont {{Trattner}},\ and\
  \citenamefont {{Webster}}}]{Young.2016}%
  \BibitemOpen
  \bibfield  {author} {\bibinfo {author} {\bibfnamefont {D.~T.}\ \bibnamefont
  {{Young}}}, \bibinfo {author} {\bibfnamefont {J.~L.}\ \bibnamefont
  {{Burch}}}, \bibinfo {author} {\bibfnamefont {R.~G.}\ \bibnamefont
  {{Gomez}}}, \bibinfo {author} {\bibfnamefont {A.}~\bibnamefont {{De Los
  Santos}}}, \bibinfo {author} {\bibfnamefont {G.~P.}\ \bibnamefont
  {{Miller}}}, \bibinfo {author} {\bibfnamefont {P.}~\bibnamefont {{Wilson}}},
  \bibinfo {author} {\bibfnamefont {N.}~\bibnamefont {{Paschalidis}}}, \bibinfo
  {author} {\bibfnamefont {S.~A.}\ \bibnamefont {{Fuselier}}}, \bibinfo
  {author} {\bibfnamefont {K.}~\bibnamefont {{Pickens}}}, \bibinfo {author}
  {\bibfnamefont {E.}~\bibnamefont {{Hertzberg}}}, \bibinfo {author}
  {\bibfnamefont {C.~J.}\ \bibnamefont {{Pollock}}}, \bibinfo {author}
  {\bibfnamefont {J.}~\bibnamefont {{Scherrer}}}, \bibinfo {author}
  {\bibfnamefont {P.~B.}\ \bibnamefont {{Wood}}}, \bibinfo {author}
  {\bibfnamefont {E.~T.}\ \bibnamefont {{Donald}}}, \bibinfo {author}
  {\bibfnamefont {D.}~\bibnamefont {{Aaron}}}, \bibinfo {author} {\bibfnamefont
  {J.}~\bibnamefont {{Furman}}}, \bibinfo {author} {\bibfnamefont
  {D.}~\bibnamefont {{George}}}, \bibinfo {author} {\bibfnamefont {R.~S.}\
  \bibnamefont {{Gurnee}}}, \bibinfo {author} {\bibfnamefont {R.~S.}\
  \bibnamefont {{Hourani}}}, \bibinfo {author} {\bibfnamefont {A.}~\bibnamefont
  {{Jacques}}}, \bibinfo {author} {\bibfnamefont {T.}~\bibnamefont
  {{Johnson}}}, \bibinfo {author} {\bibfnamefont {T.}~\bibnamefont {{Orr}}},
  \bibinfo {author} {\bibfnamefont {K.~S.}\ \bibnamefont {{Pan}}}, \bibinfo
  {author} {\bibfnamefont {S.}~\bibnamefont {{Persyn}}}, \bibinfo {author}
  {\bibfnamefont {S.}~\bibnamefont {{Pope}}}, \bibinfo {author} {\bibfnamefont
  {J.}~\bibnamefont {{Roberts}}}, \bibinfo {author} {\bibfnamefont {M.~R.}\
  \bibnamefont {{Stokes}}}, \bibinfo {author} {\bibfnamefont {K.~J.}\
  \bibnamefont {{Trattner}}}, \ and\ \bibinfo {author} {\bibfnamefont {J.~M.}\
  \bibnamefont {{Webster}}},\ }\bibfield  {title} {\enquote {\bibinfo {title}
  {{Hot Plasma Composition Analyzer for the Magnetospheric Multiscale
  Mission}},}\ }\href {\doibase 10.1007/s11214-014-0119-6} {\bibfield
  {journal} {\bibinfo  {journal} {Spa. Sci. Rev.}\ }\textbf {\bibinfo {volume}
  {199}},\ \bibinfo {pages} {407--470} (\bibinfo {year} {2016})}\BibitemShut
  {NoStop}%
\bibitem [{\citenamefont {{Lindqvist}}\ \emph {et~al.}(2016)\citenamefont
  {{Lindqvist}}, \citenamefont {{Olsson}}, \citenamefont {{Torbert}},
  \citenamefont {{King}}, \citenamefont {{Granoff}}, \citenamefont {{Rau}},
  \citenamefont {{Needell}}, \citenamefont {{Turco}}, \citenamefont {{Dors}},
  \citenamefont {{Beckman}}, \citenamefont {{Macri}}, \citenamefont {{Frost}},
  \citenamefont {{Salwen}}, \citenamefont {{Eriksson}}, \citenamefont
  {{{\AA}hl{\'e}n}}, \citenamefont {{Khotyaintsev}}, \citenamefont {{Porter}},
  \citenamefont {{Lappalainen}}, \citenamefont {{Ergun}}, \citenamefont
  {{Wermeer}},\ and\ \citenamefont {{Tucker}}}]{Lindqvist.2016}%
  \BibitemOpen
  \bibfield  {author} {\bibinfo {author} {\bibfnamefont {P.-A.}\ \bibnamefont
  {{Lindqvist}}}, \bibinfo {author} {\bibfnamefont {G.}~\bibnamefont
  {{Olsson}}}, \bibinfo {author} {\bibfnamefont {R.~B.}\ \bibnamefont
  {{Torbert}}}, \bibinfo {author} {\bibfnamefont {B.}~\bibnamefont {{King}}},
  \bibinfo {author} {\bibfnamefont {M.}~\bibnamefont {{Granoff}}}, \bibinfo
  {author} {\bibfnamefont {D.}~\bibnamefont {{Rau}}}, \bibinfo {author}
  {\bibfnamefont {G.}~\bibnamefont {{Needell}}}, \bibinfo {author}
  {\bibfnamefont {S.}~\bibnamefont {{Turco}}}, \bibinfo {author} {\bibfnamefont
  {I.}~\bibnamefont {{Dors}}}, \bibinfo {author} {\bibfnamefont
  {P.}~\bibnamefont {{Beckman}}}, \bibinfo {author} {\bibfnamefont
  {J.}~\bibnamefont {{Macri}}}, \bibinfo {author} {\bibfnamefont
  {C.}~\bibnamefont {{Frost}}}, \bibinfo {author} {\bibfnamefont
  {J.}~\bibnamefont {{Salwen}}}, \bibinfo {author} {\bibfnamefont
  {A.}~\bibnamefont {{Eriksson}}}, \bibinfo {author} {\bibfnamefont
  {L.}~\bibnamefont {{{\AA}hl{\'e}n}}}, \bibinfo {author} {\bibfnamefont
  {Y.~V.}\ \bibnamefont {{Khotyaintsev}}}, \bibinfo {author} {\bibfnamefont
  {J.}~\bibnamefont {{Porter}}}, \bibinfo {author} {\bibfnamefont
  {K.}~\bibnamefont {{Lappalainen}}}, \bibinfo {author} {\bibfnamefont {R.~E.}\
  \bibnamefont {{Ergun}}}, \bibinfo {author} {\bibfnamefont {W.}~\bibnamefont
  {{Wermeer}}}, \ and\ \bibinfo {author} {\bibfnamefont {S.}~\bibnamefont
  {{Tucker}}},\ }\bibfield  {title} {\enquote {\bibinfo {title} {{The
  Spin-Plane Double Probe Electric Field Instrument for MMS}},}\ }\href
  {\doibase 10.1007/s11214-014-0116-9} {\bibfield  {journal} {\bibinfo
  {journal} {Spa. Sci. Rev.}\ }\textbf {\bibinfo {volume} {199}},\ \bibinfo
  {pages} {137--165} (\bibinfo {year} {2016})}\BibitemShut {NoStop}%
\bibitem [{\citenamefont {{Ergun}}\ \emph
  {et~al.}(2016{\natexlab{b}})\citenamefont {{Ergun}}, \citenamefont
  {{Tucker}}, \citenamefont {{Westfall}}, \citenamefont {{Goodrich}},
  \citenamefont {{Malaspina}}, \citenamefont {{Summers}}, \citenamefont
  {{Wallace}}, \citenamefont {{Karlsson}}, \citenamefont {{Mack}},
  \citenamefont {{Brennan}}, \citenamefont {{Pyke}}, \citenamefont
  {{Withnell}}, \citenamefont {{Torbert}}, \citenamefont {{Macri}},
  \citenamefont {{Rau}}, \citenamefont {{Dors}}, \citenamefont {{Needell}},
  \citenamefont {{Lindqvist}}, \citenamefont {{Olsson}},\ and\ \citenamefont
  {{Cully}}}]{Ergun.2016a}%
  \BibitemOpen
  \bibfield  {author} {\bibinfo {author} {\bibfnamefont {R.~E.}\ \bibnamefont
  {{Ergun}}}, \bibinfo {author} {\bibfnamefont {S.}~\bibnamefont {{Tucker}}},
  \bibinfo {author} {\bibfnamefont {J.}~\bibnamefont {{Westfall}}}, \bibinfo
  {author} {\bibfnamefont {K.~A.}\ \bibnamefont {{Goodrich}}}, \bibinfo
  {author} {\bibfnamefont {D.~M.}\ \bibnamefont {{Malaspina}}}, \bibinfo
  {author} {\bibfnamefont {D.}~\bibnamefont {{Summers}}}, \bibinfo {author}
  {\bibfnamefont {J.}~\bibnamefont {{Wallace}}}, \bibinfo {author}
  {\bibfnamefont {M.}~\bibnamefont {{Karlsson}}}, \bibinfo {author}
  {\bibfnamefont {J.}~\bibnamefont {{Mack}}}, \bibinfo {author} {\bibfnamefont
  {N.}~\bibnamefont {{Brennan}}}, \bibinfo {author} {\bibfnamefont
  {B.}~\bibnamefont {{Pyke}}}, \bibinfo {author} {\bibfnamefont
  {P.}~\bibnamefont {{Withnell}}}, \bibinfo {author} {\bibfnamefont
  {R.}~\bibnamefont {{Torbert}}}, \bibinfo {author} {\bibfnamefont
  {J.}~\bibnamefont {{Macri}}}, \bibinfo {author} {\bibfnamefont
  {D.}~\bibnamefont {{Rau}}}, \bibinfo {author} {\bibfnamefont
  {I.}~\bibnamefont {{Dors}}}, \bibinfo {author} {\bibfnamefont
  {J.}~\bibnamefont {{Needell}}}, \bibinfo {author} {\bibfnamefont {P.-A.}\
  \bibnamefont {{Lindqvist}}}, \bibinfo {author} {\bibfnamefont
  {G.}~\bibnamefont {{Olsson}}}, \ and\ \bibinfo {author} {\bibfnamefont
  {C.~M.}\ \bibnamefont {{Cully}}},\ }\bibfield  {title} {\enquote {\bibinfo
  {title} {{The Axial Double Probe and Fields Signal Processing for the MMS
  Mission}},}\ }\href {\doibase 10.1007/s11214-014-0115-x} {\bibfield
  {journal} {\bibinfo  {journal} {Spa. Sci. Rev.}\ }\textbf {\bibinfo {volume}
  {199}},\ \bibinfo {pages} {167--188} (\bibinfo {year}
  {2016}{\natexlab{b}})}\BibitemShut {NoStop}%
\bibitem [{\citenamefont {{Russell}}\ \emph {et~al.}(2016)\citenamefont
  {{Russell}}, \citenamefont {{Anderson}}, \citenamefont {{Baumjohann}},
  \citenamefont {{Bromund}}, \citenamefont {{Dearborn}}, \citenamefont
  {{Fischer}}, \citenamefont {{Le}}, \citenamefont {{Leinweber}}, \citenamefont
  {{Leneman}}, \citenamefont {{Magnes}}, \citenamefont {{Means}}, \citenamefont
  {{Moldwin}}, \citenamefont {{Nakamura}}, \citenamefont {{Pierce}},
  \citenamefont {{Plaschke}}, \citenamefont {{Rowe}}, \citenamefont {{Slavin}},
  \citenamefont {{Strangeway}}, \citenamefont {{Torbert}}, \citenamefont
  {{Hagen}}, \citenamefont {{Jernej}}, \citenamefont {{Valavanoglou}},\ and\
  \citenamefont {{Richter}}}]{Russell.2016}%
  \BibitemOpen
  \bibfield  {author} {\bibinfo {author} {\bibfnamefont {C.~T.}\ \bibnamefont
  {{Russell}}}, \bibinfo {author} {\bibfnamefont {B.~J.}\ \bibnamefont
  {{Anderson}}}, \bibinfo {author} {\bibfnamefont {W.}~\bibnamefont
  {{Baumjohann}}}, \bibinfo {author} {\bibfnamefont {K.~R.}\ \bibnamefont
  {{Bromund}}}, \bibinfo {author} {\bibfnamefont {D.}~\bibnamefont
  {{Dearborn}}}, \bibinfo {author} {\bibfnamefont {D.}~\bibnamefont
  {{Fischer}}}, \bibinfo {author} {\bibfnamefont {G.}~\bibnamefont {{Le}}},
  \bibinfo {author} {\bibfnamefont {H.~K.}\ \bibnamefont {{Leinweber}}},
  \bibinfo {author} {\bibfnamefont {D.}~\bibnamefont {{Leneman}}}, \bibinfo
  {author} {\bibfnamefont {W.}~\bibnamefont {{Magnes}}}, \bibinfo {author}
  {\bibfnamefont {J.~D.}\ \bibnamefont {{Means}}}, \bibinfo {author}
  {\bibfnamefont {M.~B.}\ \bibnamefont {{Moldwin}}}, \bibinfo {author}
  {\bibfnamefont {R.}~\bibnamefont {{Nakamura}}}, \bibinfo {author}
  {\bibfnamefont {D.}~\bibnamefont {{Pierce}}}, \bibinfo {author}
  {\bibfnamefont {F.}~\bibnamefont {{Plaschke}}}, \bibinfo {author}
  {\bibfnamefont {K.~M.}\ \bibnamefont {{Rowe}}}, \bibinfo {author}
  {\bibfnamefont {J.~A.}\ \bibnamefont {{Slavin}}}, \bibinfo {author}
  {\bibfnamefont {R.~J.}\ \bibnamefont {{Strangeway}}}, \bibinfo {author}
  {\bibfnamefont {R.}~\bibnamefont {{Torbert}}}, \bibinfo {author}
  {\bibfnamefont {C.}~\bibnamefont {{Hagen}}}, \bibinfo {author} {\bibfnamefont
  {I.}~\bibnamefont {{Jernej}}}, \bibinfo {author} {\bibfnamefont
  {A.}~\bibnamefont {{Valavanoglou}}}, \ and\ \bibinfo {author} {\bibfnamefont
  {I.}~\bibnamefont {{Richter}}},\ }\bibfield  {title} {\enquote {\bibinfo
  {title} {{The Magnetospheric Multiscale Magnetometers}},}\ }\href {\doibase
  10.1007/s11214-014-0057-3} {\bibfield  {journal} {\bibinfo  {journal} {Spa.
  Sci. Rev.}\ }\textbf {\bibinfo {volume} {199}},\ \bibinfo {pages} {189--256}
  (\bibinfo {year} {2016})}\BibitemShut {NoStop}%
\bibitem [{\citenamefont {{Phan}}\ \emph {et~al.}(2016)\citenamefont {{Phan}},
  \citenamefont {{Eastwood}}, \citenamefont {{Cassak}}, \citenamefont
  {{{\O}ieroset}}, \citenamefont {{Gosling}}, \citenamefont {{Gershman}},
  \citenamefont {{Mozer}}, \citenamefont {{Shay}}, \citenamefont {{Fujimoto}},
  \citenamefont {{Daughton}}, \citenamefont {{Drake}}, \citenamefont {{Burch}},
  \citenamefont {{Torbert}}, \citenamefont {{Ergun}}, \citenamefont {{Chen}},
  \citenamefont {{Wang}}, \citenamefont {{Pollock}}, \citenamefont {{Dorelli}},
  \citenamefont {{Lavraud}}, \citenamefont {{Giles}}, \citenamefont {{Moore}},
  \citenamefont {{Saito}}, \citenamefont {{Avanov}}, \citenamefont
  {{Paterson}}, \citenamefont {{Strangeway}}, \citenamefont {{Russell}},
  \citenamefont {{Khotyaintsev}}, \citenamefont {{Lindqvist}}, \citenamefont
  {{Oka}},\ and\ \citenamefont {{Wilder}}}]{Phan.2016a}%
  \BibitemOpen
  \bibfield  {author} {\bibinfo {author} {\bibfnamefont {T.~D.}\ \bibnamefont
  {{Phan}}}, \bibinfo {author} {\bibfnamefont {J.~P.}\ \bibnamefont
  {{Eastwood}}}, \bibinfo {author} {\bibfnamefont {P.~A.}\ \bibnamefont
  {{Cassak}}}, \bibinfo {author} {\bibfnamefont {M.}~\bibnamefont
  {{{\O}ieroset}}}, \bibinfo {author} {\bibfnamefont {J.~T.}\ \bibnamefont
  {{Gosling}}}, \bibinfo {author} {\bibfnamefont {D.~J.}\ \bibnamefont
  {{Gershman}}}, \bibinfo {author} {\bibfnamefont {F.~S.}\ \bibnamefont
  {{Mozer}}}, \bibinfo {author} {\bibfnamefont {M.~A.}\ \bibnamefont {{Shay}}},
  \bibinfo {author} {\bibfnamefont {M.}~\bibnamefont {{Fujimoto}}}, \bibinfo
  {author} {\bibfnamefont {W.}~\bibnamefont {{Daughton}}}, \bibinfo {author}
  {\bibfnamefont {J.~F.}\ \bibnamefont {{Drake}}}, \bibinfo {author}
  {\bibfnamefont {J.~L.}\ \bibnamefont {{Burch}}}, \bibinfo {author}
  {\bibfnamefont {R.~B.}\ \bibnamefont {{Torbert}}}, \bibinfo {author}
  {\bibfnamefont {R.~E.}\ \bibnamefont {{Ergun}}}, \bibinfo {author}
  {\bibfnamefont {L.~J.}\ \bibnamefont {{Chen}}}, \bibinfo {author}
  {\bibfnamefont {S.}~\bibnamefont {{Wang}}}, \bibinfo {author} {\bibfnamefont
  {C.}~\bibnamefont {{Pollock}}}, \bibinfo {author} {\bibfnamefont {J.~C.}\
  \bibnamefont {{Dorelli}}}, \bibinfo {author} {\bibfnamefont {B.}~\bibnamefont
  {{Lavraud}}}, \bibinfo {author} {\bibfnamefont {B.~L.}\ \bibnamefont
  {{Giles}}}, \bibinfo {author} {\bibfnamefont {T.~E.}\ \bibnamefont
  {{Moore}}}, \bibinfo {author} {\bibfnamefont {Y.}~\bibnamefont {{Saito}}},
  \bibinfo {author} {\bibfnamefont {L.~A.}\ \bibnamefont {{Avanov}}}, \bibinfo
  {author} {\bibfnamefont {W.}~\bibnamefont {{Paterson}}}, \bibinfo {author}
  {\bibfnamefont {R.~J.}\ \bibnamefont {{Strangeway}}}, \bibinfo {author}
  {\bibfnamefont {C.~T.}\ \bibnamefont {{Russell}}}, \bibinfo {author}
  {\bibfnamefont {Y.}~\bibnamefont {{Khotyaintsev}}}, \bibinfo {author}
  {\bibfnamefont {P.~A.}\ \bibnamefont {{Lindqvist}}}, \bibinfo {author}
  {\bibfnamefont {M.}~\bibnamefont {{Oka}}}, \ and\ \bibinfo {author}
  {\bibfnamefont {F.~D.}\ \bibnamefont {{Wilder}}},\ }\bibfield  {title}
  {\enquote {\bibinfo {title} {{MMS observations of electron-scale filamentary
  currents in the reconnection exhaust and near the X line}},}\ }\href
  {\doibase 10.1002/2016GL069212} {\bibfield  {journal} {\bibinfo  {journal}
  {Geophys. Res. Lett.}\ }\textbf {\bibinfo {volume} {43}},\ \bibinfo {pages}
  {6060--6069} (\bibinfo {year} {2016})}\BibitemShut {NoStop}%
\bibitem [{\citenamefont {{Phan}}\ \emph {et~al.}(2007)\citenamefont {{Phan}},
  \citenamefont {{Drake}}, \citenamefont {{Shay}}, \citenamefont {{Mozer}},\
  and\ \citenamefont {{Eastwood}}}]{Phan.2007}%
  \BibitemOpen
  \bibfield  {author} {\bibinfo {author} {\bibfnamefont {T.~D.}\ \bibnamefont
  {{Phan}}}, \bibinfo {author} {\bibfnamefont {J.~F.}\ \bibnamefont {{Drake}}},
  \bibinfo {author} {\bibfnamefont {M.~A.}\ \bibnamefont {{Shay}}}, \bibinfo
  {author} {\bibfnamefont {F.~S.}\ \bibnamefont {{Mozer}}}, \ and\ \bibinfo
  {author} {\bibfnamefont {J.~P.}\ \bibnamefont {{Eastwood}}},\ }\bibfield
  {title} {\enquote {\bibinfo {title} {{Evidence for an Elongated (>60 Ion Skin
  Depths) Electron Diffusion Region during Fast Magnetic Reconnection}},}\
  }\href {\doibase 10.1103/PhysRevLett.99.255002} {\bibfield  {journal}
  {\bibinfo  {journal} {Physical Review Letters}\ }\textbf {\bibinfo {volume}
  {99}},\ \bibinfo {eid} {255002} (\bibinfo {year} {2007})}\BibitemShut
  {NoStop}%
\bibitem [{\citenamefont {Chen}\ \emph {et~al.}(2008)\citenamefont {Chen},
  \citenamefont {Bessho}, \citenamefont {Lefebvre}, \citenamefont {Vaith},
  \citenamefont {Fazakerley}, \citenamefont {Bhattacharjee}, \citenamefont
  {Puhl-Quinn}, \citenamefont {Runov}, \citenamefont {Khotyaintsev},
  \citenamefont {Vaivads}, \citenamefont {Georgescu},\ and\ \citenamefont
  {Torbert}}]{Chen.2008}%
  \BibitemOpen
  \bibfield  {author} {\bibinfo {author} {\bibfnamefont {L.-J.}\ \bibnamefont
  {Chen}}, \bibinfo {author} {\bibfnamefont {N.}~\bibnamefont {Bessho}},
  \bibinfo {author} {\bibfnamefont {B.}~\bibnamefont {Lefebvre}}, \bibinfo
  {author} {\bibfnamefont {H.}~\bibnamefont {Vaith}}, \bibinfo {author}
  {\bibfnamefont {A.}~\bibnamefont {Fazakerley}}, \bibinfo {author}
  {\bibfnamefont {A.}~\bibnamefont {Bhattacharjee}}, \bibinfo {author}
  {\bibfnamefont {P.~A.}\ \bibnamefont {Puhl-Quinn}}, \bibinfo {author}
  {\bibfnamefont {A.}~\bibnamefont {Runov}}, \bibinfo {author} {\bibfnamefont
  {Y.}~\bibnamefont {Khotyaintsev}}, \bibinfo {author} {\bibfnamefont
  {A.}~\bibnamefont {Vaivads}}, \bibinfo {author} {\bibfnamefont
  {E.}~\bibnamefont {Georgescu}}, \ and\ \bibinfo {author} {\bibfnamefont
  {R.}~\bibnamefont {Torbert}},\ }\bibfield  {title} {\enquote {\bibinfo
  {title} {Evidence of an extended electron current sheet and its neighboring
  magnetic island during magnetotail reconnection},}\ }\href {\doibase
  10.1029/2008JA013385} {\bibfield  {journal} {\bibinfo  {journal} {Journal of
  Geophysical Research: Space Physics}\ }\textbf {\bibinfo {volume} {113}}
  (\bibinfo {year} {2008}),\ 10.1029/2008JA013385},\ \Eprint
  {http://arxiv.org/abs/https://agupubs.onlinelibrary.wiley.com/doi/pdf/10.1029/2008JA013385}
  {https://agupubs.onlinelibrary.wiley.com/doi/pdf/10.1029/2008JA013385}
  \BibitemShut {NoStop}%
\bibitem [{\citenamefont {{Zenitani}}\ \emph {et~al.}(2011)\citenamefont
  {{Zenitani}}, \citenamefont {{Hesse}}, \citenamefont {{Klimas}},\ and\
  \citenamefont {{Kuznetsova}}}]{Zenitani.2011}%
  \BibitemOpen
  \bibfield  {author} {\bibinfo {author} {\bibfnamefont {S.}~\bibnamefont
  {{Zenitani}}}, \bibinfo {author} {\bibfnamefont {M.}~\bibnamefont {{Hesse}}},
  \bibinfo {author} {\bibfnamefont {A.}~\bibnamefont {{Klimas}}}, \ and\
  \bibinfo {author} {\bibfnamefont {M.}~\bibnamefont {{Kuznetsova}}},\
  }\bibfield  {title} {\enquote {\bibinfo {title} {{New Measure of the
  Dissipation Region in Collisionless Magnetic Reconnection}},}\ }\href
  {\doibase 10.1103/PhysRevLett.106.195003} {\bibfield  {journal} {\bibinfo
  {journal} {Physical Review Letters}\ }\textbf {\bibinfo {volume} {106}},\
  \bibinfo {eid} {195003} (\bibinfo {year} {2011})},\ \Eprint
  {http://arxiv.org/abs/1104.3846} {arXiv:1104.3846 [astro-ph.SR]} \BibitemShut
  {NoStop}%
\bibitem [{\citenamefont {{Chen}}\ \emph {et~al.}(2016)\citenamefont {{Chen}},
  \citenamefont {{Hesse}}, \citenamefont {{Wang}}, \citenamefont {{Gershman}},
  \citenamefont {{Ergun}}, \citenamefont {{Pollock}}, \citenamefont
  {{Torbert}}, \citenamefont {{Bessho}}, \citenamefont {{Daughton}},
  \citenamefont {{Dorelli}}, \citenamefont {{Giles}}, \citenamefont
  {{Strangeway}}, \citenamefont {{Russell}}, \citenamefont {{Khotyaintsev}},
  \citenamefont {{Burch}}, \citenamefont {{Moore}}, \citenamefont {{Lavraud}},
  \citenamefont {{Phan}},\ and\ \citenamefont {{Avanov}}}]{Chen.2016}%
  \BibitemOpen
  \bibfield  {author} {\bibinfo {author} {\bibfnamefont {L.-J.}\ \bibnamefont
  {{Chen}}}, \bibinfo {author} {\bibfnamefont {M.}~\bibnamefont {{Hesse}}},
  \bibinfo {author} {\bibfnamefont {S.}~\bibnamefont {{Wang}}}, \bibinfo
  {author} {\bibfnamefont {D.}~\bibnamefont {{Gershman}}}, \bibinfo {author}
  {\bibfnamefont {R.}~\bibnamefont {{Ergun}}}, \bibinfo {author} {\bibfnamefont
  {C.}~\bibnamefont {{Pollock}}}, \bibinfo {author} {\bibfnamefont
  {R.}~\bibnamefont {{Torbert}}}, \bibinfo {author} {\bibfnamefont
  {N.}~\bibnamefont {{Bessho}}}, \bibinfo {author} {\bibfnamefont
  {W.}~\bibnamefont {{Daughton}}}, \bibinfo {author} {\bibfnamefont
  {J.}~\bibnamefont {{Dorelli}}}, \bibinfo {author} {\bibfnamefont
  {B.}~\bibnamefont {{Giles}}}, \bibinfo {author} {\bibfnamefont
  {R.}~\bibnamefont {{Strangeway}}}, \bibinfo {author} {\bibfnamefont
  {C.}~\bibnamefont {{Russell}}}, \bibinfo {author} {\bibfnamefont
  {Y.}~\bibnamefont {{Khotyaintsev}}}, \bibinfo {author} {\bibfnamefont
  {J.}~\bibnamefont {{Burch}}}, \bibinfo {author} {\bibfnamefont
  {T.}~\bibnamefont {{Moore}}}, \bibinfo {author} {\bibfnamefont
  {B.}~\bibnamefont {{Lavraud}}}, \bibinfo {author} {\bibfnamefont
  {T.}~\bibnamefont {{Phan}}}, \ and\ \bibinfo {author} {\bibfnamefont
  {L.}~\bibnamefont {{Avanov}}},\ }\bibfield  {title} {\enquote {\bibinfo
  {title} {{Electron energization and mixing observed by MMS in the vicinity of
  an electron diffusion region during magnetopause reconnection}},}\ }\href
  {\doibase 10.1002/2016GL069215} {\bibfield  {journal} {\bibinfo  {journal}
  {Geophys. Res. Lett.}\ }\textbf {\bibinfo {volume} {43}},\ \bibinfo {pages}
  {6036--6043} (\bibinfo {year} {2016})}\BibitemShut {NoStop}%
\bibitem [{\citenamefont {{Chen}}\ \emph {et~al.}(2017)\citenamefont {{Chen}},
  \citenamefont {{Hesse}}, \citenamefont {{Wang}}, \citenamefont {{Gershman}},
  \citenamefont {{Ergun}}, \citenamefont {{Burch}}, \citenamefont {{Bessho}},
  \citenamefont {{Torbert}}, \citenamefont {{Giles}}, \citenamefont
  {{Webster}}, \citenamefont {{Pollock}}, \citenamefont {{Dorelli}},
  \citenamefont {{Moore}}, \citenamefont {{Paterson}}, \citenamefont
  {{Lavraud}}, \citenamefont {{Strangeway}}, \citenamefont {{Russell}},
  \citenamefont {{Khotyaintsev}}, \citenamefont {{Lindqvist}},\ and\
  \citenamefont {{Avanov}}}]{Chen.2017}%
  \BibitemOpen
  \bibfield  {author} {\bibinfo {author} {\bibfnamefont {L.~J.}\ \bibnamefont
  {{Chen}}}, \bibinfo {author} {\bibfnamefont {M.}~\bibnamefont {{Hesse}}},
  \bibinfo {author} {\bibfnamefont {S.}~\bibnamefont {{Wang}}}, \bibinfo
  {author} {\bibfnamefont {D.}~\bibnamefont {{Gershman}}}, \bibinfo {author}
  {\bibfnamefont {R.~E.}\ \bibnamefont {{Ergun}}}, \bibinfo {author}
  {\bibfnamefont {J.}~\bibnamefont {{Burch}}}, \bibinfo {author} {\bibfnamefont
  {N.}~\bibnamefont {{Bessho}}}, \bibinfo {author} {\bibfnamefont {R.~B.}\
  \bibnamefont {{Torbert}}}, \bibinfo {author} {\bibfnamefont {B.}~\bibnamefont
  {{Giles}}}, \bibinfo {author} {\bibfnamefont {J.}~\bibnamefont {{Webster}}},
  \bibinfo {author} {\bibfnamefont {C.}~\bibnamefont {{Pollock}}}, \bibinfo
  {author} {\bibfnamefont {J.}~\bibnamefont {{Dorelli}}}, \bibinfo {author}
  {\bibfnamefont {T.}~\bibnamefont {{Moore}}}, \bibinfo {author} {\bibfnamefont
  {W.}~\bibnamefont {{Paterson}}}, \bibinfo {author} {\bibfnamefont
  {B.}~\bibnamefont {{Lavraud}}}, \bibinfo {author} {\bibfnamefont
  {R.}~\bibnamefont {{Strangeway}}}, \bibinfo {author} {\bibfnamefont
  {C.}~\bibnamefont {{Russell}}}, \bibinfo {author} {\bibfnamefont
  {Y.}~\bibnamefont {{Khotyaintsev}}}, \bibinfo {author} {\bibfnamefont
  {P.~A.}\ \bibnamefont {{Lindqvist}}}, \ and\ \bibinfo {author} {\bibfnamefont
  {L.}~\bibnamefont {{Avanov}}},\ }\bibfield  {title} {\enquote {\bibinfo
  {title} {{Electron diffusion region during magnetopause reconnection with an
  intermediate guide field: Magnetospheric multiscale observations}},}\ }\href
  {\doibase 10.1002/2017JA024004} {\bibfield  {journal} {\bibinfo  {journal}
  {Journal of Geophysical Research (Space Physics)}\ }\textbf {\bibinfo
  {volume} {122}},\ \bibinfo {pages} {5235--5246} (\bibinfo {year}
  {2017})}\BibitemShut {NoStop}%
\bibitem [{\citenamefont {{Lavraud}}\ \emph {et~al.}(2016)\citenamefont
  {{Lavraud}}, \citenamefont {{Zhang}}, \citenamefont {{Vernisse}},
  \citenamefont {{Gershman}}, \citenamefont {{Dorelli}}, \citenamefont
  {{Cassak}}, \citenamefont {{Dargent}}, \citenamefont {{Pollock}},
  \citenamefont {{Giles}}, \citenamefont {{Aunai}}, \citenamefont {{Argall}},
  \citenamefont {{Avanov}}, \citenamefont {{Barrie}}, \citenamefont {{Burch}},
  \citenamefont {{Chandler}}, \citenamefont {{Chen}}, \citenamefont {{Clark}},
  \citenamefont {{Cohen}}, \citenamefont {{Coffey}}, \citenamefont
  {{Eastwood}}, \citenamefont {{Egedal}}, \citenamefont {{Eriksson}},
  \citenamefont {{Ergun}}, \citenamefont {{Farrugia}}, \citenamefont
  {{Fuselier}}, \citenamefont {{G{\'e}not}}, \citenamefont {{Graham}},
  \citenamefont {{Grigorenko}}, \citenamefont {{Hasegawa}}, \citenamefont
  {{Jacquey}}, \citenamefont {{Kacem}}, \citenamefont {{Khotyaintsev}},
  \citenamefont {{MacDonald}}, \citenamefont {{Magnes}}, \citenamefont
  {{Marchaudon}}, \citenamefont {{Mauk}}, \citenamefont {{Moore}},
  \citenamefont {{Mukai}}, \citenamefont {{Nakamura}}, \citenamefont
  {{Paterson}}, \citenamefont {{Penou}}, \citenamefont {{Phan}}, \citenamefont
  {{Rager}}, \citenamefont {{Retino}}, \citenamefont {{Rong}}, \citenamefont
  {{Russell}}, \citenamefont {{Saito}}, \citenamefont {{Sauvaud}},
  \citenamefont {{Schwartz}}, \citenamefont {{Shen}}, \citenamefont {{Smith}},
  \citenamefont {{Strangeway}}, \citenamefont {{Toledo-Redondo}}, \citenamefont
  {{Torbert}}, \citenamefont {{Turner}}, \citenamefont {{Wang}},\ and\
  \citenamefont {{Yokota}}}]{Lavraud.2016}%
  \BibitemOpen
  \bibfield  {author} {\bibinfo {author} {\bibfnamefont {B.}~\bibnamefont
  {{Lavraud}}}, \bibinfo {author} {\bibfnamefont {Y.~C.}\ \bibnamefont
  {{Zhang}}}, \bibinfo {author} {\bibfnamefont {Y.}~\bibnamefont {{Vernisse}}},
  \bibinfo {author} {\bibfnamefont {D.~J.}\ \bibnamefont {{Gershman}}},
  \bibinfo {author} {\bibfnamefont {J.}~\bibnamefont {{Dorelli}}}, \bibinfo
  {author} {\bibfnamefont {P.~A.}\ \bibnamefont {{Cassak}}}, \bibinfo {author}
  {\bibfnamefont {J.}~\bibnamefont {{Dargent}}}, \bibinfo {author}
  {\bibfnamefont {C.}~\bibnamefont {{Pollock}}}, \bibinfo {author}
  {\bibfnamefont {B.}~\bibnamefont {{Giles}}}, \bibinfo {author} {\bibfnamefont
  {N.}~\bibnamefont {{Aunai}}}, \bibinfo {author} {\bibfnamefont
  {M.}~\bibnamefont {{Argall}}}, \bibinfo {author} {\bibfnamefont
  {L.}~\bibnamefont {{Avanov}}}, \bibinfo {author} {\bibfnamefont
  {A.}~\bibnamefont {{Barrie}}}, \bibinfo {author} {\bibfnamefont
  {J.}~\bibnamefont {{Burch}}}, \bibinfo {author} {\bibfnamefont
  {M.}~\bibnamefont {{Chandler}}}, \bibinfo {author} {\bibfnamefont {L.-J.}\
  \bibnamefont {{Chen}}}, \bibinfo {author} {\bibfnamefont {G.}~\bibnamefont
  {{Clark}}}, \bibinfo {author} {\bibfnamefont {I.}~\bibnamefont {{Cohen}}},
  \bibinfo {author} {\bibfnamefont {V.}~\bibnamefont {{Coffey}}}, \bibinfo
  {author} {\bibfnamefont {J.~P.}\ \bibnamefont {{Eastwood}}}, \bibinfo
  {author} {\bibfnamefont {J.}~\bibnamefont {{Egedal}}}, \bibinfo {author}
  {\bibfnamefont {S.}~\bibnamefont {{Eriksson}}}, \bibinfo {author}
  {\bibfnamefont {R.}~\bibnamefont {{Ergun}}}, \bibinfo {author} {\bibfnamefont
  {C.~J.}\ \bibnamefont {{Farrugia}}}, \bibinfo {author} {\bibfnamefont
  {S.~A.}\ \bibnamefont {{Fuselier}}}, \bibinfo {author} {\bibfnamefont
  {V.}~\bibnamefont {{G{\'e}not}}}, \bibinfo {author} {\bibfnamefont
  {D.}~\bibnamefont {{Graham}}}, \bibinfo {author} {\bibfnamefont
  {E.}~\bibnamefont {{Grigorenko}}}, \bibinfo {author} {\bibfnamefont
  {H.}~\bibnamefont {{Hasegawa}}}, \bibinfo {author} {\bibfnamefont
  {C.}~\bibnamefont {{Jacquey}}}, \bibinfo {author} {\bibfnamefont
  {I.}~\bibnamefont {{Kacem}}}, \bibinfo {author} {\bibfnamefont
  {Y.}~\bibnamefont {{Khotyaintsev}}}, \bibinfo {author} {\bibfnamefont
  {E.}~\bibnamefont {{MacDonald}}}, \bibinfo {author} {\bibfnamefont
  {W.}~\bibnamefont {{Magnes}}}, \bibinfo {author} {\bibfnamefont
  {A.}~\bibnamefont {{Marchaudon}}}, \bibinfo {author} {\bibfnamefont
  {B.}~\bibnamefont {{Mauk}}}, \bibinfo {author} {\bibfnamefont {T.~E.}\
  \bibnamefont {{Moore}}}, \bibinfo {author} {\bibfnamefont {T.}~\bibnamefont
  {{Mukai}}}, \bibinfo {author} {\bibfnamefont {R.}~\bibnamefont {{Nakamura}}},
  \bibinfo {author} {\bibfnamefont {W.}~\bibnamefont {{Paterson}}}, \bibinfo
  {author} {\bibfnamefont {E.}~\bibnamefont {{Penou}}}, \bibinfo {author}
  {\bibfnamefont {T.~D.}\ \bibnamefont {{Phan}}}, \bibinfo {author}
  {\bibfnamefont {A.}~\bibnamefont {{Rager}}}, \bibinfo {author} {\bibfnamefont
  {A.}~\bibnamefont {{Retino}}}, \bibinfo {author} {\bibfnamefont {Z.~J.}\
  \bibnamefont {{Rong}}}, \bibinfo {author} {\bibfnamefont {C.~T.}\
  \bibnamefont {{Russell}}}, \bibinfo {author} {\bibfnamefont {Y.}~\bibnamefont
  {{Saito}}}, \bibinfo {author} {\bibfnamefont {J.-A.}\ \bibnamefont
  {{Sauvaud}}}, \bibinfo {author} {\bibfnamefont {S.~J.}\ \bibnamefont
  {{Schwartz}}}, \bibinfo {author} {\bibfnamefont {C.}~\bibnamefont {{Shen}}},
  \bibinfo {author} {\bibfnamefont {S.}~\bibnamefont {{Smith}}}, \bibinfo
  {author} {\bibfnamefont {R.}~\bibnamefont {{Strangeway}}}, \bibinfo {author}
  {\bibfnamefont {S.}~\bibnamefont {{Toledo-Redondo}}}, \bibinfo {author}
  {\bibfnamefont {R.}~\bibnamefont {{Torbert}}}, \bibinfo {author}
  {\bibfnamefont {D.~L.}\ \bibnamefont {{Turner}}}, \bibinfo {author}
  {\bibfnamefont {S.}~\bibnamefont {{Wang}}}, \ and\ \bibinfo {author}
  {\bibfnamefont {S.}~\bibnamefont {{Yokota}}},\ }\bibfield  {title} {\enquote
  {\bibinfo {title} {{Currents and associated electron scattering and bouncing
  near the diffusion region at Earth's magnetopause}},}\ }\href {\doibase
  10.1002/2016GL068359} {\bibfield  {journal} {\bibinfo  {journal} {Geophys.
  Res. Lett.}\ }\textbf {\bibinfo {volume} {43}},\ \bibinfo {pages}
  {3042--3050} (\bibinfo {year} {2016})}\BibitemShut {NoStop}%
\bibitem [{\citenamefont {Webster}\ \emph {et~al.}(2018)\citenamefont
  {Webster}, \citenamefont {Burch}, \citenamefont {Reiff}, \citenamefont
  {Daou}, \citenamefont {Genestreti}, \citenamefont {Graham}, \citenamefont
  {Torbert}, \citenamefont {Ergun}, \citenamefont {Sazykin}, \citenamefont
  {Marshall}, \citenamefont {Allen}, \citenamefont {Chen}, \citenamefont
  {Wang}, \citenamefont {Phan}, \citenamefont {Giles}, \citenamefont {Moore},
  \citenamefont {Fuselier}, \citenamefont {Cozzani}, \citenamefont {Russell},
  \citenamefont {Eriksson}, \citenamefont {Rager}, \citenamefont {Broll},
  \citenamefont {Goodrich},\ and\ \citenamefont {Wilder}}]{Webster.2018}%
  \BibitemOpen
  \bibfield  {author} {\bibinfo {author} {\bibfnamefont {J.~M.}\ \bibnamefont
  {Webster}}, \bibinfo {author} {\bibfnamefont {J.~L.}\ \bibnamefont {Burch}},
  \bibinfo {author} {\bibfnamefont {P.~H.}\ \bibnamefont {Reiff}}, \bibinfo
  {author} {\bibfnamefont {A.~G.}\ \bibnamefont {Daou}}, \bibinfo {author}
  {\bibfnamefont {K.~J.}\ \bibnamefont {Genestreti}}, \bibinfo {author}
  {\bibfnamefont {D.~B.}\ \bibnamefont {Graham}}, \bibinfo {author}
  {\bibfnamefont {R.~B.}\ \bibnamefont {Torbert}}, \bibinfo {author}
  {\bibfnamefont {R.~E.}\ \bibnamefont {Ergun}}, \bibinfo {author}
  {\bibfnamefont {S.~Y.}\ \bibnamefont {Sazykin}}, \bibinfo {author}
  {\bibfnamefont {A.}~\bibnamefont {Marshall}}, \bibinfo {author}
  {\bibfnamefont {R.~C.}\ \bibnamefont {Allen}}, \bibinfo {author}
  {\bibfnamefont {L.-J.}\ \bibnamefont {Chen}}, \bibinfo {author}
  {\bibfnamefont {S.}~\bibnamefont {Wang}}, \bibinfo {author} {\bibfnamefont
  {T.~D.}\ \bibnamefont {Phan}}, \bibinfo {author} {\bibfnamefont {B.~L.}\
  \bibnamefont {Giles}}, \bibinfo {author} {\bibfnamefont {T.~E.}\ \bibnamefont
  {Moore}}, \bibinfo {author} {\bibfnamefont {S.~A.}\ \bibnamefont {Fuselier}},
  \bibinfo {author} {\bibfnamefont {G.}~\bibnamefont {Cozzani}}, \bibinfo
  {author} {\bibfnamefont {C.~T.}\ \bibnamefont {Russell}}, \bibinfo {author}
  {\bibfnamefont {S.}~\bibnamefont {Eriksson}}, \bibinfo {author}
  {\bibfnamefont {A.~C.}\ \bibnamefont {Rager}}, \bibinfo {author}
  {\bibfnamefont {J.~M.}\ \bibnamefont {Broll}}, \bibinfo {author}
  {\bibfnamefont {K.}~\bibnamefont {Goodrich}}, \ and\ \bibinfo {author}
  {\bibfnamefont {F.}~\bibnamefont {Wilder}},\ }\bibfield  {title} {\enquote
  {\bibinfo {title} {Magnetospheric multiscale dayside reconnection electron
  diffusion region events},}\ }\href {\doibase 10.1029/2018JA025245} {\bibfield
   {journal} {\bibinfo  {journal} {Journal of Geophysical Research: Space
  Physics}\ }\textbf {\bibinfo {volume} {123}},\ \bibinfo {pages} {4858--4878}
  (\bibinfo {year} {2018})},\ \Eprint
  {http://arxiv.org/abs/https://agupubs.onlinelibrary.wiley.com/doi/pdf/10.1029/2018JA025245}
  {https://agupubs.onlinelibrary.wiley.com/doi/pdf/10.1029/2018JA025245}
  \BibitemShut {NoStop}%
\bibitem [{\citenamefont {{Li}}\ \emph {et~al.}(2020)\citenamefont {{Li}},
  \citenamefont {{Graham}}, \citenamefont {{Khotyaintsev}}, \citenamefont
  {{Vaivads}}, \citenamefont {{Andr{\'e}}}, \citenamefont {{Min}},
  \citenamefont {{Liu}}, \citenamefont {{Tang}}, \citenamefont {{Wang}},
  \citenamefont {{Fujimoto}}, \citenamefont {{Norgren}}, \citenamefont
  {{Toledo-Redondo}}, \citenamefont {{Lindqvist}}, \citenamefont {{Ergun}},
  \citenamefont {{Torbert}}, \citenamefont {{Rager}}, \citenamefont
  {{Dorelli}}, \citenamefont {{Gershman}}, \citenamefont {{Giles}},
  \citenamefont {{Lavraud}}, \citenamefont {{Plaschke}}, \citenamefont
  {{Magnes}}, \citenamefont {{Le Contel}}, \citenamefont {{Russell}},\ and\
  \citenamefont {{Burch}}}]{Li.2020}%
  \BibitemOpen
  \bibfield  {author} {\bibinfo {author} {\bibfnamefont {W.~Y.}\ \bibnamefont
  {{Li}}}, \bibinfo {author} {\bibfnamefont {D.~B.}\ \bibnamefont {{Graham}}},
  \bibinfo {author} {\bibfnamefont {Y.~V.}\ \bibnamefont {{Khotyaintsev}}},
  \bibinfo {author} {\bibfnamefont {A.}~\bibnamefont {{Vaivads}}}, \bibinfo
  {author} {\bibfnamefont {M.}~\bibnamefont {{Andr{\'e}}}}, \bibinfo {author}
  {\bibfnamefont {K.}~\bibnamefont {{Min}}}, \bibinfo {author} {\bibfnamefont
  {K.}~\bibnamefont {{Liu}}}, \bibinfo {author} {\bibfnamefont {B.~B.}\
  \bibnamefont {{Tang}}}, \bibinfo {author} {\bibfnamefont {C.}~\bibnamefont
  {{Wang}}}, \bibinfo {author} {\bibfnamefont {K.}~\bibnamefont {{Fujimoto}}},
  \bibinfo {author} {\bibfnamefont {C.}~\bibnamefont {{Norgren}}}, \bibinfo
  {author} {\bibfnamefont {S.}~\bibnamefont {{Toledo-Redondo}}}, \bibinfo
  {author} {\bibfnamefont {P.~A.}\ \bibnamefont {{Lindqvist}}}, \bibinfo
  {author} {\bibfnamefont {R.~E.}\ \bibnamefont {{Ergun}}}, \bibinfo {author}
  {\bibfnamefont {R.~B.}\ \bibnamefont {{Torbert}}}, \bibinfo {author}
  {\bibfnamefont {A.~C.}\ \bibnamefont {{Rager}}}, \bibinfo {author}
  {\bibfnamefont {J.~C.}\ \bibnamefont {{Dorelli}}}, \bibinfo {author}
  {\bibfnamefont {D.~J.}\ \bibnamefont {{Gershman}}}, \bibinfo {author}
  {\bibfnamefont {B.~L.}\ \bibnamefont {{Giles}}}, \bibinfo {author}
  {\bibfnamefont {B.}~\bibnamefont {{Lavraud}}}, \bibinfo {author}
  {\bibfnamefont {F.}~\bibnamefont {{Plaschke}}}, \bibinfo {author}
  {\bibfnamefont {W.}~\bibnamefont {{Magnes}}}, \bibinfo {author}
  {\bibfnamefont {O.}~\bibnamefont {{Le Contel}}}, \bibinfo {author}
  {\bibfnamefont {C.~T.}\ \bibnamefont {{Russell}}}, \ and\ \bibinfo {author}
  {\bibfnamefont {J.~L.}\ \bibnamefont {{Burch}}},\ }\bibfield  {title}
  {\enquote {\bibinfo {title} {{Electron Bernstein waves driven by electron
  crescents near the electron diffusion region}},}\ }\href {\doibase
  10.1038/s41467-019-13920-w} {\bibfield  {journal} {\bibinfo  {journal}
  {Nature Communications}\ }\textbf {\bibinfo {volume} {11}},\ \bibinfo {eid}
  {141} (\bibinfo {year} {2020})}\BibitemShut {NoStop}%
\bibitem [{\citenamefont {{Zhou}}\ \emph {et~al.}(2019)\citenamefont {{Zhou}},
  \citenamefont {{Deng}}, \citenamefont {{Zhong}}, \citenamefont {{Pang}},
  \citenamefont {{Tang}}, \citenamefont {{El-Alaoui}}, \citenamefont
  {{Walker}}, \citenamefont {{Russell}}, \citenamefont {{Lapenta}},
  \citenamefont {{Strangeway}}, \citenamefont {{Torbert}}, \citenamefont
  {{Burch}}, \citenamefont {{Paterson}}, \citenamefont {{Giles}}, \citenamefont
  {{Khotyaintsev}}, \citenamefont {{Ergun}},\ and\ \citenamefont
  {{Lindqvist}}}]{Zhou.2019}%
  \BibitemOpen
  \bibfield  {author} {\bibinfo {author} {\bibfnamefont {M.}~\bibnamefont
  {{Zhou}}}, \bibinfo {author} {\bibfnamefont {X.~H.}\ \bibnamefont {{Deng}}},
  \bibinfo {author} {\bibfnamefont {Z.~H.}\ \bibnamefont {{Zhong}}}, \bibinfo
  {author} {\bibfnamefont {Y.}~\bibnamefont {{Pang}}}, \bibinfo {author}
  {\bibfnamefont {R.~X.}\ \bibnamefont {{Tang}}}, \bibinfo {author}
  {\bibfnamefont {M.}~\bibnamefont {{El-Alaoui}}}, \bibinfo {author}
  {\bibfnamefont {R.~J.}\ \bibnamefont {{Walker}}}, \bibinfo {author}
  {\bibfnamefont {C.~T.}\ \bibnamefont {{Russell}}}, \bibinfo {author}
  {\bibfnamefont {G.}~\bibnamefont {{Lapenta}}}, \bibinfo {author}
  {\bibfnamefont {R.~J.}\ \bibnamefont {{Strangeway}}}, \bibinfo {author}
  {\bibfnamefont {R.~B.}\ \bibnamefont {{Torbert}}}, \bibinfo {author}
  {\bibfnamefont {J.~L.}\ \bibnamefont {{Burch}}}, \bibinfo {author}
  {\bibfnamefont {W.~R.}\ \bibnamefont {{Paterson}}}, \bibinfo {author}
  {\bibfnamefont {B.~L.}\ \bibnamefont {{Giles}}}, \bibinfo {author}
  {\bibfnamefont {Y.~V.}\ \bibnamefont {{Khotyaintsev}}}, \bibinfo {author}
  {\bibfnamefont {R.~E.}\ \bibnamefont {{Ergun}}}, \ and\ \bibinfo {author}
  {\bibfnamefont {P.~A.}\ \bibnamefont {{Lindqvist}}},\ }\bibfield  {title}
  {\enquote {\bibinfo {title} {{Observations of an Electron Diffusion Region in
  Symmetric Reconnection with Weak Guide Field}},}\ }\href {\doibase
  10.3847/1538-4357/aaf16f} {\bibfield  {journal} {\bibinfo  {journal} {Ap.
  J.}\ }\textbf {\bibinfo {volume} {870}},\ \bibinfo {eid} {34} (\bibinfo
  {year} {2019})}\BibitemShut {NoStop}%
\bibitem [{\citenamefont {{Denton}}\ \emph {et~al.}(2018)\citenamefont
  {{Denton}}, \citenamefont {{Sonnerup}}, \citenamefont {{Russell}},
  \citenamefont {{Hasegawa}}, \citenamefont {{Phan}}, \citenamefont
  {{Strangeway}}, \citenamefont {{Giles}}, \citenamefont {{Ergun}},
  \citenamefont {{Lindqvist}}, \citenamefont {{Torbert}}, \citenamefont
  {{Burch}},\ and\ \citenamefont {{Vines}}}]{Denton.2018}%
  \BibitemOpen
  \bibfield  {author} {\bibinfo {author} {\bibfnamefont {R.~E.}\ \bibnamefont
  {{Denton}}}, \bibinfo {author} {\bibfnamefont {B.~U.~{\"O}.}\ \bibnamefont
  {{Sonnerup}}}, \bibinfo {author} {\bibfnamefont {C.~T.}\ \bibnamefont
  {{Russell}}}, \bibinfo {author} {\bibfnamefont {H.}~\bibnamefont
  {{Hasegawa}}}, \bibinfo {author} {\bibfnamefont {T.~D.}\ \bibnamefont
  {{Phan}}}, \bibinfo {author} {\bibfnamefont {R.~J.}\ \bibnamefont
  {{Strangeway}}}, \bibinfo {author} {\bibfnamefont {B.~L.}\ \bibnamefont
  {{Giles}}}, \bibinfo {author} {\bibfnamefont {R.~E.}\ \bibnamefont
  {{Ergun}}}, \bibinfo {author} {\bibfnamefont {P.~A.}\ \bibnamefont
  {{Lindqvist}}}, \bibinfo {author} {\bibfnamefont {R.~B.}\ \bibnamefont
  {{Torbert}}}, \bibinfo {author} {\bibfnamefont {J.~L.}\ \bibnamefont
  {{Burch}}}, \ and\ \bibinfo {author} {\bibfnamefont {S.~K.}\ \bibnamefont
  {{Vines}}},\ }\bibfield  {title} {\enquote {\bibinfo {title} {{Determining
  L-M-N Current Sheet Coordinates at the Magnetopause From Magnetospheric
  Multiscale Data}},}\ }\href {\doibase 10.1002/2017JA024619} {\bibfield
  {journal} {\bibinfo  {journal} {Journal of Geophysical Research (Space
  Physics)}\ }\textbf {\bibinfo {volume} {123}},\ \bibinfo {pages} {2274--2295}
  (\bibinfo {year} {2018})}\BibitemShut {NoStop}%
\bibitem [{\citenamefont {{Shi}}\ \emph {et~al.}(2005)\citenamefont {{Shi}},
  \citenamefont {{Shen}}, \citenamefont {{Pu}}, \citenamefont {{Dunlop}},
  \citenamefont {{Zong}}, \citenamefont {{Zhang}}, \citenamefont {{Xiao}},
  \citenamefont {{Liu}},\ and\ \citenamefont {{Balogh}}}]{Shi.2005}%
  \BibitemOpen
  \bibfield  {author} {\bibinfo {author} {\bibfnamefont {Q.~Q.}\ \bibnamefont
  {{Shi}}}, \bibinfo {author} {\bibfnamefont {C.}~\bibnamefont {{Shen}}},
  \bibinfo {author} {\bibfnamefont {Z.~Y.}\ \bibnamefont {{Pu}}}, \bibinfo
  {author} {\bibfnamefont {M.~W.}\ \bibnamefont {{Dunlop}}}, \bibinfo {author}
  {\bibfnamefont {Q.~G.}\ \bibnamefont {{Zong}}}, \bibinfo {author}
  {\bibfnamefont {H.}~\bibnamefont {{Zhang}}}, \bibinfo {author} {\bibfnamefont
  {C.~J.}\ \bibnamefont {{Xiao}}}, \bibinfo {author} {\bibfnamefont {Z.~X.}\
  \bibnamefont {{Liu}}}, \ and\ \bibinfo {author} {\bibfnamefont
  {A.}~\bibnamefont {{Balogh}}},\ }\bibfield  {title} {\enquote {\bibinfo
  {title} {{Dimensional analysis of observed structures using multipoint
  magnetic field measurements: Application to Cluster}},}\ }\href {\doibase
  10.1029/2005GL022454} {\bibfield  {journal} {\bibinfo  {journal} {Geophys.
  Res. Lett.}\ }\textbf {\bibinfo {volume} {32}},\ \bibinfo {eid} {L12105}
  (\bibinfo {year} {2005})}\BibitemShut {NoStop}%
\bibitem [{\citenamefont {{Paschmann}}\ \emph {et~al.}(1986)\citenamefont
  {{Paschmann}}, \citenamefont {{Papamastorakis}}, \citenamefont
  {{Baumjohann}}, \citenamefont {{Sckopke}}, \citenamefont {{Carlson}},
  \citenamefont {{Sonnerup}},\ and\ \citenamefont
  {{L{\"u}hr}}}]{Paschmann.1986}%
  \BibitemOpen
  \bibfield  {author} {\bibinfo {author} {\bibfnamefont {G.}~\bibnamefont
  {{Paschmann}}}, \bibinfo {author} {\bibfnamefont {I.}~\bibnamefont
  {{Papamastorakis}}}, \bibinfo {author} {\bibfnamefont {W.}~\bibnamefont
  {{Baumjohann}}}, \bibinfo {author} {\bibfnamefont {N.}~\bibnamefont
  {{Sckopke}}}, \bibinfo {author} {\bibfnamefont {C.~W.}\ \bibnamefont
  {{Carlson}}}, \bibinfo {author} {\bibfnamefont {B.~U.~{\"O}.}\ \bibnamefont
  {{Sonnerup}}}, \ and\ \bibinfo {author} {\bibfnamefont {H.}~\bibnamefont
  {{L{\"u}hr}}},\ }\bibfield  {title} {\enquote {\bibinfo {title} {{The
  magnetopause for large magnetic shear: AMPTE/IRM observations}},}\ }\href
  {\doibase 10.1029/JA091iA10p11099} {\bibfield  {journal} {\bibinfo  {journal}
  {J. Geophys. Res.}\ }\textbf {\bibinfo {volume} {91}},\ \bibinfo {pages}
  {11099--11115} (\bibinfo {year} {1986})}\BibitemShut {NoStop}%
\bibitem [{\citenamefont {{Sonnerup}}(1987)}]{Sonnerup.1987}%
  \BibitemOpen
  \bibfield  {author} {\bibinfo {author} {\bibfnamefont {B.~U.~O.}\
  \bibnamefont {{Sonnerup}}},\ }\bibfield  {title} {\enquote {\bibinfo {title}
  {{On the stress balance in flux transfer events}},}\ }\href {\doibase
  10.1029/JA092iA08p08613} {\bibfield  {journal} {\bibinfo  {journal} {J.
  Geophys. Res.}\ }\textbf {\bibinfo {volume} {92}},\ \bibinfo {pages}
  {8613--8620} (\bibinfo {year} {1987})}\BibitemShut {NoStop}%
\bibitem [{\citenamefont {{Sonnerup}}\ and\ \citenamefont
  {{Cahill}}(1967)}]{SonnerupandCahill.1967}%
  \BibitemOpen
  \bibfield  {author} {\bibinfo {author} {\bibfnamefont {B.~U.~O.}\
  \bibnamefont {{Sonnerup}}}\ and\ \bibinfo {author} {\bibfnamefont
  {J.}~\bibnamefont {{Cahill}}, \bibfnamefont {L.~J.}},\ }\bibfield  {title}
  {\enquote {\bibinfo {title} {{Magnetopause Structure and Attitude from
  Explorer 12 Observations}},}\ }\href {\doibase 10.1029/JZ072i001p00171}
  {\bibfield  {journal} {\bibinfo  {journal} {J. Geophys. Res.}\ }\textbf
  {\bibinfo {volume} {72}},\ \bibinfo {pages} {171} (\bibinfo {year}
  {1967})}\BibitemShut {NoStop}%
\bibitem [{\citenamefont {Myers}, \citenamefont {Well},\ and\ \citenamefont
  {Lorch~Jr}(2013)}]{myers2013research}%
  \BibitemOpen
  \bibfield  {author} {\bibinfo {author} {\bibfnamefont {J.~L.}\ \bibnamefont
  {Myers}}, \bibinfo {author} {\bibfnamefont {A.~D.}\ \bibnamefont {Well}}, \
  and\ \bibinfo {author} {\bibfnamefont {R.~F.}\ \bibnamefont {Lorch~Jr}},\
  }\href@noop {} {\emph {\bibinfo {title} {Research design and statistical
  analysis}}}\ (\bibinfo  {publisher} {Routledge},\ \bibinfo {year}
  {2013})\BibitemShut {NoStop}%
\bibitem [{\citenamefont {{Bowers}}\ \emph {et~al.}(2008)\citenamefont
  {{Bowers}}, \citenamefont {{Albright}}, \citenamefont {{Yin}}, \citenamefont
  {{Bergen}},\ and\ \citenamefont {{Kwan}}}]{Bowers.2008}%
  \BibitemOpen
  \bibfield  {author} {\bibinfo {author} {\bibfnamefont {K.~J.}\ \bibnamefont
  {{Bowers}}}, \bibinfo {author} {\bibfnamefont {B.~J.}\ \bibnamefont
  {{Albright}}}, \bibinfo {author} {\bibfnamefont {L.}~\bibnamefont {{Yin}}},
  \bibinfo {author} {\bibfnamefont {B.}~\bibnamefont {{Bergen}}}, \ and\
  \bibinfo {author} {\bibfnamefont {T.~J.~T.}\ \bibnamefont {{Kwan}}},\
  }\bibfield  {title} {\enquote {\bibinfo {title} {{Ultrahigh performance
  three-dimensional electromagnetic relativistic kinetic plasma
  simulations)}},}\ }\href {\doibase 10.1063/1.2840133} {\bibfield  {journal}
  {\bibinfo  {journal} {Physics of Plasmas}\ }\textbf {\bibinfo {volume}
  {15}},\ \bibinfo {eid} {055703} (\bibinfo {year} {2008})}\BibitemShut
  {NoStop}%
\bibitem [{\citenamefont {{Fuselier}}\ \emph {et~al.}(2021)\citenamefont
  {{Fuselier}}, \citenamefont {{Webster}}, \citenamefont {{Trattner}},
  \citenamefont {{Petrinec}}, \citenamefont {{Genestreti}}, \citenamefont
  {{Pritchard}}, \citenamefont {{LLera}}, \citenamefont {{Broll}},
  \citenamefont {{Burch}},\ and\ \citenamefont {{Strangeway}}}]{Fuselier.2021}%
  \BibitemOpen
  \bibfield  {author} {\bibinfo {author} {\bibfnamefont {S.~A.}\ \bibnamefont
  {{Fuselier}}}, \bibinfo {author} {\bibfnamefont {J.~M.}\ \bibnamefont
  {{Webster}}}, \bibinfo {author} {\bibfnamefont {K.~J.}\ \bibnamefont
  {{Trattner}}}, \bibinfo {author} {\bibfnamefont {S.~M.}\ \bibnamefont
  {{Petrinec}}}, \bibinfo {author} {\bibfnamefont {K.~J.}\ \bibnamefont
  {{Genestreti}}}, \bibinfo {author} {\bibfnamefont {K.~R.}\ \bibnamefont
  {{Pritchard}}}, \bibinfo {author} {\bibfnamefont {K.}~\bibnamefont
  {{LLera}}}, \bibinfo {author} {\bibfnamefont {J.~M.}\ \bibnamefont
  {{Broll}}}, \bibinfo {author} {\bibfnamefont {J.~L.}\ \bibnamefont
  {{Burch}}}, \ and\ \bibinfo {author} {\bibfnamefont {R.~J.}\ \bibnamefont
  {{Strangeway}}},\ }\bibfield  {title} {\enquote {\bibinfo {title}
  {{Reconnection X-Line Orientations at the Earth's Magnetopause}},}\ }\href
  {\doibase 10.1029/2021JA029789} {\bibfield  {journal} {\bibinfo  {journal}
  {Journal of Geophysical Research (Space Physics)}\ }\textbf {\bibinfo
  {volume} {126}},\ \bibinfo {eid} {e29789} (\bibinfo {year}
  {2021})}\BibitemShut {NoStop}%
\bibitem [{\citenamefont {{Angelopoulos}}\ \emph {et~al.}(2019)\citenamefont
  {{Angelopoulos}}, \citenamefont {{Cruce}}, \citenamefont {{Drozdov}},
  \citenamefont {{Grimes}}, \citenamefont {{Hatzigeorgiu}}, \citenamefont
  {{King}}, \citenamefont {{Larson}}, \citenamefont {{Lewis}}, \citenamefont
  {{McTiernan}}, \citenamefont {{Roberts}}, \citenamefont {{Russell}},
  \citenamefont {{Hori}}, \citenamefont {{Kasahara}}, \citenamefont
  {{Kumamoto}}, \citenamefont {{Matsuoka}}, \citenamefont {{Miyashita}},
  \citenamefont {{Miyoshi}}, \citenamefont {{Shinohara}}, \citenamefont
  {{Teramoto}}, \citenamefont {{Faden}}, \citenamefont {{Halford}},
  \citenamefont {{McCarthy}}, \citenamefont {{Millan}}, \citenamefont
  {{Sample}}, \citenamefont {{Smith}}, \citenamefont {{Woodger}}, \citenamefont
  {{Masson}}, \citenamefont {{Narock}}, \citenamefont {{Asamura}},
  \citenamefont {{Chang}}, \citenamefont {{Chiang}}, \citenamefont {{Kazama}},
  \citenamefont {{Keika}}, \citenamefont {{Matsuda}}, \citenamefont {{Segawa}},
  \citenamefont {{Seki}}, \citenamefont {{Shoji}}, \citenamefont {{Tam}},
  \citenamefont {{Umemura}}, \citenamefont {{Wang}}, \citenamefont {{Wang}},
  \citenamefont {{Redmon}}, \citenamefont {{Rodriguez}}, \citenamefont
  {{Singer}}, \citenamefont {{Vandegriff}}, \citenamefont {{Abe}},
  \citenamefont {{Nose}}, \citenamefont {{Shinbori}}, \citenamefont {{Tanaka}},
  \citenamefont {{UeNo}}, \citenamefont {{Andersson}}, \citenamefont {{Dunn}},
  \citenamefont {{Fowler}}, \citenamefont {{Halekas}}, \citenamefont {{Hara}},
  \citenamefont {{Harada}}, \citenamefont {{Lee}}, \citenamefont {{Lillis}},
  \citenamefont {{Mitchell}}, \citenamefont {{Argall}}, \citenamefont
  {{Bromund}}, \citenamefont {{Burch}}, \citenamefont {{Cohen}}, \citenamefont
  {{Galloy}}, \citenamefont {{Giles}}, \citenamefont {{Jaynes}}, \citenamefont
  {{Le Contel}}, \citenamefont {{Oka}}, \citenamefont {{Phan}}, \citenamefont
  {{Walsh}}, \citenamefont {{Westlake}}, \citenamefont {{Wilder}},
  \citenamefont {{Bale}}, \citenamefont {{Livi}}, \citenamefont {{Pulupa}},
  \citenamefont {{Whittlesey}}, \citenamefont {{DeWolfe}}, \citenamefont
  {{Harter}}, \citenamefont {{Lucas}}, \citenamefont {{Auster}}, \citenamefont
  {{Bonnell}}, \citenamefont {{Cully}}, \citenamefont {{Donovan}},
  \citenamefont {{Ergun}}, \citenamefont {{Frey}}, \citenamefont {{Jackel}},
  \citenamefont {{Keiling}}, \citenamefont {{Korth}}, \citenamefont
  {{McFadden}}, \citenamefont {{Nishimura}}, \citenamefont {{Plaschke}},
  \citenamefont {{Robert}}, \citenamefont {{Turner}}, \citenamefont {{Weygand
  }}, \citenamefont {{Candey}}, \citenamefont {{Johnson}}, \citenamefont
  {{Kovalick}}, \citenamefont {{Liu}}, \citenamefont {{McGuire}}, \citenamefont
  {{Breneman}}, \citenamefont {{Kersten}},\ and\ \citenamefont
  {{Schroeder}}}]{spedas}%
  \BibitemOpen
  \bibfield  {author} {\bibinfo {author} {\bibfnamefont {V.}~\bibnamefont
  {{Angelopoulos}}}, \bibinfo {author} {\bibfnamefont {P.}~\bibnamefont
  {{Cruce}}}, \bibinfo {author} {\bibfnamefont {A.}~\bibnamefont {{Drozdov}}},
  \bibinfo {author} {\bibfnamefont {E.~W.}\ \bibnamefont {{Grimes}}}, \bibinfo
  {author} {\bibfnamefont {N.}~\bibnamefont {{Hatzigeorgiu}}}, \bibinfo
  {author} {\bibfnamefont {D.~A.}\ \bibnamefont {{King}}}, \bibinfo {author}
  {\bibfnamefont {D.}~\bibnamefont {{Larson}}}, \bibinfo {author}
  {\bibfnamefont {J.~W.}\ \bibnamefont {{Lewis}}}, \bibinfo {author}
  {\bibfnamefont {J.~M.}\ \bibnamefont {{McTiernan}}}, \bibinfo {author}
  {\bibfnamefont {D.~A.}\ \bibnamefont {{Roberts}}}, \bibinfo {author}
  {\bibfnamefont {C.~L.}\ \bibnamefont {{Russell}}}, \bibinfo {author}
  {\bibfnamefont {T.}~\bibnamefont {{Hori}}}, \bibinfo {author} {\bibfnamefont
  {Y.}~\bibnamefont {{Kasahara}}}, \bibinfo {author} {\bibfnamefont
  {A.}~\bibnamefont {{Kumamoto}}}, \bibinfo {author} {\bibfnamefont
  {A.}~\bibnamefont {{Matsuoka}}}, \bibinfo {author} {\bibfnamefont
  {Y.}~\bibnamefont {{Miyashita}}}, \bibinfo {author} {\bibfnamefont
  {Y.}~\bibnamefont {{Miyoshi}}}, \bibinfo {author} {\bibfnamefont
  {I.}~\bibnamefont {{Shinohara}}}, \bibinfo {author} {\bibfnamefont
  {M.}~\bibnamefont {{Teramoto}}}, \bibinfo {author} {\bibfnamefont {J.~B.}\
  \bibnamefont {{Faden}}}, \bibinfo {author} {\bibfnamefont {A.~J.}\
  \bibnamefont {{Halford}}}, \bibinfo {author} {\bibfnamefont {M.}~\bibnamefont
  {{McCarthy}}}, \bibinfo {author} {\bibfnamefont {R.~M.}\ \bibnamefont
  {{Millan}}}, \bibinfo {author} {\bibfnamefont {J.~G.}\ \bibnamefont
  {{Sample}}}, \bibinfo {author} {\bibfnamefont {D.~M.}\ \bibnamefont
  {{Smith}}}, \bibinfo {author} {\bibfnamefont {L.~A.}\ \bibnamefont
  {{Woodger}}}, \bibinfo {author} {\bibfnamefont {A.}~\bibnamefont {{Masson}}},
  \bibinfo {author} {\bibfnamefont {A.~A.}\ \bibnamefont {{Narock}}}, \bibinfo
  {author} {\bibfnamefont {K.}~\bibnamefont {{Asamura}}}, \bibinfo {author}
  {\bibfnamefont {T.~F.}\ \bibnamefont {{Chang}}}, \bibinfo {author}
  {\bibfnamefont {C.~Y.}\ \bibnamefont {{Chiang}}}, \bibinfo {author}
  {\bibfnamefont {Y.}~\bibnamefont {{Kazama}}}, \bibinfo {author}
  {\bibfnamefont {K.}~\bibnamefont {{Keika}}}, \bibinfo {author} {\bibfnamefont
  {S.}~\bibnamefont {{Matsuda}}}, \bibinfo {author} {\bibfnamefont
  {T.}~\bibnamefont {{Segawa}}}, \bibinfo {author} {\bibfnamefont
  {K.}~\bibnamefont {{Seki}}}, \bibinfo {author} {\bibfnamefont
  {M.}~\bibnamefont {{Shoji}}}, \bibinfo {author} {\bibfnamefont {S.~W.~Y.}\
  \bibnamefont {{Tam}}}, \bibinfo {author} {\bibfnamefont {N.}~\bibnamefont
  {{Umemura}}}, \bibinfo {author} {\bibfnamefont {B.~J.}\ \bibnamefont
  {{Wang}}}, \bibinfo {author} {\bibfnamefont {S.~Y.}\ \bibnamefont {{Wang}}},
  \bibinfo {author} {\bibfnamefont {R.}~\bibnamefont {{Redmon}}}, \bibinfo
  {author} {\bibfnamefont {J.~V.}\ \bibnamefont {{Rodriguez}}}, \bibinfo
  {author} {\bibfnamefont {H.~J.}\ \bibnamefont {{Singer}}}, \bibinfo {author}
  {\bibfnamefont {J.}~\bibnamefont {{Vandegriff}}}, \bibinfo {author}
  {\bibfnamefont {S.}~\bibnamefont {{Abe}}}, \bibinfo {author} {\bibfnamefont
  {M.}~\bibnamefont {{Nose}}}, \bibinfo {author} {\bibfnamefont
  {A.}~\bibnamefont {{Shinbori}}}, \bibinfo {author} {\bibfnamefont {Y.~M.}\
  \bibnamefont {{Tanaka}}}, \bibinfo {author} {\bibfnamefont {S.}~\bibnamefont
  {{UeNo}}}, \bibinfo {author} {\bibfnamefont {L.}~\bibnamefont {{Andersson}}},
  \bibinfo {author} {\bibfnamefont {P.}~\bibnamefont {{Dunn}}}, \bibinfo
  {author} {\bibfnamefont {C.}~\bibnamefont {{Fowler}}}, \bibinfo {author}
  {\bibfnamefont {J.~S.}\ \bibnamefont {{Halekas}}}, \bibinfo {author}
  {\bibfnamefont {T.}~\bibnamefont {{Hara}}}, \bibinfo {author} {\bibfnamefont
  {Y.}~\bibnamefont {{Harada}}}, \bibinfo {author} {\bibfnamefont {C.~O.}\
  \bibnamefont {{Lee}}}, \bibinfo {author} {\bibfnamefont {R.}~\bibnamefont
  {{Lillis}}}, \bibinfo {author} {\bibfnamefont {D.~L.}\ \bibnamefont
  {{Mitchell}}}, \bibinfo {author} {\bibfnamefont {M.~R.}\ \bibnamefont
  {{Argall}}}, \bibinfo {author} {\bibfnamefont {K.}~\bibnamefont {{Bromund}}},
  \bibinfo {author} {\bibfnamefont {J.~L.}\ \bibnamefont {{Burch}}}, \bibinfo
  {author} {\bibfnamefont {I.~J.}\ \bibnamefont {{Cohen}}}, \bibinfo {author}
  {\bibfnamefont {M.}~\bibnamefont {{Galloy}}}, \bibinfo {author}
  {\bibfnamefont {B.}~\bibnamefont {{Giles}}}, \bibinfo {author} {\bibfnamefont
  {A.~N.}\ \bibnamefont {{Jaynes}}}, \bibinfo {author} {\bibfnamefont
  {O.}~\bibnamefont {{Le Contel}}}, \bibinfo {author} {\bibfnamefont
  {M.}~\bibnamefont {{Oka}}}, \bibinfo {author} {\bibfnamefont {T.~D.}\
  \bibnamefont {{Phan}}}, \bibinfo {author} {\bibfnamefont {B.~M.}\
  \bibnamefont {{Walsh}}}, \bibinfo {author} {\bibfnamefont {J.}~\bibnamefont
  {{Westlake}}}, \bibinfo {author} {\bibfnamefont {F.~D.}\ \bibnamefont
  {{Wilder}}}, \bibinfo {author} {\bibfnamefont {S.~D.}\ \bibnamefont
  {{Bale}}}, \bibinfo {author} {\bibfnamefont {R.}~\bibnamefont {{Livi}}},
  \bibinfo {author} {\bibfnamefont {M.}~\bibnamefont {{Pulupa}}}, \bibinfo
  {author} {\bibfnamefont {P.}~\bibnamefont {{Whittlesey}}}, \bibinfo {author}
  {\bibfnamefont {A.}~\bibnamefont {{DeWolfe}}}, \bibinfo {author}
  {\bibfnamefont {B.}~\bibnamefont {{Harter}}}, \bibinfo {author}
  {\bibfnamefont {E.}~\bibnamefont {{Lucas}}}, \bibinfo {author} {\bibfnamefont
  {U.}~\bibnamefont {{Auster}}}, \bibinfo {author} {\bibfnamefont {J.~W.}\
  \bibnamefont {{Bonnell}}}, \bibinfo {author} {\bibfnamefont {C.~M.}\
  \bibnamefont {{Cully}}}, \bibinfo {author} {\bibfnamefont {E.}~\bibnamefont
  {{Donovan}}}, \bibinfo {author} {\bibfnamefont {R.~E.}\ \bibnamefont
  {{Ergun}}}, \bibinfo {author} {\bibfnamefont {H.~U.}\ \bibnamefont {{Frey}}},
  \bibinfo {author} {\bibfnamefont {B.}~\bibnamefont {{Jackel}}}, \bibinfo
  {author} {\bibfnamefont {A.}~\bibnamefont {{Keiling}}}, \bibinfo {author}
  {\bibfnamefont {H.}~\bibnamefont {{Korth}}}, \bibinfo {author} {\bibfnamefont
  {J.~P.}\ \bibnamefont {{McFadden}}}, \bibinfo {author} {\bibfnamefont
  {Y.}~\bibnamefont {{Nishimura}}}, \bibinfo {author} {\bibfnamefont
  {F.}~\bibnamefont {{Plaschke}}}, \bibinfo {author} {\bibfnamefont
  {P.}~\bibnamefont {{Robert}}}, \bibinfo {author} {\bibfnamefont {D.~L.}\
  \bibnamefont {{Turner}}}, \bibinfo {author} {\bibfnamefont {J.~M.}\
  \bibnamefont {{Weygand }}}, \bibinfo {author} {\bibfnamefont {R.~M.}\
  \bibnamefont {{Candey}}}, \bibinfo {author} {\bibfnamefont {R.~C.}\
  \bibnamefont {{Johnson}}}, \bibinfo {author} {\bibfnamefont {T.}~\bibnamefont
  {{Kovalick}}}, \bibinfo {author} {\bibfnamefont {M.~H.}\ \bibnamefont
  {{Liu}}}, \bibinfo {author} {\bibfnamefont {R.~E.}\ \bibnamefont
  {{McGuire}}}, \bibinfo {author} {\bibfnamefont {A.}~\bibnamefont
  {{Breneman}}}, \bibinfo {author} {\bibfnamefont {K.}~\bibnamefont
  {{Kersten}}}, \ and\ \bibinfo {author} {\bibfnamefont {P.}~\bibnamefont
  {{Schroeder}}},\ }\bibfield  {title} {\enquote {\bibinfo {title} {{The Space
  Physics Environment Data Analysis System (SPEDAS)}},}\ }\href {\doibase
  10.1007/s11214-018-0576-4} {\bibfield  {journal} {\bibinfo  {journal} {Spa.
  Sci. Rev.}\ }\textbf {\bibinfo {volume} {215}},\ \bibinfo {eid} {9} (\bibinfo
  {year} {2019})}\BibitemShut {NoStop}%
\end{thebibliography}%

\end{document}